# Scheduling in Cloud–Fog–Edge Computing: A Comparative Analysis and Decision Framework for Technique Selection

Rana Ghazali ✉, Douglas G.Down

Ghazalir@mcmaster.ca, Downd@mcmaster.ca

Department of Computing and Software, McMaster University, 1280 Main St W, Hamilton, ON, Canada

## Abstract

The increasing heterogeneity and dynamism of cloud, fog, and edge computing make scheduling a challenging problem involving diverse resource capabilities, workload characteristics, latency and energy constraints, and competing optimization objectives. Although numerous scheduling methods have been proposed, existing reviews often provide fragmented coverage of computing environments, scheduling granularities, or technique families, offering limited guidance on selecting an appropriate approach for specific requirements. This review presents a comparative analysis of task, job, and workflow scheduling across the cloud–fog–edge paradigms, covering heuristic, metaheuristic, learning-based, and hybrid techniques. Studies published between 2020 and 2026 are analyzed according to scheduling granularity, computing environment, optimization objectives, constraints, performance metrics, and evaluation tools. The analysis highlights trade-offs among computational efficiency, optimization capability, adaptability, and implementation complexity across technique families. Heuristic methods offer low computational overhead for relatively stable scheduling conditions, whereas metaheuristic methods are suited to complex and multi-objective optimization problems. Learning-based methods provide greater adaptability to dynamic and uncertain environments but introduce data, training, inference, and convergence requirements. Hybrid methods combine complementary capabilities but generally increase computational and implementation complexity. Based on these findings, a decision framework is developed to guide technique selection according to workload characteristics, environmental conditions, optimization objectives, resource constraints, and computational requirements. The review also identifies research gaps related to real-world validation, reliability and QoS, workload dynamics, scalability, and the interpretability of intelligent scheduling methods. The resulting framework provides a structured basis for researchers and practitioners to compare scheduling approaches and select techniques appropriate to specific cloud–fog–edge requirements.

## 1. Introduction

The rapid evolution of cloud, fog, and edge computing has transformed distributed computing from predominantly centralized resource provisioning toward highly heterogeneous and geographically distributed infrastructures. Cloud computing provides abundant computational and storage resources, whereas fog and edge computing extend computation closer to data sources and end users to reduce communication latency and network congestion. This computing continuum supports applications such as Internet of Things (IoT), smart cities, industrial automation, autonomous systems, healthcare, and real-time analytics. However, the increasing heterogeneity, decentralization, and dynamism of these environments make efficient scheduling increasingly challenging. Scheduling decisions must account for heterogeneous resources, workload characteristics, communication delays, execution time, energy consumption, cost, quality of service (QoS), deadlines, and task dependencies.

Scheduling can be performed at different levels of computational granularity. Task scheduling generally concerns the assignment of individual tasks to available resources, whereas job scheduling considers collections of computational requests and workflow scheduling additionally incorporates precedence relationships, data dependencies, and communication costs. These differences directly affect the suitability of scheduling techniques. An approach that is effective for independent tasks may be less appropriate for dependency-constrained workflows or large-scale jobs. Therefore, scheduling techniques should be evaluated not only according to their optimization performance but also according to the characteristics and requirements of the scheduling problem.

A broad range of scheduling approaches has been proposed, which can be generally grouped into heuristic, metaheuristic, learning-based, and hybrid techniques. These families represent different trade-offs between computational overhead, optimization capability, adaptability, scalability, and implementation complexity. Heuristics typically provide rapid decisions with relatively low computational overhead but may have limited search capability. Metaheuristics can address complex and multi-objective search spaces but may involve higher computational cost and convergence or parameter-sensitivity issues. Learning-based approaches can adapt to dynamic and uncertain environments but introduce requirements related to training data, learning stability, inference, and computational resources. Hybrid approaches combine complementary mechanisms but generally increase algorithmic and implementation complexity. Consequently, the increasing sophistication of a scheduling technique does not necessarily imply greater suitability for a particular application or environment.

Despite the large body of scheduling research, the existing review literature provides limited guidance on how to select an appropriate scheduling technique according to problem requirements**.** To examine the state of the art, this study analyzes 28 survey articles published between 2020 and 2026 and identifies several limitations in the existing literature. First, many reviews focus on a particular computing paradigm or a subset of the cloud–fog–edge continuum rather than examining scheduling across these environments from a unified perspective. Second, task, job, and workflow scheduling are frequently treated separately, making it difficult to compare their requirements and algorithmic implications. Third, existing reviews often organize literature around algorithms or technique families without systematically examining the conditions under which one family may be more appropriate than another. Finally, although existing surveys commonly report algorithms, objectives, and performance metrics, comparatively less attention is given to the relationship between scheduling requirements and technique selection**.**

To address these limitations, this review analyzes 183 primary studies, comprising 12 heuristic, 105 metaheuristic, 22 learning-based, and 44 hybrid studies, together with the 28 existing surveys used for comparison. Rather than providing another algorithm-by-algorithm catalogue, the study develops a comparative and requirement-oriented analysis of scheduling techniques across cloud, fog, and edge environments. The analysis considers the relationships among scheduling granularity, computing environment, technique family, optimization objectives, constraints, evaluation methodology, and practical deployment requirements.

The central objective is to answer a practical question that remains insufficiently addressed in the existing literature: under what conditions should a heuristic, metaheuristic, learning-based, or hybrid technique be considered for a particular scheduling problem? To this end, the review develops a decision framework for technique selection based on factors including scheduling entity, computing environment, workload characteristics, optimization objectives, resource constraints, system dynamics, and computational budget. The framework does not assume that one technique family is universally superior; instead, it identifies the principal trade-offs that should be considered when selecting a scheduling strategy. This perspective shifts the focus from simply identifying what techniques have been proposed toward understanding why and when different techniques are appropriate.

Accordingly, the review addresses the following research questions:

**RQ1.** What are the principal limitations and research gaps in recent surveys of scheduling in cloud, fog, and edge computing?

**RQ2.** How can scheduling techniques be classified and compared across task, job, and workflow scheduling and across cloud, fog, and edge environments?

**RQ3.** What are the major scheduling objectives, constraints, performance metrics, and evaluation practices associated with different technique families?

**RQ4.** What are the principal characteristics, advantages, limitations, and computational trade-offs of heuristic, metaheuristic, learning-based, and hybrid scheduling techniques?

**RQ5.** Under what workload, environmental, optimization, and computational conditions are different scheduling technique families appropriate?

**RQ6.** What unresolved challenges and research opportunities remain for developing scalable, adaptive, efficient, and practically deployable scheduling strategies?

The main contributions of this review are therefore as follows:

**Unified comparative perspective:** It provides a common analytical framework for comparing heuristic, metaheuristic, learning-based, and hybrid scheduling techniques across cloud, fog, and edge environments.

**Multi-level scheduling analysis:** It jointly considers task, job, and workflow scheduling and examines how scheduling granularity affects technique requirements and selection.

**Requirement-oriented technique selection:** It develops a decision framework that relates scheduling requirements, workload characteristics, environmental conditions, optimization objectives, and computational constraints to candidate technique families.

**Evidence-based synthesis:** It analyzes 28 recent surveys and 183 primary studies to identify research trends, dominant objectives, evaluation practices, and underexplored areas.
**Trade-off analysis:** It systematically examines the relationships among optimization capability, adaptability, computational overhead, scalability, convergence, and implementation complexity across the four technique families.
**Research gaps and future directions:** It identifies persistent challenges and opportunities, particularly regarding dynamic workloads, multi-objective optimization, realistic evaluation, scalability, reliability, QoS, and the practical deployment of learning-based and hybrid approaches.

The remainder of this paper is organized as follows. Section 2 presents the review methodology, including the search strategy, selection criteria, and taxonomy-development process. Section 3 provides a comprehensive comparison with existing surveys and highlights the distinctive contributions of the present review**.** Section 4 introduces cloud, fog, and edge computing paradigms and discusses their associated scheduling challenges. Section 5 analyzes scheduling requirements, optimization objectives, and constraints. Section 6 presents the taxonomy and comparative analysis of heuristic, metaheuristic, learning-based, and hybrid scheduling techniques with synthesizes statistical trends and provides cross-technique comparisons and practical selection guidelines. Section 7 presents a comparative analysis of simulation platforms for scheduling evaluation**.** Finally, Section 8 discusses the principal findings, research gaps, and future research directions.

## 2. Research methodology

This study adopts a systematic and comparative review methodology to analyze scheduling techniques across the cloud–fog–edge computing continuum. The methodology is designed to support the central objective of the review: examining how computing environments, scheduling requirements, and workload characteristics relate to the suitability of different scheduling technique families. Accordingly, the review jointly considers task, job, and workflow scheduling and classifies primary studies into four technique families: heuristic, metaheuristic, learning-based, and hybrid approaches**.**

The review covers studies published between 2020 and 2026. Existing surveys were first examined to establish the research landscape and identify limitations in previous reviews. Subsequently, primary studies were selected and analyzed using a common set of dimensions, including computing environment, scheduling granularity, technique family, objectives, constraints, workload characteristics, and evaluation approach. These dimensions provide the basis for a comparative synthesis rather than a study-by-study description of individual algorithms.

The methodology therefore follows three analytical stages:
(1) literature identification and selection → (2) comparative classification and synthesis → (3) technique-selection framework.

The first stage establishes the review corpus; the second identifies relationships and trade-offs among environments, scheduling requirements, and technique families; and the third translates these findings into decision guidance for selecting appropriate technique families under different scheduling conditions.

### 2.1. Literature Search and Study Selection

A structured literature search was conducted using major scientific databases, including IEEE Xplore, SpringerLink, ScienceDirect, Press, MDPI, ACM, and Wiley Online Library. The search strategy combined terms representing the three principal dimensions of the review: computing environment, scheduling entity, and scheduling technique. The core search concept was:
(Cloud OR Fog OR Edge) AND (Scheduling OR Resource Allocation) AND (Task OR Job OR Workflow) AND (Heuristic OR Metaheuristic OR Learning OR Hybrid)

Additional terminology, including reinforcement learning, deep learning, evolutionary algorithms, swarm intelligence, and intelligent scheduling, was considered where necessary to capture studies using alternative terminology for learning-based and optimization-based scheduling.

The search was limited to studies published during 2020–2026. Relevant survey papers were also examined to identify additional studies through backward reference searching. After removing duplicates and applying the eligibility criteria, the selected primary studies formed the corpus for the comparative analysis.

Studies were included when they satisfied the following criteria:

- published between 2020 and 2026;
- addressed cloud, fog, edge, or integrated cloud–fog–edge environments;
- investigated task, job, or workflow scheduling or closely related resource-allocation problems;

- proposed or evaluated a heuristic, metaheuristic, learning-based, or hybrid technique;
- provided sufficient methodological information to identify the scheduling approach and its operating context; and
- included simulation-based or experimental evaluation.

Studies were excluded when they:

- were published outside the defined period;
- did not address scheduling or resource allocation;
- concerned computing environments outside the scope of this review;
- were duplicate or substantially redundant publications;
- lacked sufficient methodological or evaluation information;
- were purely conceptual without an identifiable scheduling approach or evaluation; or
- could not be consistently assigned to one of the four technique families defined in this review.

Following this process, 28 representative survey articles were used to characterize the existing review landscape, while 183 primary studies were retained for comparative analysis. The primary studies comprise 12 heuristic, 105 metaheuristic, 22 learning-based, and 44 hybrid approaches. These studies constitute the evidence base for quantitative synthesis and the subsequent decision framework for technique selection.

## 3. Comparison with existing surveys

Scheduling in cloud, fog, and edge computing has been extensively surveyed, but existing reviews differ in computing coverage, scheduling granularity, technique classification, and analytical objectives. To position the present study, 28 representative surveys published between 2020 and 2026 were examined [1–28]. Cloud-oriented studies [1–10] mainly address task or job scheduling; fog-oriented studies [11–19] focus largely on task scheduling in resource-constrained environments; cloud–fog/cloud–edge studies [20–22] consider selected combinations of computing layers; and edge-oriented studies [23–27] increasingly examine optimization and learning-based approaches. The recent integrated survey [28] considers task scheduling and load balancing across the cloud–fog–edge continuum.

**Table 1: Comparison of existing survey articles**

| References | Taxonomy | State of the art | Comparison based on the year | Comparison based on performance metrics | Comparison based on constraints | Statistical information | Simulation tools | Future trends | Guideline |
|---|---|---|---|---|---|---|---|---|---|
| [1] | ✓ | ✓ | - | ✓ | ✓ | ✓ | ✓ | ✓ | - |
| [2] | ✓ | ✓ | ✓ | - | - | - | ✓ | ✓ | - |
| [3] | ✓ | ✓ | - | - | - | - | ✓ | - | - |
| [4] | - | - | ✓ | - | - | - | ✓ | - | - |
| [5] | ✓ | - | ✓ | ✓ | - | ✓ | - | - | - |
| [6] | - | - | - | - | - | - | - | ✓ | - |
| [7] | ✓ | ✓ | - | ✓ | ✓ | ✓ | ✓ | ✓ | - |
| {8] | ✓ | ✓ | ✓ | ✓ | ✓ | ✓ | - | ✓ | - |
| [9] | ✓ | ✓ | ✓ | ✓ | - | ✓ | - | ✓ | - |
| [10] | ✓ | ✓ | ✓ | ✓ | - | - | - | ✓ | - |
| [11] | ✓ | ✓ | ✓ | ✓ | - | ✓ | ✓ | ✓ | - |
| [12] | - | - | ✓ | ✓ | - | ✓ | ✓ | - | - |
| [13] | ✓ | ✓ | - | ✓ | - | ✓ | ✓ | ✓ | - |
| [14] | - | ✓ | - | - | - | - | - | ✓ | - |
| [15] | ✓ | ✓ | - | ✓ | - | ✓ | ✓ | ✓ | - |
| [16] | ✓ | - | - | ✓ | - | - | - | - | - |
| [17] | ✓ | ✓ | - | - | - | ✓ | - | ✓ | - |
| [18] | ✓ | ✓ | - | ✓ | - | ✓ | ✓ | ✓ | - |
| [19] | ✓ | ✓ | ✓ | ✓ | ✓ | ✓ | - | ✓ | - |
| [20] | ✓ | ✓ | - | ✓ | - | - | - | - | - |
| [21] | ✓ | ✓ | ✓ | ✓ | - | ✓ | - | - | - |
| [22] | - | ✓ | - | - | - | - | - | ✓ | - |
| [23] | ✓ | ✓ | - | ✓ | - | - | ✓ | - | - |
| [24] | ✓ | ✓ | - | - | - | - | - | ✓ | - |
| [25] | ✓ | ✓ | - | - | - | - | - | ✓ | - |
| [26] | ✓ | ✓ | - | - | - | - | - | - | - |
| [27] | - | ✓ | - | ✓ | - | - | ✓ | ✓ | - |
| [28] | ✓ | ✓ | - | ✓ | - | - | ✓ | ✓ | - |
| Proposed survey | ✓ | ✓ | ✓ | ✓ | ✓ | ✓ | ✓ | ✓ | ✓ |

Three limitations are particularly relevant to the present study. First, most reviews focus on a single scheduling granularity, especially task scheduling, rather than jointly considering task, job, and workflow scheduling. Second, many surveys are organized around a specific computing paradigm or technique family, limiting cross-environment and cross-technique comparison. Third, although objectives, performance metrics, constraints, and simulation tools are frequently reported, their relationships with workload characteristics and technique suitability are less consistently analyzed.

The present review therefore adopts a decision-oriented comparative perspective. Four technique families like heuristic, metaheuristic, learning-based, and hybrid are compared across the cloud–fog–edge continuum and across task, job, and workflow scheduling. Rather than identifying a universally superior technique, the analysis examines how computing environment, workload dynamics, objectives, constraints, scalability, adaptability, and computational overhead influence technique suitability. Table 1 presents a comparative evaluation based on performance metrics, scheduling constraints, simulation tools, future research directions, and the availability of practical guidelines.

The distinction from the recent integrated survey [28] is primarily analytical: while it emphasizes task scheduling and load balancing, this review extends the analysis to task, job, and workflow scheduling and explicitly connects scheduling requirements and environmental characteristics with technique selection.

Table 2 summarizes the main differences between representative existing surveys and the present review in terms of computing environment, scheduling granularity, technique coverage, analytical dimensions, statistical synthesis, and technique-selection guidance. As shown in Table 2, the present review adopts a broader decision-oriented perspective by jointly analyzing scheduling granularity, computing environment, technique families, and system requirements.

**Table 2. Comparison of the present review with representative existing surveys**

| Review dimension | Existing survey literature [1-28] | Present review |
|---|---|---|
| **Computing environment** | Cloud, fog, edge, or selected combinations | Unified cloud–fog–edge continuum |
| **Scheduling scope** | Often task-, job-, workflow-, or load-balancing-specific | Task, job, and workflow scheduling |
| **Technique coverage** | Usually, selected technique families | Heuristic, metaheuristic, learning-based, and hybrid |
| **Cross-technique comparison** | Generally limited to specific families or contexts | Systematic comparison across technique families |
| **Objectives and constraints** | Unevenly analyzed across surveys | Integrated analysis of objectives, constraints, and system characteristics |
| **Workload/system dynamics** | Mainly described as environmental characteristics | Considered in assessing technique applicability |
| **Quantitative synthesis** | Provided by some surveys, with limited cross-study synthesis | 183 primary studies systematically synthesized |
| **Technique-selection guidance** | Generally limited or qualitative | Decision framework linking problem characteristics to technique families |

## 4. Computing Paradigms from a Scheduling Perspective

Cloud, fog, and edge computing form a continuum of computational resources with substantially different scheduling conditions. Rather than revisiting their general architectural characteristics, this section focuses on the properties that directly influence scheduling decisions and technique selection. The main differentiating factors are resource capacity, proximity to users and data sources, resource heterogeneity, network conditions, system dynamics, and the degree of centralized control.

Cloud computing provides relatively abundant and centralized computational and storage resources, making it suitable for large-scale and compute-intensive workloads. Scheduling can therefore exploit broader resource availability and centralized information, with common objectives including makespan, execution time, cost, energy consumption, and resource utilization. However, large-scale workloads and increasing resource heterogeneity can produce complex optimization spaces, particularly when multiple objectives must be optimized simultaneously. Consequently, the availability of computational resources does not eliminate the need for efficient scheduling; rather, it changes the balance between optimization quality, scalability, and computational overhead.

Fog and edge computing move computation closer to data sources and end users, creating scheduling conditions that differ substantially from centralized cloud environments. Lower latency and reduced data transfer are advantageous for time-sensitive applications, but resources are typically more constrained, heterogeneous, and geographically distributed. Mobility, intermittent connectivity, rapidly changing workloads, and limited local computational capacity can further increase scheduling uncertainty. These characteristics increase the importance of low-overhead, latency-aware, and adaptive scheduling decisions, particularly when decisions must be made close to the data source.

The integrated cloud–fog–edge environment combines these characteristics and introduces additional coordination requirements. A scheduler may need to determine not only which resource should execute a task but also where computation should occur across the continuum. Task placement, offloading, migration, communication cost, latency, energy consumption, and resource availability can therefore become interdependent decisions. The resulting scheduling problem is often multi-objective and dynamic, requiring a balance between global optimization and local responsiveness. These conditions create different opportunities and constraints for heuristic, metaheuristic, learning-based, and hybrid techniques.

From a technique-selection perspective, the computing paradigm should therefore be considered together with the characteristics of the scheduling problem. Centralized and relatively stable environments may favor techniques that exploit global information with limited adaptation requirements, whereas resource-constrained and dynamic environments increase the value of rapid and adaptive decision-making. Similarly, heterogeneous multi-layer environments may benefit from techniques capable of handling large search spaces, multiple objectives, or changing system states. Thus, the computing paradigm does not determine the appropriate scheduling technique; rather, it establishes a set of environmental conditions that must be considered jointly with workload, objectives, constraints, and computational resources. Table 3 summarizes the principal scheduling implications of the cloud–fog–edge continuum and provides the basis for the requirement-oriented analysis in the following section. The characteristics summarized above show that scheduling technique selection cannot be based solely on algorithmic performance. The appropriate technique depends on the interaction between the computing environment, workload characteristics, optimization objectives, system constraints, and available computational resources. These requirements are examined systematically in the next section.

**Table 3. Scheduling characteristics of cloud–fog–edge computing paradigms**

| Computing paradigm | Key scheduling characteristics | Dominant scheduling concerns | Implications for technique selection |
|---|---|---|---|
| **Cloud** | High resource capacity; centralized control; relatively stable resource availability | Makespan, cost, energy, utilization, and scalability | Efficient optimization and scalability are important. |
| **Fog** | Distributed intermediate resources; heterogeneous nodes; closer to users | Latency, energy, resource utilization, and coordination | Low-overhead and adaptive decisions become more important. |
| **Edge** | Resource-constrained and highly distributed devices; proximity to data sources | Latency, responsiveness, energy, and mobility | Fast, lightweight, and adaptive techniques are particularly relevant. |
| **Cloud–Fog–Edge** | Multi-layer heterogeneity; distributed resources; variable network conditions | Placement, offloading, latency, migration, energy, and QoS | Multi-objective, scalable, and adaptive techniques may be required. |

## 5. Scheduling Requirements and Optimization Objectives

Scheduling in cloud–fog–edge environments is inherently a multi-objective decision problem because a scheduling strategy must satisfy several performance, resource, economic, and service requirements simultaneously. The relative importance of these requirements depends on the workload, scheduling entity, computing environment, and application constraints. Therefore, optimization objectives should be considered not only as performance measures but also as selection criteria for determining the suitability of different scheduling techniques.

The principal scheduling objectives can be grouped into five categories: performance, economic, energy, QoS, and user-centric requirements. Performance objectives commonly include makespan, execution time, response time, throughput, and resource utilization. Economic objectives concern execution and resource costs, whereas energy-oriented objectives address energy consumption and, increasingly, carbon emissions. QoS-related requirements include reliability, availability, SLA satisfaction, and deadline compliance. In latency-sensitive applications, particularly at fog and edge levels, response time and communication delay may become more important than global computational efficiency. These objectives can conflict; for example, reducing execution time may require additional resources and energy, while minimizing cost may increase latency or execution time.

In addition to optimization objectives, scheduling is subject to system and application constraints. These include resource capacity, task dependencies, deadlines, budget limits, data locality, communication bandwidth, mobility, fault tolerance, and workload variability. Such constraints can determine whether a scheduling solution is feasible before its optimization quality is considered. The combination of multiple objectives and constraints therefore transforms scheduling from a simple resource-allocation problem into a constrained decision problem involving trade-offs among competing requirements.

The requirements also differ according to the characteristics of the computing environment. Cloud environments can generally exploit relatively abundant resources and centralized information, whereas fog and edge environments impose stronger constraints on latency, energy, resource capacity, and responsiveness. Integrated cloud–fog–edge systems additionally require coordination across heterogeneous layers, making task placement, offloading, migration, and resource allocation interdependent decisions. However, these environmental differences should not be interpreted as fixed rules for technique selection. The appropriate scheduling approach depends on the combined characteristics of the environment, workload, objectives, constraints, and available computational resources.

From the perspective of this review, the most important issue is therefore not simply identifying which objectives are frequently optimized but understanding how requirements influence the choice of scheduling technique. Problems dominated by rapid decision requirements may place greater emphasis on computational overhead, whereas large and complex search spaces may justify techniques with stronger global search capabilities. Highly dynamic environments may increase the value of adaptive methods, while resource-constrained systems may favor approaches with lower execution and inference overhead. Multi-objective problems may additionally require mechanisms capable of balancing conflicting criteria rather than optimizing a single metric.

Table 4 summarizes the principal scheduling objectives and constraints considered in literature and their implications for technique selection. In particular, technique selection should be driven by the requirements that a scheduling problem must satisfy rather than by the algorithmic popularity or complexity of a particular method. This requirement-oriented perspective is subsequently used to compare heuristic, metaheuristic, learning-based, and hybrid techniques and to develop the proposed decision framework.

**Table 4. Major scheduling objectives and constraints and their implications for technique selection**

| Category | Representative metrics or constraints | Implication for technique selection |
|---|---|---|
| **Performance** | Makespan, execution time, throughput, resource utilization | Efficient optimization and low scheduling overhead |
| **Economic** | Execution cost, resource cost, budget | Cost-aware resource allocation |
| **Energy** | Energy consumption, power usage, carbon footprint | Energy-aware optimization |
| **Latency** | Response time, communication delay, end-to-end latency | Fast and locality-aware decisions |
| **QoS** | Reliability, availability, SLA satisfaction | Constraint- and service-aware scheduling |
| **Temporal** | Deadline, deadline violation, waiting time | Time-constrained decision-making |
| **Resource** | Capacity, load balance, heterogeneity | Effective resource allocation and scalability |
| **Structural** | Task dependency, precedence, data locality | Dependency- and data-aware scheduling |
| **Dynamic** | Workload variation, resource fluctuation, mobility | Adaptive and context-aware scheduling |
| **Scalability** | Number of tasks, users, and resources | Computationally scalable techniques |
| **Multi-objective** | Conflicting performance, cost, energy, and QoS objectives | Trade-off and Pareto-oriented optimization |

## 6. Taxonomy of Scheduling techniques

Scheduling techniques for cloud–fog–edge computing have evolved from simple rule-based strategies toward optimization, learning, and integrated approaches capable of addressing heterogeneous resources, dynamic workloads, and multiple scheduling objectives. To provide a unified basis for comparing these approaches, this review classifies scheduling techniques into four major families: heuristic, metaheuristic, learning-based, and hybrid techniques, as illustrated in Figure 1. The classification emphasizes the underlying decision mechanism rather than individual algorithms, allowing techniques developed for different computing paradigms and scheduling scenarios to be compared within a common framework.

Heuristic techniques rely primarily on predefined rules and problem-specific decision criteria. Metaheuristic techniques employ systematic search and optimization mechanisms to explore large solution spaces. Learning-based techniques derive scheduling decisions from data, learned representations, or interactions with the computing environment. Hybrid techniques integrate two or more complementary approaches to combine their respective advantages. This taxonomy also considers the major subcategories within each family, including task-priority, resource-aware, and load-balancing heuristics; nature-inspired and non-nature-inspired metaheuristics; supervised, unsupervised, deep learning, and reinforcement learning (RL) approaches; and different combinations of heuristic, metaheuristic, and learning-based methods.

The taxonomy is designed to support comparative analysis and technique selection rather than simply cataloguing algorithms. The suitability of a scheduling family depends on several interacting factors, including the degree of resource heterogeneity, workload dynamics, scheduling objectives, computational budget, availability of historical data, decision latency, and the need for adaptation. Accordingly, the four families and their subcategories are discussed below.

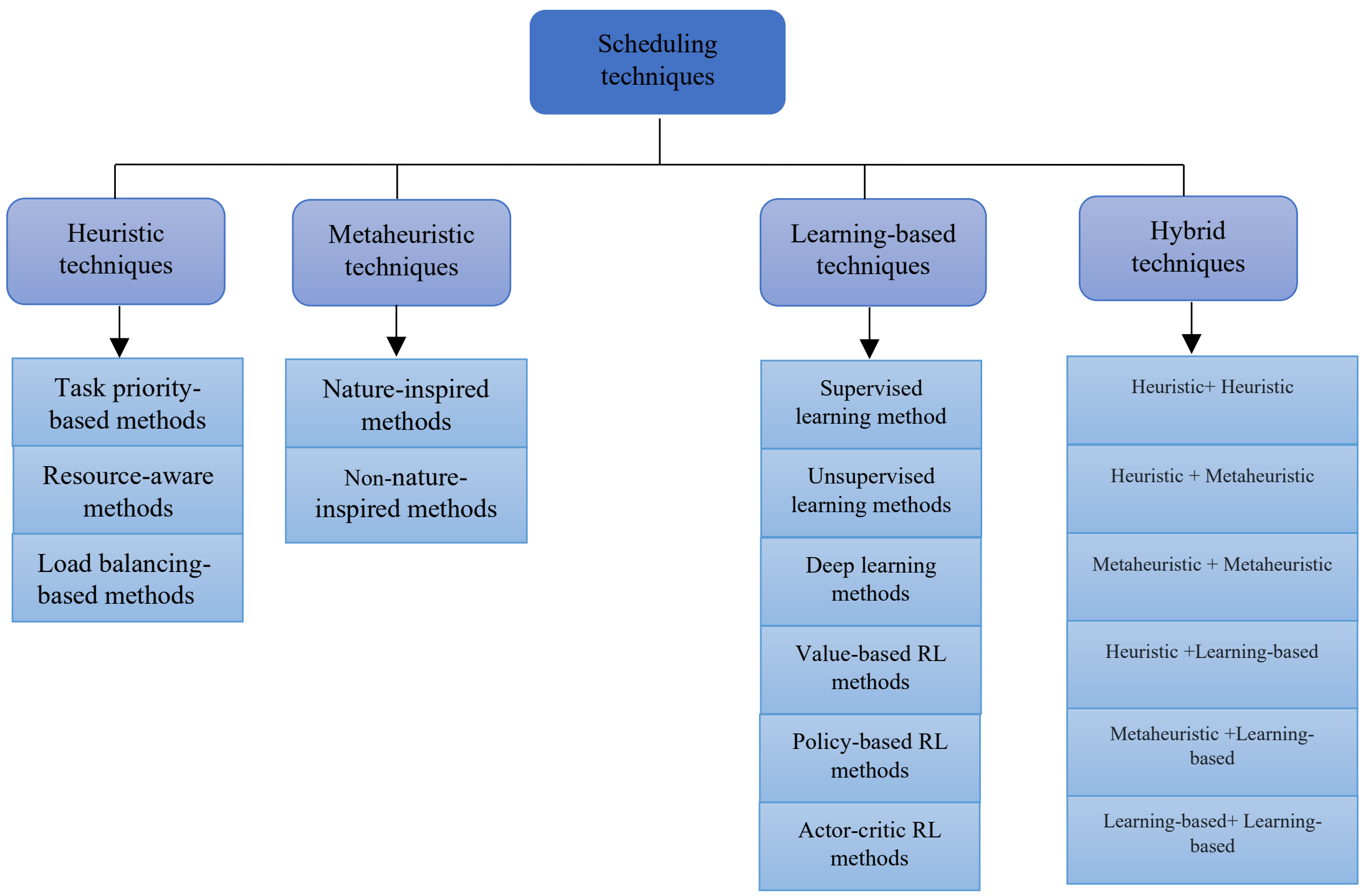

**Figure 1: Classification of scheduling techniques**

### 6.1 Heuristic techniques

Heuristic techniques are problem-specific rule-based strategies that exploit domain knowledge, approximations, and simplified decision mechanisms to obtain near-optimal solutions within polynomial or acceptable computational time. Their main characteristics are low computational overhead, relatively simple implementation, scalability, and rapid decision making, making them attractive for large scale scheduling problems and environments where decisions must be generated quickly or computational resources are limited. However, their effectiveness depends strongly on the assumptions and problem characteristics for which the rules are designed. As a result, conventional heuristics may perform poorly in highly dynamic, heterogeneous, or complex scheduling environments and may fail to simultaneously address multiple conflicting objectives.

#### 6.1.1 Classification of Heuristic Scheduling Techniques

As shown in Figure 2, heuristic scheduling techniques in this study are classified into three major families according to the primary information used to make scheduling decisions: task priority-based, resource-aware, and load-balancing-based methods. Task priority-based methods determine the execution order according to task characteristics such as arrival time, execution time, priority, or deadline. Representative techniques include First-In-First-Out (FIFO),

Shortest Job First (SJF), Longest Job First (LJF), Earliest Deadline First (EDF), Heterogeneous Earliest Finish Time (HEFT), and Priority-Based Fair Scheduling (PBFS). Resource-aware methods incorporate resource availability, processing capability, or estimated completion time into task-to-resource assignment and include Minimum Completion Time (MCT), Minimum Execution Time (MET), Min–Min, Max–Min, and Suffrage. Load-balancing-based methods primarily aim to distribute workloads across available resources and include Round Robin (RR) and Opportunistic Load Balancing (OLB).

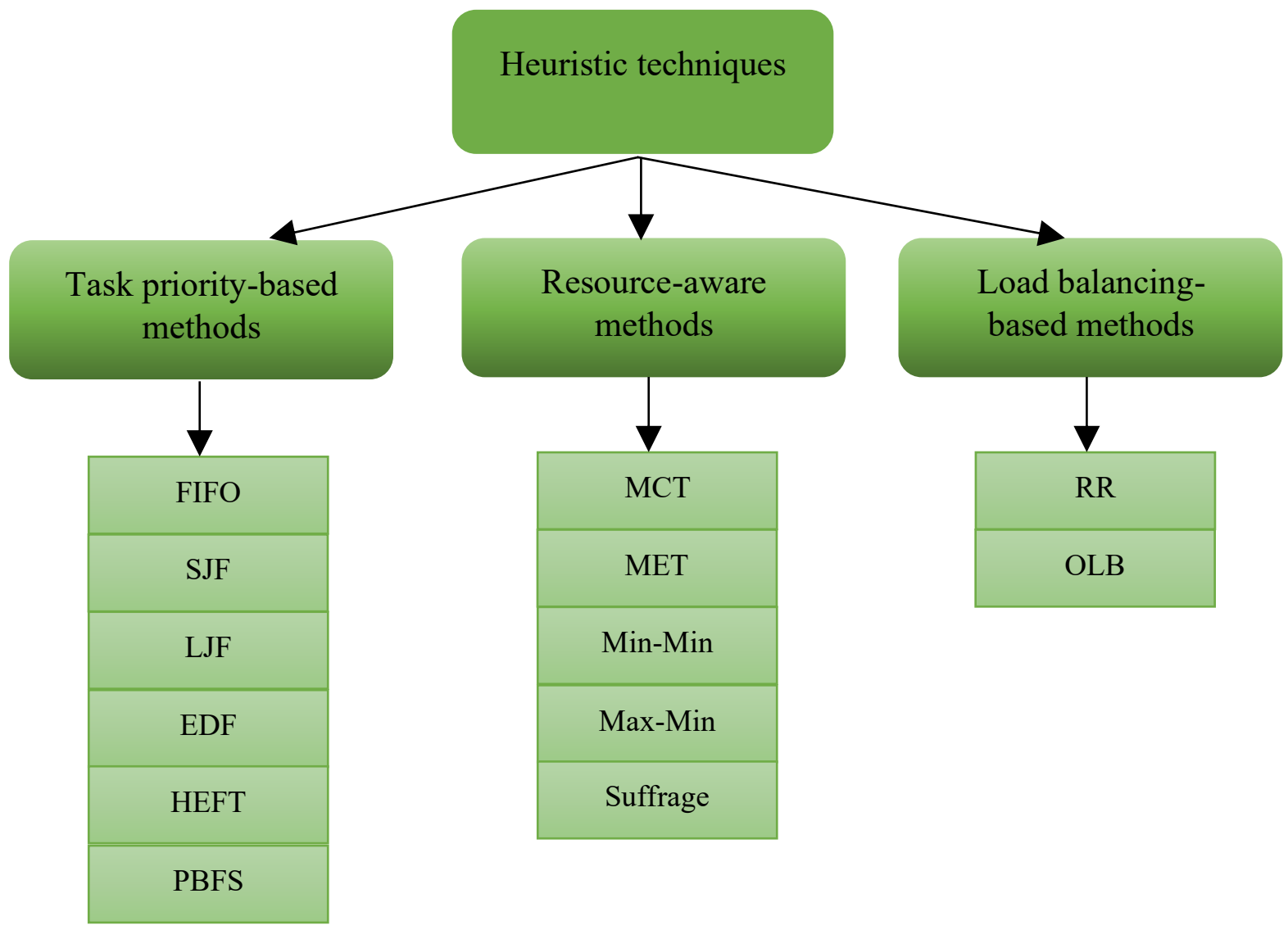

**Figure 2: Classification of heuristic techniques**

#### 6.1.1.1Task priority-based scheduling

Task-priority heuristics determine the scheduling order according to task characteristics such as arrival time, execution time, deadline, fairness requirements, or workflow dependencies. Representative techniques include First In First Out (FIFO)/ First Come First Service (FCFS), Shortest Job First (SJF), Longest Job First (LJF), Earliest Deadline First (EDF), Heterogeneous Earliest Finish Time (HEFT), and Priority-Based Fair Scheduling (PBFS). However, these methods may have limitations such as starvation of low-priority tasks, disregard for resource capabilities, and poor load balancing.

- ***First In First Out (FIFO)*****:** Its resource-aware variant *Resources Then First in First Out (RTFIFO)* [29] prioritize tasks according to queue order while considering resource availability, whereas SJF and LJF use estimated task execution time to determine scheduling order.
- ***Earliest Deadline First (EDF)***: Prioritizes tasks according to their deadlines and is therefore applicable to latency- and deadline-sensitive workloads. Its frequency-aware extension, *Frequency-aware Shortest Remaining Deadline First (FSRDF)* [30], incorporates additional urgency-related information into deadline-based scheduling.
- ***Heterogeneous Earliest Finish Time (HEFT)***: Prioritizes tasks according to precedence constraints and estimated finish times on heterogeneous resources. Several variants extend this principle by incorporating additional scheduling objectives. *Greedy Resource Provisioning and modified HEFT (GRP-HEFT)* [31] integrates resource provisioning and budget constraints, *Enhanced Heterogeneous Earliest Finish time based on Rule (EHEFT-R)* [32] incorporates energy, makespan, and QoS considerations, and *Load Balancing HEFT (LB-HEFT)* [33] introduces load-balancing considerations into workflow scheduling.
- ***Priority-Based Fair Scheduling (PBFS)***: Considers task and resource-related parameters to improve scheduling fairness, while *Earliest Gap Priority-Based Fair Scheduling (EG-PBFS)* [34] exploits available scheduling gaps to reduce waiting time and improve resource utilization.

Overall, task-priority heuristics provide simple and interpretable scheduling decisions and are particularly applicable when task order, deadlines, or workflow dependencies are dominant requirements. Their main limitation is that priority information alone may not adequately capture current resource conditions, workload distribution, or multiple conflicting objectives. Consequently, enhanced variants increasingly combine priority information with resource awareness, cost, energy, or load-balancing mechanisms.

#### 6.1.1.2 Resource-aware scheduling

Resource-aware heuristics explicitly incorporate resource capabilities and estimated execution or completion times into scheduling decisions. These scheduling mechanisms focus on optimizing resource utilization to improve performance and efficiency. Representative methods include Minimum Completion Time (MCT), Minimum Execution Time (MET), Min-Min, Max-Min, and Suffrage. MCT selects resources according to estimated task completion time, whereas MET emphasizes the shortest estimated execution time. Min-Min initially favors tasks with short completion times, while Max-Min gives greater priority to tasks with longer completion times. Suffrage uses the difference between the best and second-best resource completion times to identify tasks for which resource selection is more critical. These approaches can improve resource utilization and completion performance in heterogeneous environments, but their effectiveness depends on the accuracy of execution-time estimates and the availability of up-to-date resource information. They may also exhibit workload imbalance or task starvation under particular task distributions. Recent variants address these limitations by incorporating additional objectives.

- ***Max-Min***: *Multi-Objective Normalization Workflow Scheduling (MONWS)* [35] extends Max-Min by considering task categories and multiple scheduling objectives, including makespan, cost, and resource utilization. The *Advanced Cost-Aware Max-Min Algorithm* [36] incorporates cost considerations into task allocation
- ***Suffrage****: Balanced Candidate Suffrage Value (BCSV)* [37] extends Suffrage by incorporating multiple suffrage-related criteria to improve resource distribution.

Thus, resource-aware heuristics are particularly relevant when resource heterogeneity, execution time, resource utilization, or cost are important scheduling considerations. Their principal trade-off is the additional dependence on resource monitoring and runtime estimation.

#### 6.1.1.3 Load-balancing-based scheduling

Load-balancing heuristics primarily aim to distribute workloads across available resources and prevent resource hotspots. Round Robin (RR) assigns tasks according to a predefined or dynamically adjusted time quantum, while Opportunistic Load Balancing (OLB) assigns tasks to available resources with low scheduling overhead. RR provides a simple mechanism for distributing workloads, but its effectiveness depends on the selection of the time quantum. Large or small quantum values can respectively increase FCFS-like behavior or context-switching overhead.

- ***Round Robin (RR)***: Several adaptive variants modify the time quantum according to workload characteristics. *Dynamic Round-Robin Heuristic Algorithm (DRRHA)* [38] dynamically adjusts the quantum according to task burst times, *Ameliorated Round-Robin Algorithm (ARRA)* [39] combines fixed and dynamic approaches for quantum selection, and *New Round-Robin Task Scheduling Algorithm (NRRTSA)* [40] derives the quantum from characteristics of the largest tasks in the ready queue. These variants improve the adaptability of RR but introduce additional scheduling calculations.

Load-balancing heuristics are therefore useful when workload distribution, scalability, and low scheduling complexity are primary concerns. However, because they generally focus on workload distribution rather than application-specific objectives, they may provide limited support for strict deadlines, energy optimization, cost constraints, or workflow dependencies. Adaptive or distributed load-balancing mechanisms are consequently more appropriate for highly dynamic fog and edge environments.

### 6.1.2 Cross-technique comparison of heuristic scheduling families across computing paradigms

The three heuristic families differ mainly in the information used to make scheduling decisions. Task-priority heuristics emphasize task ordering and are appropriate when deadlines, precedence constraints, or service priorities dominate the scheduling problem. Resource-aware heuristics explicitly consider resource capabilities and estimated execution or completion times and are therefore useful for heterogeneous environments where resource selection strongly affects performance. Load-balancing heuristics focus on distributing workloads across resources and are particularly relevant when preventing resource overload and maintaining scalable resource utilization are the primary requirements.

Recent research shows a general transition from simple, single-criterion heuristics toward enhanced variants that incorporate multiple scheduling factors. For example, HEFT-based methods incorporate cost, energy, QoS, or load balancing; Max-Min variants incorporate cost and multiple objectives; and Round-Robin variants dynamically adapt scheduling parameters to workload conditions. This evolution improves the ability of heuristic techniques to address heterogeneous and multi-objective scheduling problems, but it also increases scheduling complexity and dependence on runtime information.

Table 5 summarizes representative heuristic techniques according to their scheduling problem, computing environment, principal objectives, and constraints. Rather than describing every algorithm individually, the table emphasizes the characteristics that are relevant to cross-technique comparison and subsequent technique selection.

**Table 5: Comparative analysis of heuristic-based scheduling algorithms**

| Heuristic-based technique | Publish year | Baseline technique | Scheduling issue | Computing environment | Performance parameters | User constraint | Improvement strategy | Simulation tool | Use case/Selection context |
|---|---|---|---|---|---|---|---|---|---|
| Resources Then First in First Out (RTFIFO) [29] | 2021 | FIFO | Task scheduling | Fog computing | Execution time | Deadline | Incorporates resource availability | N/A | Suitable for optimizing the number of tasks rejected by a fog node. |
| Frequency-aware Shortest Remaining Deadline First (FSRDF) [30] | 2020 | EDF | Task scheduling | Fog computing | Deadline | Deadline and frequency | Incorporates deadline/frequency information | MATLAB | Suitable for real-time dynamic task schedulers that consider deadlines & frequency constraints in the smart factories. |
| Greedy Resource Provisioning and modified HEFT (GRP-HEFT) [31] | 2020 | HEFT | Workflow scheduling | Cloud computing | Makespan, resource utilization, and cost | Budget | Adds resource provisioning and budget awareness | CloudSim and JAVA | Suitable for budget constraints for the hourly-based cost model of modern IaaS clouds. |
| Enhanced Heterogeneous Earliest Finish time based on Rule (EHEFT-R) [32] | 2021 | HEFT | Workflow scheduling | Cloud computing | Execution time, makespan, and energy consumption | QOS | Adds energy and QoS considerations | CloudSim | Suitable for synchronization of task sequencing and resource allocation |
| Load Balancing HEFT (LB-HEFT) [33] | 2021 | HEFT | Workflow scheduling | Cloud computing | Makespan, resource utilization, and load balancing | N/A | Incorporates load balancing | CloudSim and WorkflowSim | Suitable for heterogeneous cloud computing. |
| Earliest Gap Priority-Based Fair Scheduling (EG-PBFS) [34] | 2024 | PBFS | Job scheduling | Cloud computing | Delay and makespan | Priority, reliability | Incorporates execution-gap information into priority-based scheduling | CloudSim | Suitable for priority-based job scheduling in cloud environments with constrained resources. |
| Multi-Objective Normalization Workflow Scheduling (MONWS) [35] | 2023 | Max-Min | Workflow scheduling | Cloud computing | Makespan, cost, and resource utilization | N/A | Extends Max-Min for multiple objectives | CloudSim | Suitable for the dynamic priority of tasks |
| Advanced Cost-Aware Max-Min Algorithm [36] | 2024 | Max-Min | Workflow scheduling | Cloud computing | makespan, response time, resource utilization, and cost | QOS | Adds explicit cost awareness | Python | Suitable for adjusting task allocation based on cost considerations dynamically in cloud computing. |
| Balanced Candidate Suffrage Value (BCSV) [37] | 2023 | Suffrage | Task scheduling | Cloud computing | Makespan, load balancing, and resource utilization | N/A | Uses multiple suffrage criteria | JAVA | Suitable for the load balancing problem in a heterogeneous environment. |
| Dynamic Round-Robin Heuristic Algorithm (DRRHA) [38] | 2021 | RR | Task scheduling | Cloud computing | Response time and waiting time | N/A | Dynamically adjusts scheduling quantum | CloudSim | Suitable for dynamically adjusting the quantum time based on the mean time quantum |
| Ameliorated Round-Robin Algorithm (ARRA) [39] | 2023 | RR | Task scheduling | Cloud computing | Response time and waiting time | N/A | Estimates quantum according to workload | C++ | Suitable for adjusting quantum time based on the average of tasks' burst time using fixed and dynamic manner |
| New Round-Robin Task Scheduling Algorithm (NRRTSA) [40] | 2023 | RR | Task scheduling | Cloud computing | Response time and waiting time | N/A | Incorporates ready-queue characteristics | C++ and Python | Suitable for dynamically adjusting quantum time by understanding the environment & surroundings of the ready queue |

The comparison indicates that heuristic scheduling has progressively incorporated additional decision criteria beyond basic task ordering. Traditional methods such as FIFO, EDF, Min-Min, Max-Min, HEFT, and RR provide low-complexity scheduling decisions, whereas their enhanced variants address specific limitations involving resource availability, workload distribution, cost, energy, QoS, deadlines, and fairness. This progression reflects a trade-off between decision simplicity and scheduling awareness: incorporating more system and workload information can improve applicability to complex environments but generally requires additional computation and more accurate runtime information.

### 6.1.3 Statistical analysis of heuristic-based strategies

In Figure 3, statistical evidence demonstrates that heuristic-based scheduling methods are predominantly applied in cloud computing environments (79%), whereas their adoption in fog computing remains limited (21%). This imbalance arises because most heuristic algorithms, such as HEFT, Max–Min, and Round Robin variants, assume relatively stable resource availability, predictable execution behavior, and centralized management, characteristics that closely match cloud infrastructures. In contrast, fog and edge environments are highly dynamic and heterogeneous, with resource constraints, user mobility, and fluctuating workloads that reduce the effectiveness of static scheduling rules. Furthermore, heuristic techniques are mainly employed for task scheduling (63%), followed by workflow scheduling (31%) and job scheduling (6%), owing to their low computational complexity and ability to generate near-optimal solutions rapidly for independent tasks. As scheduling complexity increases because of workflow dependencies and multi-user requirements, purely heuristic approaches become less effective, encouraging the adoption of hybrid and adaptive methods. The distribution of scheduling objectives further reflects this transition: while only 16% of studies address single-objective problems, bi-objective and multi-objective optimization account for 37% and 47%, respectively, highlighting the growing need to simultaneously optimize makespan, resource utilization, response time, cost, and QoS. Consistent with this trend, RR and HEFT remain the most widely adopted heuristics because of their simplicity and adaptability, although their enhanced variants increasingly incorporate resource-awareness and load-balancing mechanisms. Finally, experimental evaluations rely predominantly on simulations (63%), particularly CloudSim, whereas only 37% of studies report real-world implementations, reflecting the practical challenges and costs associated with validating scheduling algorithms in large-scale distributed systems.

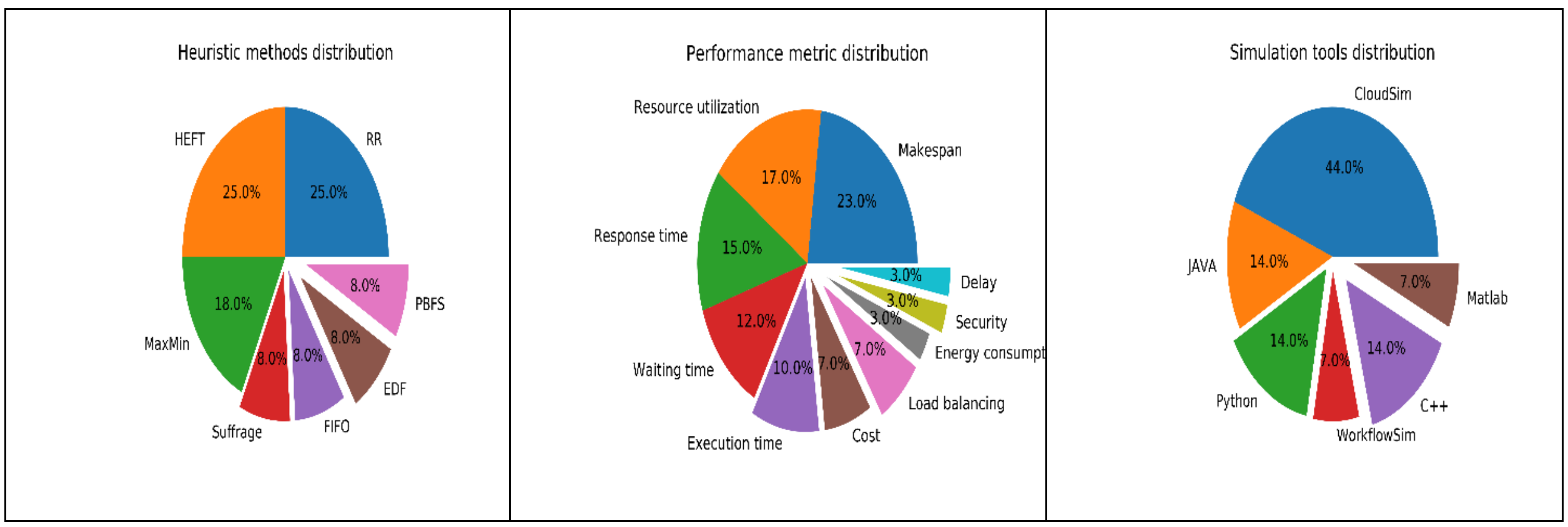

**Figure 3: Statistical information of heuristic-based techniques**

### 6.1.4 Decision framework for selecting heuristic techniques

The suitability of heuristic scheduling techniques depends on the interaction among scheduling objectives, workload characteristics, resource heterogeneity, environmental dynamics, and user constraints. Consequently, selecting a heuristic solely on the basis of its reported performance is insufficient; instead, the choice should reflect the information available to the scheduler and the characteristics of the target computing environment. Resource-aware techniques, such as Min–Min, Max–Min, Suffrage, and HEFT, are particularly suitable for cloud and workflow scheduling when resource heterogeneity, execution time, utilization, or cost are primary concerns. Task-priority approaches, including FIFO and EDF variants, are more appropriate for fog and real-time environments in which deadlines, priorities, workflow dependencies, and latency are critical scheduling considerations. Load-balancing techniques, such as Round Robin and Opportunistic Load Balancing, are useful in large-scale distributed and edge environments where workload distribution and scalability are dominant requirements, although they may provide limited consideration of application-specific requirements and QoS constraints.

No single heuristic family is therefore appropriate across all cloud–fog–edge scheduling conditions. In highly dynamic and heterogeneous environments, combining priority, resource, and load-balancing information can provide a broader basis for scheduling decisions; for example, priority information can be integrated with resource availability, while workflow scheduling can incorporate load balancing or cost-aware resource allocation. Such combinations, however, increase computational and implementation requirements. Accordingly, the proposed decision framework emphasizes the trade-off among simplicity, information requirements, adaptability, and optimization scope when selecting a scheduling technique. Table 6 summarizes these selection guidelines by relating heuristic families to their principal scheduling requirements and target computing environments. This comparative perspective also establishes a basis for examining how heuristic approaches differ from the metaheuristic, learning-based, and hybrid techniques discussed in the subsequent sections.

**Table 6: Decision framework for choosing heuristic scheduling algorithms**

| Scheduling requirement | Suitable heuristic family | Representative techniques | Appropriate context | Main consideration |
|---|---|---|---|---|
| Minimize makespan | Resource-aware / task-priority | HEFT, Min-Min, Max-Min | Cloud, heterogeneous workflows | Requires reliable execution-time estimates |
| Maximize resource utilization | Resource-aware | MCT, HEFT, Suffrage | Heterogeneous cloud | Requires resource information |
| Meet strict deadlines | Task-priority | EDF, FSRDF, PBFS | Fog, real-time IoT | Requires accurate deadlines and timely updates |
| Handle workflow dependencies | Task-priority | HEFT, LB-HEFT, EHEFT-R | Scientific workflows, cloud | Dependency and communication information required |
| Improve workload distribution | Load-balancing | RR, OLB, DRRHA | Cloud, fog, edge | Basic variants may overlook application objectives |
| Handle dynamic workloads | Adaptive load-balancing | DRRHA, ARRA, NRRTSA | Fog, edge, dynamic systems | Additional scheduling overhead |
| Address cost constraints | Resource-aware / task-priority | GRP-HEFT, Cost-Aware Max-Min | Budget-constrained cloud | Requires cost and resource-pricing information |
| Consider energy and QoS | Task-priority / resource-aware | EHEFT-R | Cloud, heterogeneous systems | Requires energy and QoS information |
| Improve fairness | Task-priority / load-balancing | PBFS, EG-PBFS, RR | Multi-user systems | Fairness may conflict with efficiency objectives |
| Large-scale workload distribution | Load-balancing | RR, OLB | Distributed cloud, edge | May require hierarchical or distributed coordination |
| Multiple conflicting objectives | Enhanced or hybrid heuristics | HEFT variants, MONWS, Cost-Aware Max-Min | Cloud and heterogeneous systems | Higher decision complexity |

## 6.2 Metaheuristic Techniques

As discussed in the previous section, heuristic techniques are problem-specific methods designed to provide efficient solutions by following predefined rules or strategies. However, these methods often encounter the challenge of local minimum, limiting their effectiveness in complex optimization problems. In contrast, metaheuristic algorithms are general-purpose optimization techniques inspired by natural processes or mathematical concepts. They are highly adaptable and applicable to a wide range of optimization problems, including scheduling. Metaheuristic approaches possess several advantageous characteristics: 1) they are problem-independent, making them suitable for various domains while maintaining acceptable performance; 2) they effectively explore the search space to identify suboptimal or near-optimal solutions for NP-complete problems within a reasonable computational time; and 3) they are typically non-deterministic and rely on approximation-based strategies.

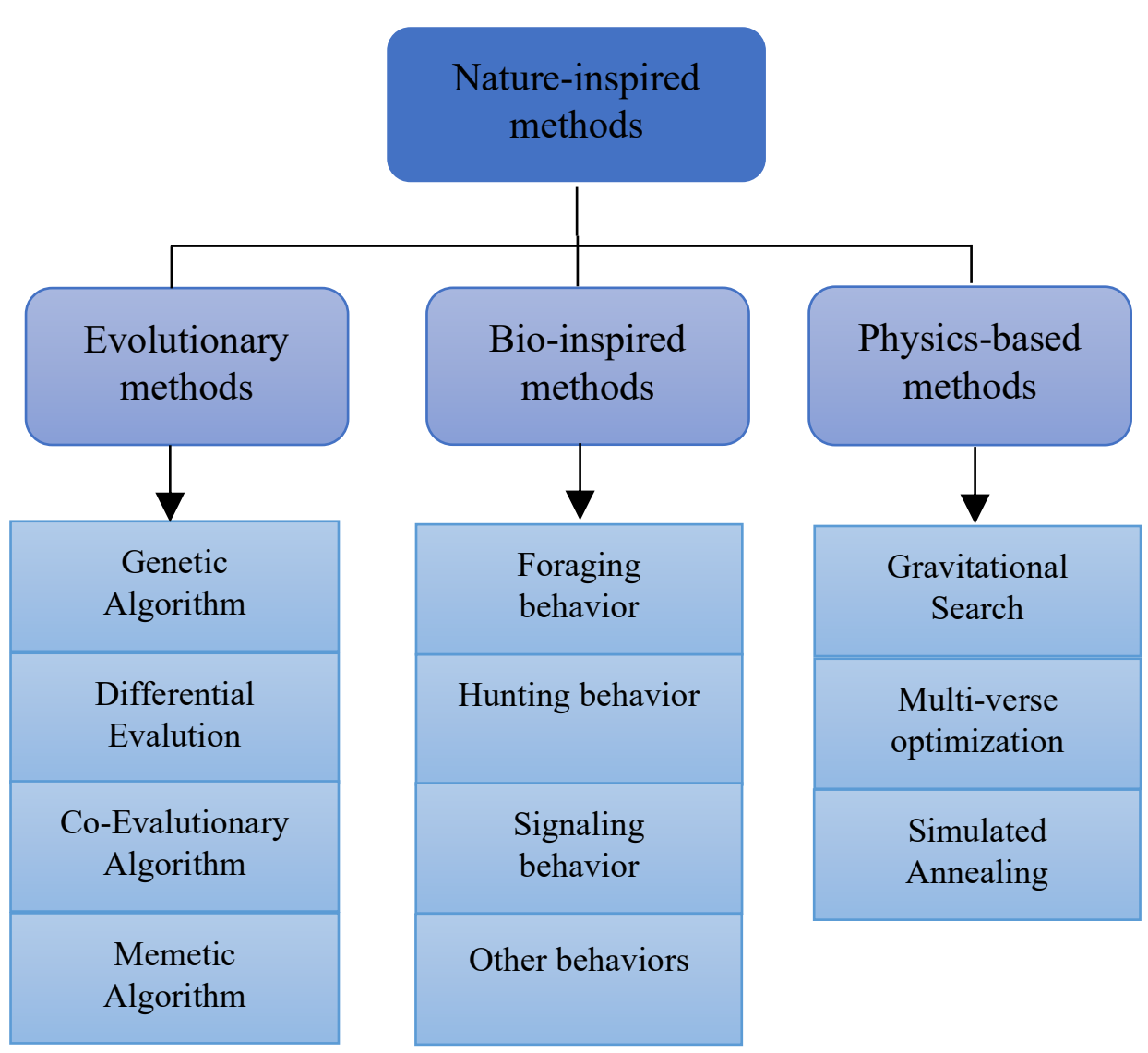

**Figure 4: Classification of metaheuristic techniques**

### 6.2.1 Classification of metaheuristic-based techniques

Metaheuristic techniques can be broadly categorized into two main groups: nature-inspired and non-nature-inspired methods. In the following subsections, each category is systematically classified, and representative algorithms are examined, highlighting their advantages and limitations.

#### 6.2.1.1 Nature-inspired techniques

Nature-inspired algorithms are optimization methodologies that address complex optimization problems by mimicking natural systems. These algorithms effectively handle multi-dimensional and multi-modal optimization tasks, demonstrating robustness and adaptability in dynamic environments. Nature-inspired techniques can be further classified into six distinct categories: 1) evolutionary-based methods, 2) swarm-based methods, 3) physics- or chemistry-based methods, 4) human-based methods, 5) ecological- or biological system-based methods, and 6) emerging methods. Figure 4 presents an overview of these classifications, followed by a detailed discussion of each category, including examples of prominent algorithms and an analysis of their strengths and weaknesses.

##### 6.2.1.1.1 Evolutionary-based methods

Evolutionary algorithms (EAs) are population-based optimization methods that iteratively improve candidate solutions through selection, variation, and survival mechanisms. Their ability to explore large and complex search spaces makes them suitable for cloud–fog–edge scheduling, particularly when task–resource assignment involves multiple objectives, precedence constraints, heterogeneous resources, or nonlinear decision spaces. However, their effectiveness depends strongly on the encoding strategy, objective formulation, parameter control, and computational overhead. The reviewed studies can be grouped into four major families: Genetic Algorithms (GA), Differential Evolution (DE), Co-Evolutionary Algorithms (CEA), and Memetic Algorithms (MA). Their main differences are therefore more relevant for technique selection than their common evolutionary structure.

- ***Genetic Algorithm (GA):*** In its standard form, GA begins with a randomly initialized population, where each chromosome encodes potential task–resource mappings or execution sequences, and their fitness is evaluated according to scheduling objectives such as minimizing makespan, balancing workloads, or maximizing resource utilization. Through iterative selection, crossover, and mutation, GA explores and exploits the solution space, selection favors high-quality solutions, crossover recombines beneficial traits, and mutation preserves diversity and prevents stagnation. Despite its robustness, flexibility, and suitability for multi-objective optimization, the standard GA is limited by issues such as premature convergence, sensitivity to parameter tuning, and slow convergence in large-scale or dynamic environments. Consequently, numerous enhanced GA variants have been proposed for scheduling across cloud, fog, and edge paradigms, incorporating adaptive operators, elitism, hybridization, and diversity maintenance mechanisms to overcome these inherent limitations and improve convergence speed and solution quality.

  For cloud scheduling, The *Multi-objective Parallel Genetic Algorithm (MPGA)* [41], *Nondominated Sorting Genetic Algorithm III (NSGA-III)* [42], and *Genetic Algorithm-based Load Balancing (GALB)* [43] illustrate different selection requirements. MPGA uses parallel subpopulations to improve scalability and jointly considers makespan, cost, and load balance, whereas NSGA-III emphasizes diversity in multi-objective Pareto optimization. GALB introduces adaptive crossover and mutation to improve workload distribution and convergence. Thus, GA variants in cloud environments are particularly relevant when large search spaces, multiple objectives, scalability, or load balancing are dominant concerns.

  For fog and fog–cloud scheduling, *Cost-Aware Genetic Algorithm (CAG)* [44], *Improved Elitism Genetic Algorithm (IEGA)* [45], *Priority-Based Genetic Algorithm (PGA)* [46], *Improved Genetic Algorithm for Permutation-Based Optimization Problems (IGA-POP)* [47], and *Cost, Load, Energy, and Makespan-based Optimized scheduler (CLEMO)* [48] target different scheduling requirements. CAG focuses on execution cost and deadline violations; IEGA improves convergence through elitism and adaptive operators; PGA incorporates task priorities to support latency-sensitive scheduling; IGA-POP preserves task precedence and improves makespan–cost optimization; and CLEMO addresses workflow scheduling using execution time, energy, cost, and load as joint objectives. These studies indicate that GA is particularly adaptable to latency-, deadline-, cost-, workflow-, and multi-objective scheduling, provided that the chromosome representation reflects the characteristics of the target environment.

For edge and edge–cloud environments, *Catastrophic Genetic Algorithm (CGA)* [49] and *Genetic Algorithm for Edge Computing (GA-EC)* [50] address the stronger resource and latency constraints of edge systems. CGA uses catastrophe-based population diversification to reduce stagnation and improve global exploration, whereas GA-EC dynamically adjusts evolutionary operators while emphasizing energy efficiency and load balancing. Their differences suggest that GA selection in edge environments should consider the trade-off between global search capability and computational practicality**.**

Overall, the GA literature demonstrates that the choice of a GA variant should be linked to the scheduling requirement rather than to the algorithm name alone. Parallel or diversity-oriented variants are appropriate when scalability and multi-objective search are important, priority- or deadline-aware variants when latency constraints dominate, and adaptive or load-aware variants when workloads and resource conditions vary dynamically.

- ***Differential Evolution (DE)****:* It differs from GA primarily in its differential mutation mechanism, which generates candidate solutions from scaled differences between population members. Its relatively small number of control parameters and strong global-search capability make it attractive for continuous and multi-objective optimization; however, discrete scheduling requires appropriate encoding and problem-specific operators. The reviewed studies therefore mainly enhance DE through adaptive parameter control, discrete initialization, hybridization, constraint handling, and energy-aware optimization.

  In cloud environments, *Modified DEA-based scheduling algorithm (MDEA-SA)* [51], *Hybrid Differential Evolution (HDE)* [52], *Adaptive Guided Differential Evolution-based Slime Mould Algorithm (AGDESMA)* [53], *Multi-Objective Multi-Factor Evolutionary Algorithm (MO-MFEA) [54]*, and *Self-Adaptive Energy-Aware Differential Evolution (SAEADE)* [55] address different priorities. MDEA-SA improves discrete scheduling and convergence through swap-based initialization, jumping mutation, and adaptive parameters. HDE enhances convergence around promising solutions while preserving diversity. AGDESMA combines DE with Slime Mould Algorithm to strengthen global search and multi-objective optimization, although with additional computational overhead. MO-MFEA targets multi-task and multi-objective optimization through knowledge transfer, whereas SAEADE adaptively selects mutation–crossover strategies for energy, makespan, and throughput optimization. Collectively, these studies show that DE is particularly useful when adaptive convergence, multi-objective optimization, energy efficiency, or multi-task scheduling is required, although increasingly sophisticated variants may increase computational complexity.

  For fog–cloud scheduling, *Modified Strength Pareto Evolutionary Algorithm II (SPEA-II)* [56] and *Energy-efficient Time and Cost (ETC)* [57] demonstrate two distinct selection priorities. SPEA-II incorporates diversity-preserving and constraint-handling mechanisms, making it suitable for schedules with strong deadline and feasibility requirements. ETC combines improved multi-objective DE with deadline-aware DVFS to reduce energy consumption and cost. Therefore, evolutionary scheduling under fog–cloud constraints may favor constraint-oriented DE variants when feasibility is critical, or energy/cost-aware variants when resource efficiency is the primary objective.

  For edge scheduling, DE-Edge [58] combines task clustering with DE-based resource scheduling to address heterogeneous task characteristics, latency, and workload distribution. This design illustrates the importance of problem-specific preprocessing and encoding for applying population-based optimization to resource-constrained edge environments. Its additional computational cost and parameter sensitivity also indicate that DE should be selected cautiously for highly dynamic or resource-limited edge scenarios.

  Compared with GA, DE provides an alternative when adaptive numerical search, diversity maintenance, and multi-objective optimization are central requirements. Nevertheless, for discrete scheduling, its suitability depends heavily on the design of mutation, crossover, and encoding mechanisms.
- ***Co-Evolutionary Algorithm (CEA):*** CEAs decompose a complex optimization problem into interacting subproblems, allowing multiple populations to evolve task assignment, resource allocation, or other scheduling decisions cooperatively. This structure is particularly relevant to heterogeneous scheduling problems in which several interdependent decisions must be optimized simultaneously.

The reviewed applications are mainly workflow-oriented cloud scheduling. *Ch-PICEA-g* [59] uses chaotic population initialization and improved fitness assignment to improve Pareto-front convergence for makespan, cost, and energy optimization. *Cluster-Based Cooperative Co-Evolutionary (CBCC)* [60], in contrast, decomposes the problem into cooperative subpopulations and incorporates cluster-based archive management and dynamic resource allocation, while also considering security-related requirements. Thus, CEA is most relevant when scheduling can naturally be decomposed into interacting decision components, particularly for complex workflow, multi-objective, heterogeneous, or security-aware environments.

Compared with single-population evolutionary methods, CEA provides greater flexibility for decomposing complex scheduling decisions, but this benefit comes with additional coordination and implementation complexity. Its selection is therefore most justified when the scheduling problem contains strongly coupled subproblems that benefit from explicit decomposition.

- ***Memetic Algorithm (MA):*** This strategy combines population-based evolutionary search with problem-specific local search. This combination strengthens exploitation while retaining the global exploration capability of evolutionary optimization and is therefore attractive for scheduling problems with complex and highly constrained search spaces. The principal limitation is additional computational overhead caused by local refinement.

  *Improved Memetic Algorithm (IMA)* [61] and the *Knowledge-Driven Memetic Algorithm (KDMA)* [62] illustrate two different MA designs. IMA employs multiple populations, diversity preservation, multi-layer resource encoding, and local search to improve scheduling across IoT–fog–cloud environments, with particular emphasis on latency and energy efficiency. KDMA incorporates domain knowledge, task-accumulation heuristics, and adaptive reproduction to guide the search toward promising resource allocations and improve multi-objective convergence. Thus, IMA is more oriented toward diversity preservation and heterogeneous layered infrastructures, whereas KDMA emphasizes knowledge-guided search and adaptive multi-objective optimization.

  MAs are consequently most suitable when evolutionary exploration alone is insufficient and problem-specific local refinement or domain knowledge can substantially improve solution quality. Their additional computational cost, however, should be considered when scheduling decisions must be generated rapidly at resource-constrained fog or edge nodes.

#### 6.2.1.1.2 Bio-inspired methods

Bio-inspired scheduling techniques have emerged as a powerful class of optimization approaches, drawing inspiration from the intelligent behaviors observed in nature. Among these, animal behavior–based algorithms have gained significant attention due to their adaptability, decentralized decision-making, and capability to efficiently explore complex solution spaces in heterogeneous computing environments. These techniques can be broadly categorized into four groups based on animal behavior including: 1-foraging behavior algorithms, which focus on resource discovery and exploitation guided by collective intelligence or parasitic strategies, 2-hunting behavior algorithms, which emulate predatory strategies such as chasing, trapping, and cooperative hunting, 3-flashing or signaling behavior algorithms, which rely on communication cues like light or scent to coordinate movement and improve search effectiveness, and 4- other behavioral algorithms, including breeding behavior that mimics biological reproduction mechanisms such as mating, parasitism, or offspring replacement, where fitter individuals propagate their traits to improve population quality over successive iterations and migration behavior which simulate seasonal and large-scale migrations to maintain global diversity and avoid local optima. By leveraging these biologically plausible mechanisms, animal behavior–inspired metaheuristics provide robust, scalable, and energy-efficient solutions for dynamic scheduling challenges across cloud, fog, and edge computing environments. In the following, we discuss these methods with their subclasses.

- ***Foraging behavior algorithms:*** Foraging behavior–based scheduling techniques emulate the natural strategies animals use to locate and exploit food resources efficiently, resulting in strong global search capability and adaptive decision-making in dynamic computing environments. The famous techniques in this category are Ant Colony Optimization (ACO) uses pheromone trails to guide optimal task–resource mappings while continuously updating solution quality based on collective experiences, making it effective for workflow scheduling with dependency constraints. Particle Swarm Optimization (PSO) models cooperative movement toward the best solutions found by individuals and the swarm, offering fast convergence and low computational complexity for large-scale scheduling. Crow Search Algorithm (CSA) mimics the intelligent food-hiding and food-following behavior of crows, where each crow acts as a candidate task–resource allocation and iteratively improves schedules by following promising solutions while using awareness probability to avoid local optima and enhance exploration. Butterfly Optimization Algorithm (BOA) leverages fragrance-based attraction to balance exploration and exploitation, improving resource allocation, especially in multi-objective scheduling. Here, we investigate some related works that applied these algorithms as an approach for solving scheduling problems in various computing environments.

1. ***Ant Colony Optimization (ACO):*** A pheromone-based metaheuristic in which candidate solutions are constructed through task–resource paths and iteratively reinforced according to solution quality. Its main strengths for cloud–fog–edge scheduling are its ability to represent discrete allocation decisions, incorporate problem-specific heuristics, and adapt the search through pheromone updates. However, computational overhead, parameter sensitivity, and slow convergence can become significant for large scheduling spaces. Consequently, ACO variants increasingly focus on multi-objective optimization, adaptive search control, scalability, and explicit constraint handling.

   For cloud scheduling, the reviewed methods address different requirements. *Multi-Objective Task Scheduling model based on ACO (MOTS-ACO)* [61] incorporates Pareto dominance and adaptive pheromone reinforcement for simultaneous optimization of makespan, energy, and load balance, making it suitable for multi-objective scheduling. *Load Balancing ACO (LBACO)* [62] explicitly incorporates load information into the transition mechanism and is therefore oriented toward resource-load balancing. *Improved Adaptive Ant Colony (IAC)* [63] uses adaptive pheromone evaporation and reinforcement to improve convergence for large task sets, whereas *Hybrid Weighted Ant Colony Optimization (HWACO)* [64] combines weighted heuristics with adaptive pheromone control to emphasize cost–time optimization. *Inverted Ant Colony Optimization (IACO)* [65] adopts a VM-oriented search perspective and multi-objective pheromone updating, supporting heterogeneous cloud environments. *Deadline-Budget Constrained Ant Colony Optimization (DB-ACO)* [66], in contrast, explicitly incorporates deadline and budget constraints and is therefore particularly relevant to QoS-constrained scheduling.

   For fog computing, *Improved Multi-Objective Ant Colony Optimization (IMOACO)* [67] employs multiple colonies and inter-colony communication together with adaptive heuristics to strengthen global exploration and address execution-time and energy trade-offs. *Task Scheduling Strategy with Ant Colony (TSSAC)* [68] combines multi-objective optimization with nonlinear pheromone control and container migration, making it more suitable for dynamic load redistribution and resource management in fog clusters.

   In edge computing, *Ant-Colony based Energy-Aware Scheduling (ACEAS)* [69] focuses on energy-aware DAG scheduling by exploiting task slack and DVFS under deadline constraints, whereas *Multi-objective Application Scheduling and Placement based on Ant Colony Optimization (MaaCO)* [70] addresses hybrid edge–cloud application placement using workflow decomposition, critical-path information, DVFS, and communication-aware scheduling.

   Collectively, these studies demonstrate that ACO is particularly appropriate when scheduling involves discrete task–resource decisions, workflow dependencies, or explicit QoS constraints. The variants primarily differ in their decision emphasis: MOTS-ACO targets multi-objective trade-offs; LBACO, IAC, and HWACO emphasize load balancing or convergence and cost–time efficiency; IACO addresses heterogeneous VM allocation; DB-ACO handles deadline–budget constraints; IMOACO and TSSAC target adaptive fog scheduling; and ACEAS and MaaCO address energy-aware edge and edge–cloud workflows. For the proposed decision framework, ACO should therefore be selected according to the dominant constraint rather than treated as a universally applicable optimizer. Adaptive pheromone control, problem-specific encoding, hybrid heuristics, and dynamic resource-management mechanisms remain important for large-scale and highly dynamic environments.

2. ***Practical Swarm Optimization (PSO)****:* Models candidate schedules as particles that iteratively adjust their positions using individual and swarm-level experience. Its simple formulation, relatively low computational overhead, and fast convergence have made PSO widely applicable to task and workflow scheduling. Nevertheless, standard PSO can suffer from premature convergence and is less effective when scheduling requires discrete decisions, many conflicting objectives, or strong workflow constraints. Consequently, the reviewed studies mainly improve diversity preservation, discrete encoding, adaptive search control, workflow representation, and energy awareness.

For cloud scheduling, *Improved Many-Objective PSO (I-MaOPSO)* [71] addresses many-objective optimization through heuristic initialization, adaptive velocity control, and diversity-oriented leader selection, making it suitable when scheduling involves numerous QoS objectives. *Improved Dynamic PSO (IDPSO)* [72] converts particle positions into discrete task–VM mappings and uses dynamic inertia control to improve convergence. *Enhanced PSO (EPSO)* [73] reduces the search space through task aggregation, indirect encoding, and quality-aware initialization, particularly benefiting large-scale workflow scheduling. *Multi-Hierarchy PSO (MHPSO)* [74] introduces workflow aggregation, topology-aware initialization, and hierarchical evolution for deadline-constrained workflows. *Intelligent Energy-Aware Efficient Task Scheduling (IEETS)* [75] extends PSO toward energy-aware scheduling by jointly considering energy consumption, workload balance, VM utilization, and makespan.

For cloud–fog environments, *PSO Optimized Leader (PSO-OL)* [76] improves particle guidance through diversity-preserving leader selection, alternative best-particle mechanisms, and dynamic crossover, targeting premature convergence in heterogeneous scheduling spaces. *Adaptive PSO (APSO)* [77] instead emphasizes adaptive control of inertia and cognitive/social coefficients and supports simultaneous optimization of makespan, cost, and energy.

For edge computing, *Integer PSO (IPSO)* [78] employs integer encoding and EDF-based task ordering for discrete task allocation and offloading in device–edge–cloud environments. *Task Scheduling–Multi-Objective PSO (TS-MOPSO)* [79] uses adaptive search control and objective ranking for multi-objective edge scheduling, addressing execution time, completion time, and cost.

The PSO studies therefore reveal several selection dimensions. I-MaOPSO is oriented toward many-objective problems; IDPSO and IPSO are appropriate when discrete task/resource or offloading decisions dominate; EPSO and MHPSO target large and deadline-constrained workflows; IEETS emphasizes energy efficiency; and PSO-OL and APSO improve adaptability and search stability in heterogeneous environments. TS-MOPSO further targets multi-objective edge workloads. Thus, PSO is particularly attractive when rapid search and relatively lightweight optimization are required, but its representation and update mechanisms should be adapted to discrete allocation, workflow dependencies, energy objectives, and dynamic resource conditions.

3. ***Crow Search Algorithm (CSA):*** A population-based metaheuristic inspired by the food-hiding behavior of crows. Its relatively simple search structure and small number of control parameters make it attractive for task scheduling, but standard CSA may experience premature convergence, limited local exploitation, and reduced scalability for large scheduling spaces. The reviewed extensions therefore focus on resource-aware allocation, adaptive parameter control, intensified search, and multi-objective optimization**.**

   *Enhanced Crow Search Algorithm (ECSA)* [80] replaces less-informed random task and VM selection with completion-time- and load-aware mechanisms, improving task allocation, makespan, and resource utilization. *Linear-Scaling Crow Search Optimization (LS-CSO)* [81] introduces linear scaling of the awareness probability to adapt the exploration–exploitation balance without substantially changing the scheduling representation. *Imperialist Competitive Algorithm (ICA)* [82] incorporates an imperialist–colony competitive mechanism to strengthen intensification and guide the search toward promising scheduling solutions. *Crow Search Algorithm for Multi-Objective Makespan and Cost optimization (CSAMOMC)* [83] extends CSA to multi-objective makespan–cost optimization through intelligent initialization and hypervolume-based Pareto selection, improving both convergence and solution diversity.

   These variants illustrate complementary use cases: ECSA is primarily resource- and load-aware; LS-CSO emphasizes adaptive search control; ICA strengthens competitive exploitation; and CSAMOMC addresses explicit multi-objective scheduling. Accordingly, CSA is most appropriate when a simple population-based optimizer with resource-aware or adaptive enhancements is required. For large-scale or highly dynamic scheduling, adaptive parameters, diversity preservation, hybrid local search, and problem-specific task–VM representations are important to maintain scalability and avoid premature convergence.

4. ***Butterfly Optimization Algorithm (BOA):*** Uses global and local search mechanisms inspired by butterfly foraging behavior. In scheduling, candidate butterflies represent task–resource assignments, while scent-based attraction guides the search toward promising solutions. BOA provides a flexible population-based framework, but its conventional form can suffer from an inadequate exploration–exploitation balance, attraction toward already identified optima, and limited suitability for discrete and highly constrained scheduling. Consequently, the reviewed variants introduce discrete representations, adaptive control, hybridization, and swarm-level search strategies.

For edge and MEC scheduling, *Discrete Butterfly Optimization Algorithm (DBOA)* [84] converts BOA into a discrete search framework and combines chaotic initialization, Lévy flights, and genetic operators to address task prioritization, resource allocation, and energy-aware workflow decisions. *Fuzzy Butterfly Optimization Algorithm (FBOA)* [85] instead employs fuzzy control to adapt the global–local search balance and a virtual-butterfly mechanism to improve collective guidance, making it suitable for dynamic and heterogeneous scheduling conditions.

For fog–cloud scheduling, *Improved Butterfly Optimization Algorithm (IBOA)* [86] introduces adaptive step-size and cost-aware guidance to improve search efficiency while considering computation and communication costs. *Butterfly Spotted Hyena Optimization Algorithm (BSHOA)* [87] hybridizes BOA with Spotted Hyena Optimization to combine complementary exploration and exploitation mechanisms, with particular relevance to energy-aware IoT scheduling. *Butterfly Swarm Optimization (BSO)* [88] uses coordinated global and local search phases to maintain population diversity while addressing makespan and cost in large-scale scheduling.

The reviewed BOA variants consequently address different decision requirements. DBOA is oriented toward discrete and energy-constrained edge workflows; FBOA emphasizes adaptive search under dynamic conditions; IBOA incorporates cost-aware scheduling in fog–cloud environments; BSHOA strengthens exploration through hybridization for energy-sensitive workloads; and BSO emphasizes scalable swarm coordination for large-scale QoS-driven scheduling. BOA is therefore most useful when its basic search mechanism is complemented by problem-specific encoding and adaptive or hybrid search strategies, particularly for heterogeneous environments where discrete allocation, energy, communication cost, and workload dynamics must be considered simultaneously.

- **Hunting behavior algorithms:** These techniques draw on predation dynamics such as tracking, encircling, and coordinated attacks to optimize task allocation in distributed computing infrastructures. Cat Swarm Optimization (CSO), inspired by feline hunting behavior and mode switching between seeking (watchful tracking) and tracing (active chase), supports flexible and intelligent scheduling decisions while maintaining population diversity. Grey Wolf Optimizer (GWO) simulates the hierarchical leadership and cooperative hunting strategies of wolf packs, enabling strong convergence and balanced exploration–exploitation for efficient resource mapping in cloud–fog environments. Whale Optimization Algorithm (WOA) models bubble-net hunting to perform spiral and encircling search operations, improving scheduling accuracy under dynamic workloads. Harris Hawks Optimization (HHO) leverages surprise pounce strategies driven by hawk–prey interactions, enhancing adaptability and global search effectiveness in rapidly changing task conditions. Here, we investigate some related works that applied these algorithms as an approach for solving scheduling problems in various computing environments.

  1. ***Cat Swarm Optimization (CSO):*** Combines seeking and tracing behaviors to balance local refinement and global exploration. In scheduling, cats represent candidate task–resource mappings, making CSO applicable to heterogeneous and dynamic environments. Its main limitations are parameter sensitivity, scalability, and possible premature convergence, which have motivated problem-specific and multi-objective extensions.

     The reviewed variants address different scheduling requirements. *Workflow Cat Swarm Optimization (WSCSO)* [89] focuses on task prioritization according to task size and processing capacity, but its effectiveness can depend on population initialization and scalability**.** *Modified Distance Cat Swarm Optimization (MDCSO)* [90] modifies the distance calculation used during search to improve adaptability and convergence under dynamic IoT–cloud workloads. *Prioritized Cat Swarm Optimization (PCSO)* [91] incorporates task and VM priorities to support energy-efficient and SLA-aware scheduling. *Multi-objective Cat Swarm Optimization-based Fault-Tolerant Load Balancing (MCSOFLB)* [92] extends CSO toward multi-objective, fault-tolerant, and dual-level load balancing in heterogeneous multi-cloud environments. Thus, CSO variants can be differentiated by their primary decision requirement: WSCSO emphasizes workflow mapping, MDCSO search adaptability, PCSO priority- and SLA-aware allocation, and MCSOFLB multi-objective fault-tolerant load balancing.

For technique selection, CSO is most relevant when a scheduling problem requires a balance between local refinement and global exploration, particularly for heterogeneous resource allocation. Its application should use problem-specific encoding and fitness functions, while adaptive phase control or hybrid local search can improve robustness for large and dynamic workloads.

2. ***Grey wolf search optimization (GWO)***: Uses a hierarchical leader structure to guide population search and provides a relatively simple mechanism for balancing exploration and exploitation. In scheduling, candidate wolves represent task–resource assignments, while the leading solutions guide subsequent search. Although GWO is computationally simple and readily extendable to multi-objective scheduling, its diversity may decline during later iterations, increasing the risk of premature convergence.

   Four variants illustrate different improvement strategies. *Task Scheduling using Multi-Objective Grey Wolf Optimizer (TSMGWO)* [93] introduces Pareto-based optimization and an enriched leader hierarchy for conflicting scheduling objectives. *Multi-Objective Task Scheduling Grey Wolf Optimization (MOTSGWO)* [94] incorporates workload heterogeneity, energy cost, and deadline information to improve context-aware scheduling decisions. *GWO with Encoding Mechanism (GWOEM)* [95] uses a compact one-dimensional representation and greedy replacement to reduce encoding overhead and stabilize convergence. *Multi-Objective Discrete Grey Wolf Optimization (MoDGWA)* [96] adapts GWO to discrete fog scheduling through chromosome-based encoding, discrete operators, NSGA-II crowding distance, and load balancing.

   These studies position GWO variants according to the scheduling structure: TSMGWO and MOTSGWO address multi-objective decision making, GWOEM reduces representation and convergence overhead, while MoDGWA is designed for discrete and load-sensitive fog scheduling. Accordingly, GWO is most suitable when hierarchical guidance can effectively exploit promising schedules, provided that diversity preservation and constraint-aware encoding are incorporated for large or discrete search spaces.

3. ***Whale Optimization Algorithm (WOA):*** Combines encircling, spiral exploitation, and exploratory search to optimize task–resource mappings. Its relatively simple formulation and strong early-stage convergence make it attractive for cloud, fog, and edge scheduling, but limited diversity and later-stage convergence can lead to local optima. The reviewed variants therefore primarily improve adaptive convergence, diversity, multi-objective optimization, trust awareness, and cross-layer scheduling**.**

   For cloud scheduling, *Improved WOA for Cloud Task Scheduling (IWC)* [97] uses a nonlinear convergence factor and adaptive population resizing to improve convergence while maintaining diversity. *Gaussian Whale-Cloud Optimization (GCWOA)* [98] combines opposition-based initialization, adaptive mobility, and a Gaussian cloud model to strengthen global exploration. *Multi-objective Trust-aware Task Scheduling (MOTSWOA)* [99] integrates SLA-based trust, task–VM priorities, and multiple objectives including makespan, energy, and reliability. *W-Schedule* [100] uses priority-aware task–VM mapping to improve early solution quality, load balancing, and makespan, but has a narrower multi-objective scope. *Modified Chimp-Whale Optimization (MCWOA)* [101] hybridizes WOA with Chimp Optimization, Sobol initialization, hierarchical grouping, and task clustering to improve robustness for dynamic and high-dimensional workloads. *WOA-Scheduler* [102] combines adaptive coefficients, Pareto-based exploitation, and weighted objectives to address cost, execution time, and load.

   For fog and edge environments, *Golden Eagle Whale Optimization (GEWO)* [103] combines Golden Eagle and Whale search mechanisms across cloud, fog, and edge layers, with emphasis on heterogeneous resource allocation and energy efficiency. *Enhanced WOA (EWOA)* [104] uses chaotic initialization and a nonlinear convergence factor to improve diversity and multi-objective search under resource-constrained edge conditions.

   Thus, IWC emphasizes convergence adaptation, GCWOA exploration, MOTSWOA trust and QoS, W-Schedule priority-aware mapping, MCWOA hybrid exploration, and WOA-Scheduler multi-objective scheduling. At the infrastructure level, GEWO targets cross-layer heterogeneous and energy-aware scheduling, whereas EWOA focuses on adaptive search in constrained edge environments.

For the proposed decision framework, WOA is particularly relevant when rapid search must be combined with adaptive exploration in heterogeneous environments. Selection among variants should depend on the dominant requirement: trust/QoS, cost–time–load trade-offs, dynamic high-dimensional search, or cross-layer cloud–fog–edge heterogeneity. Adaptive parameter control, diversity preservation, workload prediction, and mobility- or fault-aware extensions remain important for real-time deployment.

4. ***Harris Hawks Optimization (HHO):*** Alternates between exploration and exploitation according to the estimated state of the best solution. In scheduling, its flexible search mechanisms can refine task–resource mappings, but sensitivity to initialization, premature convergence, stochastic variability, and the computational cost of multiple hunting strategies limit the classical formulation. Recent variants therefore target priority-aware scheduling, exploration enhancement, trust and fault tolerance, SLA compliance, energy efficiency, and discrete multi-objective optimization.

   For cloud scheduling, *Priority-based HHO (PHHO)* [105] combines AHP-based task prioritization with a multi-objective fitness function covering makespan, energy, resource utilization, and throughput. *Multi-strategy HHO (MHHO)* [106] strengthens exploration through chaotic and opposition-based initialization, adaptive energy decay, and elite guidance. *Fault-Tolerant Trust-Aware Task Scheduler (FTTATS)* [107] incorporates task/VM priorities, trust metrics, and adaptive HHO exploitation to address failure and SLA-related requirements. *SLA-aware HHO (SLA-HHO)* [108] explicitly models workflow dependencies and SLA constraints. *Multi-Tactic HHO (MTHHO)* [109] combines Sobol initialization, elite opposition-based learning, Gaussian-walk learning, and adaptive energy updating to improve diversity, energy efficiency, and load balancing. *Enhanced HHO (EHHO)* [ 110] introduces dynamic random walks with Lévy-flight and siege mechanisms to reduce stagnation and improve scalable resource allocation.

   For fog scheduling, *HHO with Local Search (HHOLS)* [111] combines discrete mapping, mutation, and local search for task assignment, with emphasis on makespan and energy. *Discrete Opposition-based HHO (DO-HHO)* [112] combines discrete genetic operators with opposition-based learning to improve convergence and address SLA and offloading decisions. *Multi-objective HHO-based Task Scheduling (MoHHOTS)* [113] introduces adaptive mutation, Pareto selection, and crowding distance to jointly optimize transmission delay and energy while preserving solution diversity.

   The HHO literature therefore demonstrates several distinct selection paths: PHHO and SLA-HHO emphasize task priorities and contractual constraints; MHHO, MTHHO, and EHHO strengthen search diversity and convergence; FTTATS addresses trust and fault tolerance; HHOLS and DO-HHO improve discrete scheduling; and MoHHOTS targets multi-objective fog scheduling. HHO is consequently most useful when adaptive exploration and exploitation are required, but practical deployment should incorporate discrete encoding, constraint handling, diversity preservation, and lightweight adaptive mechanisms.

- **Flashing/signal behavior algorithms:** These techniques are inspired by the way biological agents communicate and coordinate actions through light, signals, or intensity variations, and they have been effectively adapted to scheduling problems. In this paradigm, each agent represents a candidate schedule and emits a signal whose intensity is proportional to its fitness, enabling indirect communication among agents. In the *Firefly Optimization Algorithm (FOA)*, less attractive fireflies move toward brighter ones, modeling exploitation through attractiveness-driven convergence, while random perturbations maintain exploration and help escape local optima; this mechanism has been widely applied to task and workflow scheduling in cloud, fog, and edge environments. Similarly, *Sunflower Optimization Algorithm (SFO)* models heliotropic signaling, where candidate solutions orient and move toward a sun representing the best solution based on perceived signal intensity, enabling directed exploitation while maintaining exploration through adaptive step sizes. In the following, we investigate some methods that apply these techniques for scheduling in different environments.

   1. ***Firefly Optimization Algorithm (FOA):*** A population-based metaheuristic that uses attraction among candidate solutions to explore complex scheduling search spaces. In cloud–fog–edge scheduling, fireflies can represent task–resource mappings, with solution quality determined by objectives such as makespan, resource utilization, energy consumption, or SLA compliance. Its flexible search mechanism makes FOA applicable to nonlinear and heterogeneous scheduling problems, although its performance is sensitive to parameter settings and population size and may deteriorate through premature convergence or high computational cost in large search spaces.

Several variants adapt FOA to specific scheduling requirements. *Trust-Aware Firefly-based Task Scheduling Algorithm (TAFFA)* [114] incorporates task and VM priorities and SLA-related trust measures, including availability, success rate, and turnaround efficiency, making it particularly relevant when trust and SLA compliance are important scheduling criteria. *Modified Firefly Optimization Algorithm (ModFOA)* [115] integrates uniform crossover, Gaussian mutation, quasi-reflection-based learning, and adaptive step sizes to improve population diversity and the exploration–exploitation balance for hybrid cloud–edge workflow scheduling. In contrast, *Adaptive Firefly Algorithm (AFA)* [116] addresses dependency-aware IoT–fog scheduling by incorporating execution time, communication delay, and synchronization overhead into the fitness function and dynamically adapting the light absorption coefficient and discrete distance metric. These studies therefore extend FOA in different directions: TAFFA emphasizes trust and priority, ModFOA emphasizes search diversification and convergence, and AFA emphasizes workflow dependencies and heterogeneous fog resources.

From a technique-selection perspective, FOA is appropriate when the scheduling problem has a large or nonlinear search space and can benefit from population-based exploration. The choice of variant should reflect the dominant scheduling requirements: TAFFA for trust/SLA-aware cloud scheduling, ModFOA for hybrid cloud–edge workflow optimization, and AFA for dependency-aware and heterogeneous IoT–fog environments. Effective implementations require an appropriate task–resource representation, constraint handling, and fitness formulation. Adaptive parameter control, diversity preservation, and hybrid local-search or evolutionary operators can further improve robustness, particularly for dynamic and large-scale workloads.

2. ***Sunflower Optimization Algorithm (SFO):*** A population-based metaheuristic that models directional movement toward promising solutions. For scheduling, candidate sunflowers represent alternative task–resource assignments, while movement and pollination mechanisms provide global exploration and local refinement. SFO has been applied to cloud scheduling because of its relatively simple formulation and ability to address objectives such as execution time, energy consumption, and load balancing. However, convergence speed, parameter sensitivity, and computational overhead can become limiting factors as the scheduling space and resource heterogeneity increase.

   Two variants have been proposed to improve these limitations. *Enhanced Sunflower Optimization (ESFO)* [117] introduces an enhanced pollination strategy to strengthen the exploration–exploitation balance and accelerate convergence, although the additional search operations increase computational cost. *Opposition-Based Sunflower Optimization (OSFO)* [118] combines opposition-based learning with improved pollination to enlarge the explored search space and reduce premature convergence. The resulting search mechanism is particularly useful for large-scale scheduling problems, but it also introduces additional computational requirements and parameter-tuning considerations.

   Accordingly, SFO variants represent different trade-offs rather than superior alternatives. SFO provides a relatively simple baseline for less complex scheduling problems; ESFO is suitable when improved convergence and search balance are required; and OSFO is more appropriate when large search spaces and premature convergence are major concerns. For cloud–fog–edge environments, future adaptations should emphasize adaptive parameter control, constraint-aware solution representations, and lightweight hybrid search mechanisms. In decentralized fog and edge systems, distributed or localized implementations may also reduce the computational burden associated with centralized population-based optimization.

- **Other behavioral algorithms:** A distinct subgroup of bio-inspired metaheuristic algorithms can be identified by other biological behaviors, particularly breeding (reproductive) behavior and migration (immigration) behavior. Algorithms in the breeding-behavior subgroup, such as *Cuckoo Search (CS)*, mimic reproductive strategies like brood parasitism, where new candidate solutions (eggs) replace inferior ones through survival-of-the-fittest selection, often combined with random long-range exploration like Lévy flights. In contrast, the migration-behavior subgroup, exemplified by the *Monarch Butterfly Optimization (MBO)* algorithm, models seasonal migration and population redistribution, where individuals probabilistically move between subpopulations or regions, enabling information exchange between high-quality and low-quality areas of the search space. Unlike reproduction-centered methods, migration-based algorithms emphasize spatial relocation and population flow as the primary search operators. Together, these behavior-driven metaheuristics broaden the biological metaphor beyond foraging and social cooperation, offering alternative mechanisms for balancing exploration and exploitation, particularly in large-scale and dynamic optimization problems. In the following, we investigate some methods that apply these techniques for scheduling in different environments.

1. ***Cuckoo Search (CS)****:* A population-based metaheuristic inspired by brood parasitism and Lévy-flight exploration. In scheduling, candidate nests represent task-to-resource mappings, while Lévy flights promote global exploration and abandonment of inferior solutions that maintains population diversity. Its relatively simple search mechanism makes CS applicable to heterogeneous cloud, fog, and edge scheduling; however, solution encoding, parameter settings, and the balance between exploration and exploitation strongly affect its performance.

   Several studies have extended CS to multi-objective and discrete scheduling. *Multi-Objective Cuckoo Search (MOCS)* [119] provides a basic multi-objective formulation by combining multiple objectives through randomly assigned weights. Although computationally simple, this formulation has limited ability to preserve diverse Pareto solutions and may converge prematurely. *Improved Multi-Objective Cuckoo Search (IMOCS)* [120] addresses these limitations through an external archive, Pareto dominance, and fast non-dominated sorting, providing better control of the trade-off between execution latency and energy consumption for DAG-based applications, at the expense of additional computational overhead. *Multi-Objective Discrete Cuckoo Search Algorithm (MoDCSA)* [121] further adapts CS to discrete task scheduling in IoT–fog environments by introducing discrete search operators, local search, and reliability-aware scheduling. These mechanisms improve exploration of combinatorial solution spaces and explicitly incorporate reliability, making the approach suitable for scheduling problems in which discrete task placement and reliability are important considerations.

   Thus, the CS variants address different scheduling requirements rather than representing interchangeable improvements: MOCS is a lightweight baseline for multi-objective optimization, IMOCS is oriented toward Pareto-based latency–energy trade-offs in DAG applications, and MoDCSA targets discrete and reliability-aware scheduling in fog/IoT environments. Effective application of CS therefore requires an appropriate task–resource representation, constraint handling, and parameter control. For multi-objective problems, Pareto dominance, external archives, and diversity-preservation mechanisms are particularly useful for avoiding premature convergence and obtaining a well-distributed set of scheduling alternatives.

2. ***Monarch Butterfly Optimization (MBO)****:* Models the migration and adjustment behavior of monarch butterflies. Candidate solutions are distributed among subpopulations, with migration supporting global exploration and adjustment providing local exploitation. In scheduling, these operators can be used to modify task–resource assignments and improve workload allocation. However, the classical MBO formulation is primarily oriented toward single-objective optimization and can be sensitive to parameter settings, which may limit its effectiveness for large-scale, heterogeneous, and dynamically changing cloud–fog–edge environments.

   Multi-objective extensions have subsequently addressed these limitations. *Multi-Objective Monarch Butterfly Optimization (MOMBO)* [122] incorporates Pareto dominance, self-adaptive population control, and a greedy strategy to optimize conflicting objectives such as makespan, cost, and energy consumption. The self-adaptive mechanism adjusts the search population to improve the exploration–exploitation balance, while greedy retention preserves high-quality solutions across iterations. *Multi-Objective Improved Monarch Butterfly Optimization (MOIMBO)* [123] extends this framework using Epsilon-fuzzy dominance sorting and an elite archive, improving the management of non-dominated solutions and reducing the computational burden associated with multi-objective selection. Its structured population division and task–resource mapping further support search efficiency in complex scheduling spaces.

   For scheduling applications, MBO is most effective when the solution representation explicitly captures task–resource assignments and the migration and adjustment operators are adapted to the characteristics of the scheduling problem. MOMBO is suitable when several objectives must be optimized simultaneously using Pareto-based selection, whereas MOIMBO is more appropriate when greater control over Pareto-solution diversity and selection efficiency is required. Self-adaptive parameter control and elite-solution preservation can improve robustness, while hybridization with complementary search strategies may further enhance local refinement. Nevertheless, parameter sensitivity, computational cost, and adaptability to dynamic workloads should be considered when applying MBO-based methods to heterogeneous cloud–fog–edge scheduling.

#### 6.2.1.1.3 Physics-based methods

These techniques leverage principles from physical laws and natural phenomena to guide the search for optimal solutions in complex computing environments. *Gravitational Search Algorithm (GSA)* draws inspiration from Newtonian gravity, where candidate solutions act as masses attracting each other proportionally to their fitness, enabling exploration and exploitation of the scheduling space. *Multi-Verse Optimization (MVO)* models the concepts of black holes, white holes, and wormholes from cosmology, exchanging solution characteristics between candidate universes to balance global search and local refinement. *Simulated Annealing (SA)* mimics the annealing process in metallurgy, probabilistically accepting suboptimal schedules early to escape local minima and gradually converging to near-optimal task allocations. In the following, we investigate some methods utilized these techniques and compare them.

- ***Gravitational Search Algorithm (GSA)****:* Models candidate schedules as interacting agents whose movement is guided by their fitness. Its relatively simple structure and limited parameter requirements make it applicable to NP-hard scheduling problems, but the dominance of high-quality agents can reduce population diversity and cause premature convergence, particularly in large search spaces.

  Three reviewed variants illustrate different strategies for addressing these limitations. *Improved Multi-Objective Gravitational Search Algorithm (IMOGSA)* [124] combines non-dominated sorting, crowding distance, and Sine Cosine-based parameter adaptation to improve convergence and diversity in multi-objective cloud scheduling. *Gravitational Search Algorithm for Cloud Load Distribution (GSACLD)* [125] focuses on cloud load distribution by incorporating cost-aware fitness and cloudlet-to-VM assignment, targeting VM utilization and data-transfer efficiency. *Star-Quake Multi-Objective Gravitational Search Algorithm (SQMOGSA)* [126] introduces a star-quake mechanism to improve exploration and escape local optima in fog–cloud multi-objective scheduling.

  From a selection perspective, GSA variants are therefore most relevant when global search, load distribution, or multi-objective optimization is required. IMOGSA is oriented toward Pareto-based multi-objective scheduling, GSACLD toward load and cost considerations, and SQMOGSA toward maintaining exploration in complex fog–cloud search spaces. Their additional mechanisms, however, increase computational or parameter complexity and should be considered when applying them to large-scale or time-sensitive scheduling.
- ***Multi-Verse Optimization (MVO)***: Uses interacting candidate solution populations to balance global exploration and local exploitation. In scheduling, it has been applied to task–VM mapping and optimization of makespan, cost, energy, and resource utilization. Although MVO can address heterogeneous and multi-objective cloud scheduling, the original method may suffer from reduced diversity and premature convergence.

  The reviewed variants address these limitations in different ways. *Modified Multi-Verse Optimization (MMVO)* [127] incorporates grid-based dominance, Pareto archiving, and trust-aware fitness evaluation, making it suitable for multi-objective and QoS-oriented workflow scheduling. *Enhanced Multi-Verse Optimization (EMVO)* [128] uses an elitist memory mechanism to preserve high-quality solutions and accelerate convergence, with emphasis on makespan reduction. *Improved Multi-Objective Multi-Verse Optimization (IMoMVO)* [129] modifies the position-update mechanism using the best and second-best solutions and integrates multi-objective optimization to improve the exploration–exploitation balance.

  These studies suggest that MVO is particularly applicable when multi-objective optimization, QoS, trust, or convergence improvement is central to the scheduling problem. MMVO is oriented toward reliability and Pareto-based decision making, EMVO toward convergence acceleration, and IMoMVO toward maintaining search diversity in complex scheduling spaces. However, the increasing sophistication of these variants also introduces additional implementation and computational overhead.
- ***Simulated Annealing (SA)****:* A trajectory-based metaheuristic that accepts some inferior neighboring solutions during the search, allowing it to escape local optima. Its simplicity and flexibility make it suitable for NP-hard scheduling across cloud, fog, and edge environments. However, its effectiveness is sensitive to the initial temperature, cooling strategy, neighborhood generation, and stopping criteria.

Three reviewed variants demonstrate how SA can be adapted to different scheduling requirements. *Priority-Sensitive Simulated Annealing (PSA)* [130] incorporates priority-aware penalties to improve the allocation of high-priority tasks while jointly considering makespan, energy, cost, and reliability. *Dynamic Cooling Simulated Annealing (DCSA)* [131] adapts the cooling rate and neighborhood generation according to solution quality, improving convergence while retaining the ability to escape local optima. *Parallel Simulated Annealing with Time Arrival Scheduling (PANTAS)* [132] employs multiple parallel SA processes and considers task arrival timing and offloading, improving solution diversity and scheduling efficiency at the expense of greater computational cost.

Accordingly, SA is particularly appropriate when local-search flexibility, priority handling, adaptive convergence, or offloading decisions are important. PSA is suited to priority-sensitive scheduling, DCSA to convergence-sensitive optimization, and PANTAS to parallel and time-aware edge scheduling. The additional computational requirements of parallel or adaptive variants should nevertheless be considered for resource-constrained edge environments.

#### 6.2.1.2 Non-nature-inspired techniques

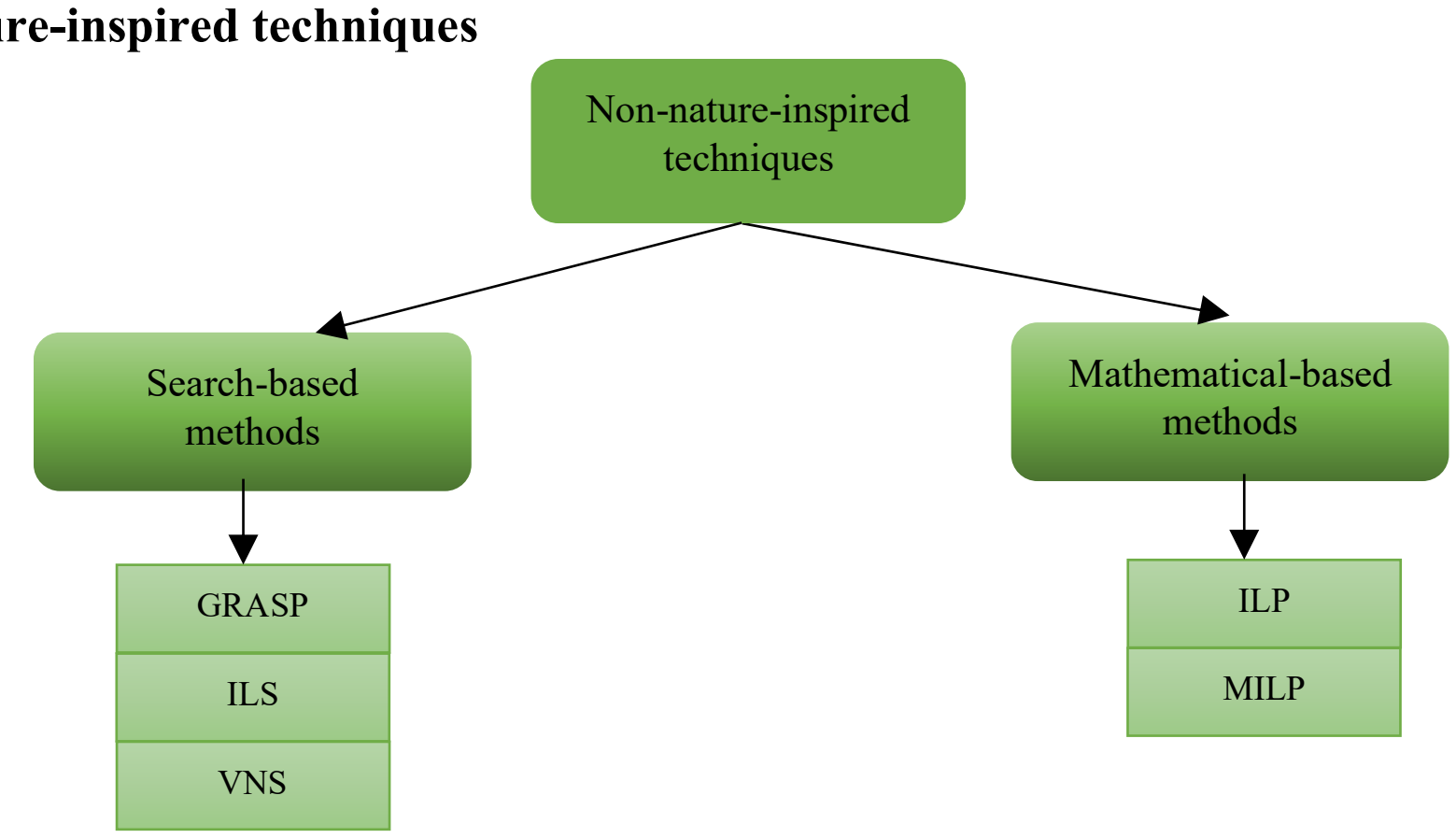

**Figure 5: Classification of non-nature-inspired techniques**

Non-nature-inspired scheduling techniques are grounded in algorithmic principles derived from operations research and computer science, rather than analogies to biological or physical systems. These methods employ mathematical and logical formulations to address scheduling problems and are generally simpler and computationally efficient, particularly for small-scale or well-structured scenarios. Non-nature-inspired techniques can be broadly classified into two categories: 1) search-based methods and 2) mathematical-based methods. Figure 5 illustrates this classification, which is subsequently elaborated through a systematic discussion of representative algorithms from each category, along with an analysis of their respective strengths and limitations.

##### 6.2.1.2.1 Search-based methods

Search-based methods iteratively explore the solution space using constructive, local-search, or neighborhood-based mechanisms. They are suitable for large and combinatorial scheduling problems in which exhaustive optimization is impractical. Their main distinction lies in how they balance diversification and intensification. *Greedy Randomized Adaptive Search Procedure (GRASP)* combines randomized greedy construction with local search, *Iterated Local Search (ILS)* repeatedly perturbs and improves a candidate solution, and *Variable Neighborhood Search (VNS)* systematically changes neighborhood structures. These methods can produce high-quality schedules with relatively flexible problem formulations, but they do not generally guarantee global optimality and may incur substantial computational cost as problem size and search complexity increase.

- **Greedy Randomized Adaptive Search Procedure (GRASP):** For cloud scheduling, *GRASP-based scheduling* [133] uses randomized greedy construction followed by local improvement to explore alternative task–VM assignments. Its effectiveness depends strongly on the quality of the greedy function and restricted candidate list. *Improved GRASP (IGRASP)* [134] extends this framework by retaining frequently occurring

job–VM assignments from elite solutions through a fixed-set search mechanism. Compared with the classical GRASP approach, IGRASP strengthens intensification by directing subsequent searches toward promising regions, although the additional learning mechanism introduces parameter and computational overhead. Thus, GRASP is appropriate when broad exploration and flexible local refinement are required, whereas IGRASP is more suitable when faster convergence through reuse of high-quality assignment patterns is important.

- **Iterated Local Search (ILS):** In workflow scheduling, *ILS* [135] iteratively improves a schedule through local search and controlled perturbation. Its relatively simple structure and low memory requirement make it attractive for large scheduling spaces, but its effectiveness depends on the quality of the perturbation and acceptance mechanisms. *Adaptive Iterated Local Search (AILS)* [136] addresses these limitations by combining a HEFT-based initial solution, greedy resource provisioning, adaptive diversification, level-based perturbation, and simulated-annealing-based acceptance. The comparison indicates that AILS is preferable when workflow dependencies, budget constraints, and complex search spaces require stronger exploration–exploitation control, whereas conventional ILS is more appropriate when implementation simplicity and low overhead are priorities.
- **Variable Neighborhood Search (VNS)**: Several VNS-based studies further illustrate how search design can be adapted to different scheduling conditions. *Earliest Finish Time–General Variable Neighborhood Search (EFT-GVNS)* [137] combines Earliest Finish Time initialization with General VNS and composite local search, improving neighborhood exploration while reducing costly restarts. *Estimation of Distribution with Composite Neighborhoods (EDP-CN)* [138] introduces probabilistic modeling of elite solutions to guide neighborhood sampling toward promising task–resource assignments, thereby reducing inefficient random exploration at the expense of additional modeling overhead. *Tolerance-Aware Neighborhood Search (TANS)* [139] incorporates tolerance-aware acceptance and fuzzy time quantification to support scheduling under execution uncertainty and soft deadlines. These studies represent three different selection patterns: *EFT-GVNS* emphasizes structurally efficient search and convergence, *EDP-CN* favors learning-guided exploration, and TANS is directed toward uncertainty- and deadline-sensitive scheduling.

From a technique-selection perspective, search-based methods are most appropriate when the scheduling problem is large, combinatorial, and difficult to solve exactly, while sufficient computational time is available for iterative exploration. The choice among GRASP, ILS, and VNS should primarily depend on the dominant scheduling requirement: randomized exploration and flexible intensification favor GRASP, repeated local improvement favors ILS, and problems requiring systematic adaptation of neighborhood structures favor VNS. Adaptive and learning-enhanced variants are particularly relevant when faster convergence, constrained search, or uncertainty handling is required. For cloud–fog–edge environments, these methods are most useful when their neighborhood structures, initialization strategies, and search objectives explicitly reflect task dependencies, heterogeneous resources, latency, energy, cost, or mobility rather than relying on generic search operators.

#### 6.2.1.2.2 Mathematical-based methods

Mathematical-based methods formulate scheduling as constrained optimization problems and provide an explicit representation of resource allocation, task dependencies, and optimization objectives. *Integer Linear Programming (ILP)* and *Mixed-Integer Linear Programming (MILP)* can model objectives such as makespan, energy, cost, latency, and reliability together with resource, precedence, deadline, and communication constraints. Their main advantage is modeling precision and, for tractable instances, the ability to obtain optimal solutions. Their principal limitation is computational scalability, since solution complexity increases rapidly with the number of tasks, resources, and constraints. Consequently, these methods are particularly useful for small or moderate problem instances, offline optimization, benchmarking, and hybrid frameworks.

- **Integer Linear Programming (ILP)**: It provides a precise formulation of task assignment and resource allocation using integer or binary decision variables. In fog–cloud scheduling, it can explicitly represent heterogeneous resources, task precedence, communication delays, deadlines, and QoS requirements. *CASSIA-INT* [140] applies exact ILP to obtain optimal task placement and makespan, but its computational cost limits its application to small-scale or offline scenarios. *CASSIA-RR* [141] addresses this limitation through linear relaxation and randomized rounding, producing near-optimal schedules with lower execution time and therefore offering greater suitability for dynamic environments. *Data-Locality Aware Scheduling in Fog–Cloud (DLSFC)* [142] further incorporates data locality, load balancing, selective migration, shortest-task-first ordering, and resource-aware distribution. Together, these studies illustrate a progression from exact optimization toward approximation and problem-specific strategies that trade a small amount of optimality for improved scalability and practical applicability.

- **Mixed-Integer Linear Programming (MILP):** It extends ILP by allowing both discrete and continuous decision variables, enabling more detailed representations of resource allocation, execution time, data transfer, and resource sharing. Compared with pure ILP, MILP is better suited to scheduling problems involving both discrete placement decisions and continuous allocation variables; however, exact solution remains computationally demanding for large cloud–fog–edge systems. *MINLP* [143] further incorporates nonlinear relationships, allowing more realistic representations of processing time, energy consumption, network latency, and task dependencies. In comparison, *MILP-based scheduling* [144] retains a linear optimization structure that is generally easier to solve, while potentially approximating nonlinear system behavior through additional variables and constraints. These studies therefore highlight a fundamental trade-off between modeling realism and computational tractability: MINLP can represent more complex system interactions, whereas MILP is generally more practical when scalability is important.

For the decision framework, mathematical-based methods are most appropriate when the scheduling problem has clearly defined objectives and constraints and when exactness or a reliable benchmark is required. ILP is particularly useful for relatively small and structured problems or for generating reference solutions; MILP is preferable when both discrete and continuous resource decisions must be represented; and MINLP becomes relevant when nonlinear system relationships are essential to the scheduling model. For large-scale or highly dynamic cloud–fog–edge environments, relaxation, decomposition, randomized rounding, or hybridization with heuristics and metaheuristics can provide a practical compromise between optimization quality and computational cost. Thus, mathematical methods should generally be selected when formulation precision and constraint representation are more important than unrestricted scalability, whereas search-based or hybrid approaches become more appropriate as problem size and dynamism increase.

### 6.2.2 Cross-technique comparison of metaheuristic-based scheduling families across computing paradigms

Tables 7 demonstrates that metaheuristic-based scheduling techniques should not be regarded as interchangeable optimization methods. Although most techniques aim to improve scheduling quality, they differ substantially in their search mechanisms, ability to handle conflicting objectives, convergence behavior, computational requirements, and suitability for different computing environments. Table 7 shows that the reviewed methods address task, workflow, and job scheduling across cloud, fog, fog–cloud, edge, edge–cloud environments, and application-specific use cases. Therefore, the relevant question is not which metaheuristic is universally superior, but under which workload, objective, and infrastructure conditions a particular technique is more appropriate.

**Table 7: Summary of metaheuristic-based scheduling algorithms**

| Method | Publish year | Baseline technique | Scheduling issue | Computing environment | Performance parameters | User constraint | Improvement strategy | Simulation tool | Use case |
|---|---|---|---|---|---|---|---|---|---|
| Multi-objective Parallel Genetic Algorithm (MPGA)[41] | 2020 | GA | Workflow scheduling | Cloud computing | Makespan, cost, and load balancing | QOS | Parallel subpopulations + integer encoding + specialized crossover/mutation | WorkflowSim | Suitable for scheduling scientific workflows in a cloud computing environment. |
| Nondominated Sorting Genetic Algorithm III (NSGA-III) [42] | 2022 | GA | Task scheduling | Cloud computing | Execution time, cost, and energy consumption | N/A | Dominance-based selection + diversity preservation | CloudSim | Suitable for multi-objective task scheduling in cloud computing. |
| Genetic Algorithm-based Load Balancing (GALB) [43] | 2022 | GA | Task scheduling | Cloud computing | Load balancing, resource utilization, and makespan | N/A | Tournament selection + adaptive crossover/mutation | CloudSim | Suitable for load-balancing-aware task scheduling in cloud computing. |
| Cost-Aware Genetic Algorithm (CAG) [44] | 2020 | GA | Task scheduling | Fog-cloud computing | Throughput, delay, and cost | Deadline | Integer resource encoding + cost-based fitness | iFogSim | Suitable for real-time applications with hard deadlines in fog-cloud environments. |
| Improved Elitism Genetic Algorithm (IEGA) [45] | 2021 | GA | Task scheduling | Fog computing | Makespan and energy consumption | QOS | Elitism + adaptive operators | JAVA | Suitable for task scheduling for large-scale IOT applications in Fog computing. |
| Priority-Based Genetic Algorithm (PGA) [46] | 2021 | GA | Task scheduling | Fog-cloud computing | Response time, energy consumption | Deadline | Task classification + latency-aware selection | MATLAB | Suitable for scheduling delay-sensitive IoT tasks in heterogeneous fog-cloud computing environments |
| Improved Genetic Algorithm for | 2022 | GA | Task scheduling | Fog-cloud computing | Execution time, cost, and makespan | Deadline | Precedence-preserving ordered encoding | MATLAB | Suitable for scheduling real-time tasks of IoT applications in the heterogeneous fog-cloud environment. |
| Permutation-Based Optimization Problems (IGA-POP) [47] | | | | | | | | | |
| Makespan-based Optimized scheduler (CLEMO) [48] | 2025 | GA | Workflow scheduling | Fog-cloud computing | Cost, load balancing, energy consumption, and makespan | N/A | Dependency-aware encoding + multi-objective fitness + diversity operators | Python | Suitable for finding the optimal schedule plan to process tasks in cloud and fog nodes. |
| Catastrophic Genetic Algorithm (CGA) [49] | 2020 | GA | Task scheduling | Edge-cloud computing | Delay, completion time | Deadline | Similarity-aware crossover + adaptive mutation + catastrophe mechanism | CloudSim | Suitable for task scheduling in edge-cloud collaborative scenarios. |
| Genetic Algorithm for Edge Computing (GA-EC) [50] | 2022 | GA | Task scheduling | Edge computing | Energy consumption, delay, execution time, and load balancing | N/A | Adaptive crossover/mutation for energy and load balancing | N/A | Suitable for scheduling resource-intensive tasks in edge computing. |
| Modified DEA-based scheduling algorithm (MDEA-SA) [51] | 2021 | DE | Task scheduling | Cloud computing | Execution time and makespan | N/A | Swap initialization + jumping mutation + adaptive parameters | N/A | Suitable for independent tasks in a cloud environment with heterogeneous and distributed resources. |
| Hybrid Differential Evolution (HDE) [52] | 2022 | DE | Task scheduling | Cloud computing | Makespan and total execution time | N/A | Adaptive scaling factor + convergence enhancement | CloudSim | Suitable for better outcomes in a smaller number of iterations in cloud computing. |
| Adaptive Guided Differential Evolution-based Slime Mould Algorithm (AGDESMA) [53] | 2023 | DE | Task scheduling | Cloud computing | computational overhead, cost, and makespan | N/A | SMA hybridization + adaptive mutation/crossover | CloudSim | Suitable for multi-objective large-scale task scheduling optimization over Infrastructure-as-a-Service (IaaS) in the cloud computing environment. |
| Multi-Objective Multi-Factor Evolutionary Algorithm (MO-MFEA) [54] | 2023 | DE | Task scheduling | Cloud computing | Execution time, load balancing, and cost | N/A | Skill-factor knowledge transfer + decomposition-based allocation | CloudSim | Suitable for multi-objective cloud task scheduling optimization. |
| Self-Adaptive Energy-Aware Differential Evolution (SAEADE) [55] | 2025 | DE | Task scheduling | Cloud computing | Makespan and energy consumption | N/A | Self-adaptive mutation/crossover strategy selection | N/A | Suitable for energy-aware scheduling for parallel tasks in a multi-cloud environment. |
| Modified Strength Pareto Evolutionary Algorithm II (SPEA-II) [56] | 2024 | DE | Task scheduling | Fog-Cloud computing | Response time, load balancing, and energy consumption | Deadline | Chaotic-opposition initialization + DE mutation + density-based fitness | Python | Suitable for optimal distribution of independent tasks among heterogeneous computing nodes. |
| Energy-efficient Time and Cost (ETC) [57] | 2024 | DE | Workflow scheduling | Fog computing | Execution time, cost, and energy consumption | QOS | Adaptive operators + deadline-aware DVFS | iFogSim | Suitable for latency-sensitive applications. |
| Differential Evolution for Edge Computing (DE-Edge) [58] | 2024 | DE | Task scheduling | Edge computing | Execution time | N/A | Task clustering + workload-aware scheduling + parameter tuning | N/A | Suitable for IoT edge task clustering and scheduling. |
| ch-PICEA-g [59] | 2021 | CEA | Workflow scheduling | Cloud computing | Makespan, cost, and energy consumption | N/A | Chaotic initialization + logistic diversification + improved fitness | WorkflowSim | Suitable for multi-objective workflow scheduling in cloud computing. |
| Cluster-based Cooperative Co-evolutionary (CBCC) [60] | 2023 | CEA | Workflow scheduling | Cloud computing | Execution time and cost | N/A | Cooperative decomposition + cluster-based archive + dynamic allocation | CloudSim | Suitable for security-aware workflow scheduling in public cloud computing. |
| Improved Memetic Algorithm (IMA) [61] | 2024 | MA | Workflow scheduling | Fog computing | Execution time, cost, and resource utilization | N/A | Dual populations + multi-layer encoding + structured local search | iFogSim | Suitable for scheduling of parallel IoT workflows across fog and cloud servers. |
| Knowledge-Driven Memetic Algorithm (KDMA) [62] | 2025 | MA | Workflow scheduling | Cloud computing | Cost and makespan | N/A | Knowledge-driven heuristics + adaptive reproduction | MATLAB | Suitable for large-scale, real-world workflows with complex dependencies. |
| Multi-Objective Task Scheduling model based on ACO (MOTS-ACO) [63] | 2021 | ACO | Task scheduling | Cloud computing | Makespan, turnaround time, energy consumption, and load balancing | N/A | Pareto dominance + adaptive pheromone updates | CloudSim | Suitable for multi-objective task scheduling in a cloud data center. |
| Load Balancing ACO (LBACO) [64] | 2022 | ACO | Task scheduling | Cloud computing | Load balancing, cost, throughput, and execution time | QOS | Load-aware transition rule | CloudSim | Suitable for QoS-based task scheduling in the cloud computing environment. |
| Improved Adaptive Ant | 2022 | ACO | Task scheduling | Cloud computing | Cost, completion time, and load balancing | N/A | Adaptive pheromone | CloudSim | Suitable for solving large-scale task scheduling problems. |
| Colony (IAC) [65] | | | | | | | evaporation/reinforcement | | |
| Hybrid Weighted Ant Colony Optimization (HWACO) [66] | 2023 | ACO | Task scheduling | Cloud computing | Cost, and makespan | N/A | Weight-based heuristic + adaptive pheromone control | CloudSim | Suitable for finding the optimal solution for task scheduling in cloud computing. |
| Inverted Ant Colony Optimization (IACO) [67] | 2023 | ACO | Task scheduling | Cloud computing | Energy consumption, makespan, execution time, and resource utilization | QOS | VM-centric search + multi-objective pheromone updates | MATLAB | Suitable for an energy-efficient task scheduling method for heterogeneous cloud computing. |
| Deadline-Budget Constrained Ant Colony Optimization (DB-ACO) [68] | 2024 | ACO | Workflow scheduling | Cloud computing | Cost and execution time | Deadline | Deadline/budget-aware pheromone and service selection | JAVA | Suitable for deadline-budget-constrained workflow scheduling in cloud computing. |
| Improved Multi-Objective Ant Colony Optimization (IMOACO) [69] | 2024 | ACO | Task scheduling | Fog computing | makespan, cost, and energy consumption | N/A | Multiple colonies + inter-colony communication | iFogSim | Suitable for a large number of tasks and heterogeneous nodes. |
| Task Scheduling Strategy with Ant Colony (TSSAC) [70] | 2025 | ACO | Task scheduling | Fog computing | Completion time, load balancing, and energy consumption | N/A | Nonlinear pheromone updates + container migration | iFogSim | Suitable for searching optimal task scheduling solution based on the DAG task model. |
| Ant-Colony based Energy-Aware Scheduling (ACEAS) [71] | 2023 | ACO | Task scheduling | Edge computing | Energy consumption and completion time | Deadline | Priority + adaptive pheromone + slack reclamation + DVFS | C | Suitable for time-constrained applications in heterogeneous edge computing. |
| Multi-objective Application Scheduling and Placement based on Ant Colony Optimization (MaaCO) [72] | 2024 | ACO | Workflow scheduling | Edge-cloud computing | Delay, resource utilization, load balancing, and energy consumption | QOS | Workflow decomposition + critical-path analysis + DVFS | CloudSim | Suitable for scheduling scientific workflows in green DVFS-enabled edge-cloud computing |
| Improved Many Objective PSO (I_MaOPSO) [73] | 2020 | PSO | Workflow scheduling | Cloud computing | Cost, makespan, reliability, and energy consumption | QOS | Heuristic initialization + adaptive velocity + leader selection | WorkflowSim | Suitable for multi-objective scientific workflow scheduling in cloud computing. |
| Improved Discrete PSO (IDPSO) [74] | 2023 | PSO | Task scheduling | Cloud computing | Execution time | N/A | Discrete encoding + dynamic sine-based inertia | CloudSim | Suitable for cloud computing task scheduling. |
| Enhanced PSO (EPSO) [75] | 2023 | PSO | Workflow scheduling | Cloud computing | Cost and completion time | N/A | Task aggregation + indirect encoding + quality-aware initialization | CloudSim | Suitable for large-scale, workflow-based cloud scheduling problems. |
| Multi-Hierarchy PSO (MHPSO) [76] | 2024 | PSO | Workflow scheduling | Cloud computing | Cost, Completion time, and resource utilization, | Deadline | Workflow aggregation + topology-aware initialization + hierarchical evolution | JAVA, WorkflowSim | Suitable for market-driven workflow scheduling on heterogeneous cloud resources with deadline constraints. |
| Intelligent Energy-Aware Efficient Task Scheduling (IEETS) [77] | 2025 | PSO | Task scheduling | Cloud computing | Response time, makespan, waiting time, load balancing, and resource utilization | N/A | Energy/workload-aware multi-metric fitness | N/A | Suitable for dynamic and heterogeneous cloud workloads |
| PSO Optimized Leader (PSO-OL) [78] | 2024 | PSO | Task scheduling | Fog-cloud computing | Makespan, cost, resource utilization, and throughput | QOS | Enhanced leader selection + dynamic crossover window | CloudSim | Suitable for IoT and Mobile Crowdsensing applications in fog-cloud computing |
| Adaptive PSO (APSO) [79] | 2024 | PSO | Workflow scheduling | Fog-cloud computing | Makespan, cost, and energy consumption | N/A | Adaptive inertia + cognitive/social coefficients | WorkflowSim | Suitable for workflow scheduling in heterogeneous cloud resources. |
| Integer PSO (IPSO) [80] | 2022 | PSO | Task scheduling | Edge computing | Resource utilization | SLA satisfaction | Integer encoding + EDF ordering + discrete operators | DE3C environment | Suitable for device-edge-cloud cooperative computing to improve SLA satisfaction. |
| Task Scheduling–Multi-Objective PSO (TS-MOPSO) [81] | 2022 | PSO | Task scheduling | Edge computing | Execution time, maximum completion time, and total | N/A | Adaptive inertia + shrinkage factor + objective ranking | CloudSim | Suitable for large-scale, heterogeneous, and computationally intensive edge environments. |
| Enhanced Crow Search Optimization (ECSA) [82] | 2021 | CSA | Task scheduling | Cloud computing | Makespan, resource utilization, and load balancing | N/A | Completion-time selection + load-aware VM allocation | CloudSim | Suitable for optimal task scheduling in cloud computing. |
| Linear Scaling-Crow Search Optimization (LS-CSO) [83] | 2022 | CSA | Workflow scheduling | Cloud computing | Makespan, resource utilization, cost, and energy consumption | N/A | Adaptive awareness probability | CloudSim | Suitable for scientific workflow scheduling in heterogeneous cloud computing |
| Imperialist Competitive Algorithm (ICA) [84] | 2023 | CSA | Task scheduling | Cloud computing | Execution time, makespan, and load balancing | N/A | Hierarchical imperialist–colony assimilation | CloudSim | Suitable for cloud computing under controlled dynamics. |
| Crow Search Algorithm for Multi-Objective Makespan and Cost Optimization (CSAMOMC) [85] | 2024 | CSA | Workflow scheduling | Cloud computing | Makespan and cost | N/A | BFS–DFS initialization + hypervolume-based Pareto selection | Python | Suitable for scientific workflow scheduling with well-defined task dependencies. |
| Discrete Butterfly Optimization Algorithm (DBOA) [86] | 2021 | BOA | Workflow scheduling | Edge computing | Energy consumption | N/A | Discrete encoding + chaotic initialization + Lévy flights + genetic operators | iFogSim | Suitable for DVFS-based scientific workflow Scheduling. |
| Fuzzy Butterfly Optimization Algorithm (FBOA) [87] | 2024 | BOA | Task scheduling | Cloud computing | Makespan, throughput, and resource utilization | QOS | Fuzzy global/local search regulation | CloudSim | Suitable for complex and scalable task scheduling across diverse resources in cloud computing. |
| Improved Butterfly Optimization Algorithm (IBOA) [88] | 2024 | BOA | Task scheduling | Fog-cloud computing | Cost | N/A | Adaptive step size + cost-aware guidance | CloudSim | Suitable for application in real-world cloud-fog environments. |
| Butterfly Spotted Hyena Optimization Algorithm (BSHOA) [89] | 2024 | BOA | Task scheduling | Fog-cloud computing | Throughput, energy consumption, makespan, and execution time | QOS | BOA–SHO hybridization + probabilistic operator switching | CloudSim | Suitable for enhancing the service quality for the IoT in cloud-fog computing environments. |
| Butterfly Swarm Optimization (BSO) [90] | 2025 | BOA | Task scheduling | Fog-cloud computing | Cost | QOS | Probabilistic global/local search switching | MATLAB | Suitable for meeting user needs for cost efficiency or high performance. |
| Workflow scheduling based on cat swarm optimization (WSCSO) [91] | 2022 | CSO | Workflow scheduling | Cloud computing | makespan and resource utilization | Task priority | Task prioritization based on size and processing capacity | WorkflowSim | |
| Modified Distance Cat Swarm Optimization (MDCSO) [92] | 2023 | CSO | Job scheduling | Cloud computing | Execution time | N/A | Modified distance calculation | CloudSim, Java | Suitable for IOT cloud job scheduling. |
| Prioritized Task Scheduling Algorithm Using Cat Swarm Optimization (PCSO) [93] | 2023 | CSO | Task scheduling | Cloud computing | Makespan, energy consumption | SLA violations, QOS | Task/VM prioritization | CloudSim, Java | Suitable for prioritized task scheduling and priorities based on unit electricity cost at the data center locations. |
| Multi-objective Cat Swarm Optimization algorithm (MCSOFLB) [94] | 2024 | CSO | Task scheduling | Cloud computing | Makespan, throughput, cost, resource utilization, fault-tolerant, and load balancing | N/A | Multi-objective + dual-level load balancing | CloudSim | Suitable for fault-tolerant, load-balanced task scheduling in a multi-cloud environment. |
| Task Schedule using a Multi-objective Grey Wolf Optimizer (TSMGWO) [95] | 2021 | GWO | Task scheduling | Cloud computing | Resource utilization, makespan, load balancing, and throughput | N/A | Pareto optimization + enriched leader hierarchy | CloudSim | Suitable for task scheduling in cloud data centers. |
| Multi-objective task scheduling Grey wolf optimization (MOTSGWO) [96] | 2022 | GWO | Task scheduling | Cloud computing | makespan, resource utilization, and energy consumption | N/A | Context-aware task prioritization | CloudSim | Suitable for resolving the issue of overutilization and underutilization of cloud resources |
| Grey Wolf Optimization using a new Encoding Mechanism (GWOEM) [97] | 2024 | GWO | Task scheduling | Cloud computing | Makespan | N/A | Compact encoding + greedy replacement | MATLAB | Suitable for scheduling optimization of cloud tasks. |
| Multi-Objective cost-aware Discrete Grey Wolf optimization-based Algorithm (MoDGWA) [98] | 2024 | GWO | Task scheduling | Fog computing | Makespan. reliability, and cost | SLA | Discrete encoding + NSGA-II crowding + load balancing | MATLAB | Suitable for real-time applications in an IoT-Fog scenario in heterogeneous fog computing. |
| Improved WOA for Cloud task scheduling (IWC) [99] | 2020 | WOA | Task scheduling | Cloud computing | Cost, execution time, and resource utilization | N/A | Nonlinear convergence factor + adaptive population sizing | MATLAB | Suitable for both small and large-scale tasks. |
| Gaussian Whale-Cloud Optimization in Cloud Computing (GCWOA) [100] | 2021 | WOA | Task scheduling | Cloud computing | Completion time, load balancing, and resource utilization | N/A | Opposition initialization + adaptive mobility + Gaussian-cloud search | MATLAB | Suitable for processing small tasks and large tasks in cloud computing |
| Multi-objective trust-aware task scheduling in cloud computing (MOTSWOA) [101] | 2023 | WOA | Task scheduling | Cloud computing | Makespan, energy consumption, and execution time | QOS | SLA-aware trust + dynamic task/VM prioritization | CloudSim | Suitable for trust-aware task scheduling in cloud computing |
| W-Scheduler [102] | 2023 | WOA | Task scheduling | Cloud computing | Makespan | SLA | Priority-aware mapping + shrinking/spiral search | CloudSim | Suitable for SLA aware task-scheduling in cloud computing |
| Modified Chimp-Whale Optimization Algorithm (MCWOA) [103] | 2024 | WOA | Task scheduling | Cloud computing | Makespan, energy consumption, and delay | QOS | COA hybridization + Sobol initialization + task clustering | CloudSim | Suitable for guaranteeing cloud-based systems perform at maximum effectiveness while meeting strict QoS requirements |
| WOA-Scheduler [104] | 2024 | WOA | Task scheduling | Cloud computing | Cost, execution time, and resource utilization | N/A | Adaptive coefficients + Pareto exploitation + weighted objectives | CloudSim | Suitable for the complexities of modern cloud services |
| Golden Eagle Whale Optimization (GEWO) [105] | 2024 | WOA | Task scheduling | Fog-cloud computing | Completion time, energy consumption, and resource utilization | N/A | Golden Eagle–Whale hybrid + hierarchical search | CloudSim | Suitable for prioritizing task scheduling in a heterogeneous fog-cloud environment. |
| Enhanced Whale Optimization Algorithm (EWOA) [106] | 2024 | WOA | Task scheduling | Edge computing | completion time, cost, and resource utilization | N/A | Chaotic initialization + nonlinear convergence factor | JAVA | Suitable for limited resources and demanding applications in edge computing environments |
| Priority-based HHO (PHHO) [107] | 2021 | HHO | Task scheduling | Cloud computing | Makespan, energy consumption, resource utilization, and throughput | N/A | AHP task prioritization + multi-objective fitness | MATLAB | Suitable for Priority-Oriented Task Scheduling in cloud computing. |
| Multi-strategy HHO (MHHO) [108] | 2023 | HHO | Task scheduling | Cloud computing | Makespan, and resource utilization | N/A | Tent chaos + opposition learning + adaptive energy decay + GWO guidance | CloudSim | Suitable for efficient task scheduling in cloud computing. |
| Fault-Tolerant Trust-Aware Task Scheduler (FTTATS) [109] | 2023 | HHO | Task scheduling | Cloud computing | Makespan, throughput, and fault tolerance | SLA | Task/VM prioritization + adaptive HHO + trust/fault awareness | CloudSim | Suitable for preserving trust in the cloud provider while minimizing the rate of failures in the cloud paradigm. |
| SLA-aware HHO (SLA-HHO) [110] | 2023 | HHO | Workflow scheduling | Cloud computing | Makespan | SLA | Task prioritization + SLA-aware fitness | WorkflowSim | Suitable for SLA based Workflow Scheduling in Cloud Computing |
| Multi-Tactic HHO (MTHHO) [111] | 2024 | HHO | Task scheduling | Cloud computing | Makespan, energy consumption, and execution time | QOS | Sobol initialization + elite opposition + Gaussian walk | SimPy | Suitable for energy-efficient VMs, considering the job dependencies |
| Enhanced HHO (EHHO) [112] | 2025 | HHO | Task scheduling | Cloud computing | Makespan, execution time, and cost | QOS | Dynamic random walk + Lévy flights + siege strategy | CloudSim | Suitable for handling complex scheduling challenges in dynamic cloud computing environments. |
| HHO with Local Search (HHOLS) [113] | 2021 | HHO | Task scheduling | Fog computing | Makespan, energy consumption, and cost | QOS | Discrete mapping + swap mutation + local search | JAVA | Suitable for industrial IOT task scheduling in fog computing applications. |
| Discrete Opposition-based HHO (DO-HHO) [114] | 2022 | HHO | Workflow scheduling | Fog computing | MakeSpan | Deadline | Discrete genetic operators + opposition learning | iFogSim | Suitable for efficient workflow scheduling in multi-fog computing. |
| Multiobjective HHO-based Task Scheduling (MoHHOTS) [115] | 2024 | HHO | Task scheduling | Fog -cloud computing | Delay, energy consumption, and load balancing | N/A | Adaptive mutation + Pareto selection + Lévy flights | MATLAB | Suitable for multi-objective task scheduling in cloud–fog computing |
| Trust-aware scheduling algorithm using firefly optimization (TAFFA) [116] | 2023 | FOA | Task scheduling | Cloud computing | Makespan, availability, and throughput | SLA | Task/VM prioritization + SLA-based trust | CloudSim | Suitable for heterogeneous resources with different types and different runtime processing capacities. |
| Modified Firefly Optimization Algorithm (ModFOA) [117] | 2024 | FOA | Workflow scheduling | Edge-cloud computing | Makespan, resource utilization, and energy consumption | N/A | Crossover + Gaussian mutation + quasi-reflection + dynamic steps | CloudSim | Suitable for scientific workflow in hybrid cloud edge environments. |
| Adaptive Firefly Algorithm (AFA) [118] | 2025 | FOA | Task scheduling | Fog computing | Execution time | N/A | Dependency-aware fitness + adaptive | JAVA | Suitable for dependent task scheduling in IoT-Fog computing. |
| | | | | | | | absorption + discrete distance | | |
| Enhanced Sunflower Optimization (ESFO) [119] | 2021 | SFO | Task scheduling | Cloud computing | Makespan, and energy consumption | N/A | Enhanced pollination strategy | CloudSim, MATLAB | Suitable for finding an optimal task scheduling problem with polynomial time complexity. |
| Opposition-based sunflower optimization (OSFO) [120] | 2022 | SFO | Task scheduling | Cloud computing | Makespan, cost, energy consumption, and throughput, | N/A | Opposition-based learning + improved pollination | CloudSim | |
| Improved Multi-Objective Cuckoo Search (IMOCS) [121] | 2022 | CS | Task scheduling | Edge computing | Energy consumption and delay | QOS | Weighted multi-objective aggregation | N/A | Suitable for task scheduling in heterogeneous mobile edge computing. |
| Multi-Objective Cuckoo Search (MOCS) [122] | 2023 | CS | Task scheduling | Cloud computing | Execution time, completion time, and energy consumption | Priority | External archive + Pareto dominance + fast non-dominated sorting | C++ | Suitable for multi-objective task scheduling in cloud computing. |
| Multi-objective Discrete Cuckoo Search Algorithm (MoDCSA) [123] | 2024 | CS | Task scheduling | Fog computing | Makespan and reliability | QOS | Discrete operators + local search + reliability awareness | MATLAB | Suitable for scheduling BoT IoT applications in heterogeneous fog computing. |
| Multi-Objective Improved Monarch Butterfly Optimization (MOIMBO) [124] | 2021 | MBO | Task scheduling | Cloud computing | Makespan, reliability, load balancing | QOS | Pareto dominance + adaptive subpopulations + greedy selection | CloudSim | Suitable for large-scale cloud data centers with conflicting objectives |
| Multi-Objective Monarch Butterfly Optimization Algorithm (MO-MBO) [125] | 2024 | MBO | Workflow scheduling | Edge-fog-cloud computing | Makespan, cost, and energy consumption | QOS | Epsilon-fuzzy dominance + elite archive + self-adaptation | WorkflowSim | Suitable for heterogeneous edge–fog–cloud scheduling with multiple QoS objectives. |
| Improved Multi-Objective Gravitational Search Algorithm (IMOGSA) [126] | 2021 | GSA | Task scheduling | Cloud computing | Execution time, completion time, and makespan | N/A | NSGA-II non-dominated sorting + crowding distance + SCA-based adaptive acceleration weights | CloudSim | Suitable for multi-objective scheduling in heterogenous cloud computing. |
| Gravitational Search Algorithm for Cloud Load Distribution (GSACLD) [127] | 2024 | GSA | Task scheduling | Cloud computing | Cost | N/A | Cloudlet–VM assignment modeling + cost-aware fitness function | CloudSim | Suitable for optimizing the allocation of tasks between cloudlets and virtual machines and cost-aware scheduling. |
| Star-Quake Multi-Objective Gravitational Search Algorithm (SQMOGSA) [128] | 2024 | GSA | Task scheduling | Fog-cloud computing | Makespan, energy consumption, and cost | QOS | Star-quake operator for escaping local optima and balancing exploration/exploit ation | MATLAB | Suitable for task scheduling in IoT-based cloud–fog computing. |
| Modified Multi-Verse Optimization (MMVO) [129] | 2020 | MVO | Workflow scheduling | Cloud computing | Makespan, resource utilization, and reliability | QOS | Grid-based dominance + Pareto archive + trust-aware fitness | CloudSim | Suitable for trust-based workflow scheduling in heterogenous cloud computing. |
| Enhanced Multi-Verse Optimization (EMVO) [130] | 2021 | MVO | Task scheduling | Cloud computing | Makespan, resource utilization, and throughput | N/A | Elitism-based memory and reinjection of high-quality solutions | MATLAB | Suitable for static scheduling in cloud computing. |
| Improved Multi-Objective Multi-Verse Optimization (IMoMVO) [131] | 2022 | MVO | Task scheduling | Cloud computing | Makespan, execution time, and throughput | N/A | Dynamic average position using best and second-best solutions + multi-objective optimization | CloudSim | Suitable for QOS-aware scheduling in cloud computing. |
| Priority-Sensitive Simulated Annealing (PSA) [132] | 2025 | SA | Task scheduling | Fog computing | Makespan, energy consumption, cost, and reliability | Priority | Priority-aware penalty function for priority-sensitive resource allocation | SimPy | Suitable for priority-aware task scheduling in dynamic fog computing. |
| Dynamic Cooling Simulated Annealing (DCSA) [133] | 2025 | SA | Task scheduling | Edge computing | Execution time | N/A | Adaptive cooling rate + iteration-based neighborhood generation | MATLAB | Suitable for the latency-sensitive scheduling in IOT and heterogenous edge computing. |
| Parallel Simulated Annealing with Time Arrival Scheduling (PANTAS) [134] | 2025 | SA | Task scheduling | Edge-cloud computing | Makespan, energy consumption, and reliability | N/A | Parallel SA instances + time-arrival scheduling for joint offloading and timing optimization | CloudSim | Suitable for energy-aware task scheduling in edge-cloud computing. |
| GRASP-based scheduling [135] | 2023 | GRASP | Task scheduling | Fog-cloud computing | Response time, energy consumption, and makespan | Deadline | Randomized greedy construction + local search for | CloudSim | Suitable for deadline-aware IoT tasks scheduling in virtualized fog-cloud |
| | | | | | | | exploration and intensification | | computing with resource constarint. |
| Improved GRASP (IGRASP) [136] | 2023 | GRASP | Task scheduling | Cloud computing | Makespan | N/A | Fixed-set learning from elite solutions to reduce search space and accelerate convergence | MATLAB | Suitable for optimal resource allocation in cloud computing. |
| ILS-based scheduling [137] | 2022 | ILS | Workflow scheduling | Cloud computing | Makespan | N/A | Perturbation + local search + acceptance criterion to escape local optima | WorkflowSim | Suitable for scientific workflow scheduling in cloud computing. |
| Adaptive Iterated Local Search (AILS) [138] | 2022 | ILS | Workflow scheduling | Cloud computing | Execution time and cost | Budget | HEFT initialization + adaptive perturbation + SA-based acceptance | WorkflowSim and Java | Suitable for budget-constrained scientific workflow scheduling in cloud computing. |
| EFT-GVNS [139] | 2024 | VNS | Task scheduling | Cloud computing | Execution time and makespan | Precedence | EFT initialization + composite neighborhoods for efficient diversification and intensification | CloudSim | Suitable for precedence-constrained tasks optimization on heterogeneous systems. |
| EDA-CN [140] | 2025 | VNS | Task scheduling | Cloud computing | Energy consumption and completion time | N/A | Elite-solution distribution learning to guide neighborhood sampling | Python | Suitable for energy-efficient and fault-aware task scheduling in homogeneous cloud computing. |
| Tolerance-Aware Neighborhood Search (TANS) [141] | 2025 | VNS | Workflow scheduling | Cloud computing | Delay | Deadline | Tolerance-aware acceptance + fuzzy time quantification + composite neighborhoods | Java | Suitable for tolerance-ware workflow scheduling with uncertainties in cloud manufacturing. |
| CASSIA-INT [142] | 2021 | ILP | Task scheduling | Fog-cloud computing | Execution time and makespan | QOS | Exact ILP formulation for optimal task placement and makespan minimization | Java | Suitable for offline and static scheduling in small scale fog-cloud computing. |
| Data-Locality Aware Job Scheduling in Fog-Cloud (DLSFC) [143] | 2025 | ILP | Job scheduling | Fog-cloud computing | Makespan and response time. | N/A | Linear relaxation + randomized rounding for near-optimal, lower-cost scheduling | CloudSim | Suitable for location-aware job scheduling for IoT application in cloud and fog computing. |
| MINLP [144] | 2021 | MILP | Task scheduling | Cloud computing | Completion time | N/A | Data locality + load balancing + selective migration for efficient task placement | CloudSim | Suitable for real-time fog computing-based applications. |
| MILP-based scheduling [145] | 2025 | MILP | Workflow scheduling | Fog computing | Cost | Deadline | Nonlinear modeling of processing, energy, latency, and task dependencies | Optimization Programming Language (OPL) | Suitable for deadline-constrained security-aware workflow scheduling in a hybrid cloud architecture. |

A fundamental distinction among metaheuristic families concerns the balance between global exploration and local exploitation. Population-based evolutionary and bio-inspired techniques such as GA, PSO, ACO, WOA, HHO, and CSA maintain multiple candidate solutions or search trajectories, allowing them to explore complex scheduling spaces. This property is particularly valuable for cloud scheduling, where a large number of tasks and resources creates a highly combinatorial search space. For example, GALB uses GA to jointly improve load balancing, makespan, and resource utilization, while NSGA-III addresses execution time, cost, and energy as a multi-objective optimization problem. However, broad exploration does not necessarily imply rapid decision-making. Population-based algorithms generally require repeated fitness evaluations and iterative population updates, which can become costly as the number of tasks and resources increases. This limitation becomes more important in fog and edge environments, where scheduling decisions must often be produced under tighter latency and computational constraints. Consequently, techniques with more controlled search, such as simulated annealing, local search, GRASP, ILS, and VNS, can be attractive when computational efficiency or rapid refinement is more important than maintaining a large population. The reviewed SA variants illustrate this trade-off: DCSA focuses on improving convergence through adaptive cooling, whereas PANTAS uses parallel search to improve solution diversity while jointly optimizing offloading and scheduling timing. Thus, population-based techniques are preferable when search-space coverage is critical, whereas local-search and adaptive techniques become more attractive when convergence efficiency and computational responsiveness dominate. This distinction is more informative than simply stating that one family "converges faster," because convergence quality depends on workload size, initialization, stopping criteria, and the complexity of the objective function.

Another important distinction is the ability to accommodate dynamic resource availability and heterogeneous workloads. Conventional metaheuristics can optimize a static representation of the scheduling problem effectively, but their effectiveness may decrease when task arrivals, resource states, or network conditions change faster than the optimization process. This limitation is particularly relevant to fog and edge systems. Adaptive variants attempt to address this problem by modifying search parameters or incorporating problem-specific information. For example, DCSA dynamically adjusts its cooling process, while AILS introduces adaptive local-search behavior for workflow scheduling. These mechanisms improve responsiveness, but they also introduce additional design and parameter-tuning requirements. Therefore, adaptive metaheuristics are most beneficial when workload or resource states vary significantly, whereas conventional variants can remain competitive in stable or offline scheduling scenarios.

Scalability represents a particularly important trade-off that is insufficiently captured by average performance metrics alone. Metaheuristics can address NP-hard scheduling problems without requiring exact optimization, but iterative population-based search can itself become computationally expensive as the problem size increases. The reviewed mathematical approaches provide an instructive comparison: ILP can provide optimal solutions for small-scale problems but its computational complexity increases rapidly with system size and heterogeneity, motivating the use of heuristic or metaheuristic approximations for larger scenarios. Metaheuristics therefore occupy an intermediate position between exact optimization and lightweight heuristics. They generally provide greater search flexibility than simple constructive heuristics but do not provide the optimality guarantees of exact formulations. This trade-off explains their strong presence in cloud scheduling and their more selective application in resource-constrained edge environments. For large-scale cloud scheduling, population-based parallel or hybrid approaches can exploit abundant computational resources; for edge scheduling, reducing population size, iteration count, or search complexity may be necessary even if this sacrifices some solution quality.

#### 6.2.3 Statistical analysis of metaheuristic-based strategies

Figure 6 provides a statistical overview of metaheuristic-based scheduling research across computing paradigms, algorithmic families, simulation environments, and evaluation metrics. Cloud computing represents the dominant computing paradigm (60%), followed by fog (28%) and edge computing (12%). This distribution reflects the suitability of centralized cloud infrastructures for large-scale metaheuristic optimization, whereas fog and edge environments introduce greater resource heterogeneity, latency, decentralization, and workload dynamics. Consequently, fog-based approaches increasingly employ adaptive, migration-aware, and multi-objective mechanisms, while the limited adoption of metaheuristics in edge computing highlights the challenges posed by constrained resources, mobility, and communication overhead.

The strong dominance of bio-inspired techniques (60%) over evolutionary methods (21%) indicates that the research community favors algorithms whose search mechanisms naturally support exploration of highly nonlinear and heterogeneous scheduling spaces. Bio-inspired methods imitate behaviors such as foraging, hunting, signaling, and collective decision-making, providing flexible mechanisms for balancing exploration and exploitation. Evolutionary techniques nevertheless remain important because population evolution, selection, crossover, and mutation provide systematic mechanisms for progressively improving candidate schedules; however, their lower proportion suggests that researchers increasingly favor more behaviorally diverse search mechanisms when addressing heterogeneous and dynamic environments. Among bio-inspired approaches, ACO (17%), PSO (15%), HHO (15%), and WOA (13%) are the most prominent due to their flexible exploration–exploitation mechanisms and suitability for task–resource mapping. ACO is particularly effective for dependency-aware and workflow scheduling, while PSO offers simple implementation and rapid convergence for large-scale problems; HHO and WOA provide adaptive search capabilities for dynamic environments. However, their susceptibility to computational overhead, premature convergence, and parameter sensitivity has motivated the development of enhanced and hybrid variants. Thus, their prevalence reflects adaptability and methodological maturity rather than universal superiority.

The evaluation results indicate that makespan (23.7%) is the most frequently optimized metric, followed by cost and energy consumption (15.5% each) and execution time (12.4%). The dominance of makespan reflects its direct relevance to overall scheduling efficiency, while the increasing emphasis on cost and energy demonstrates a shift toward resource- and sustainability-aware optimization. Consequently, many studies formulate scheduling as a multi-objective problem that jointly considers time, cost, and energy, often incorporating additional criteria such as resource utilization, load balancing, reliability, and QoS. Overall, the findings suggest a transition from single-objective optimization toward multi-objective scheduling that balances temporal, economic, and energy-efficiency requirements.

CloudSim is the predominant evaluation environment (49%), followed by MATLAB (17%), reflecting the widespread reliance on controllable and reproducible simulation platforms in scheduling research. CloudSim facilitates systematic modeling of cloud resources, workloads, and scheduling policies, whereas MATLAB is particularly suitable for rapid implementation and evaluation of optimization algorithms. However, the strong reliance on simulation indicates a potential limitation, as results obtained under controlled settings may not fully capture real-world workload variability, resource contention, communication delays, and infrastructure dynamics.

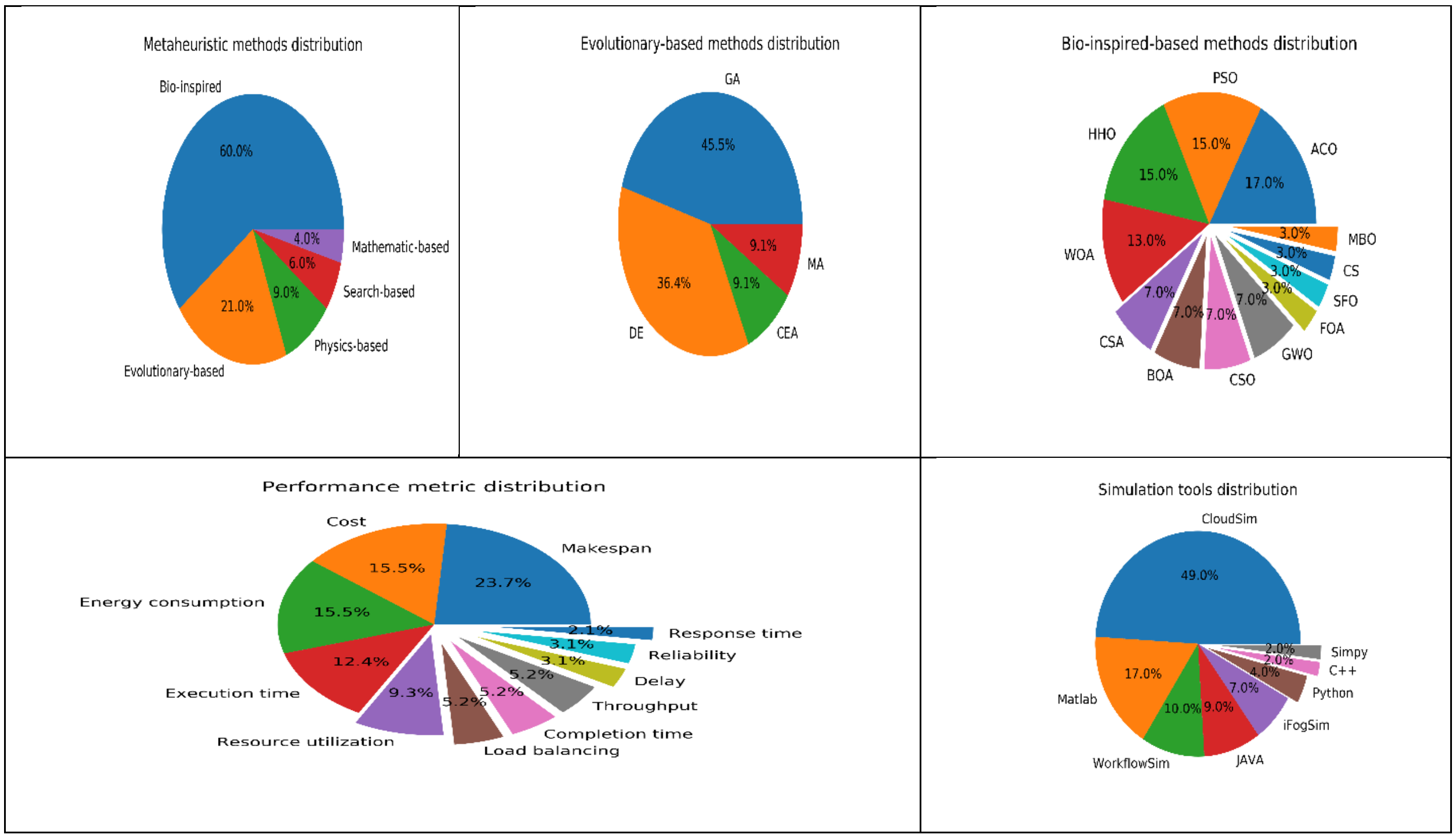

**Figure 6: Statistical information of metaheuristic-based techniques**

### 6.2.4 Guideline for selecting metaheuristic-based techniques

A practical metaheuristic selection process should consider four factors: computing infrastructure, scheduling structure, optimization objectives, and workload dynamics. Cloud environments generally prioritize scalability and multi-objective resource allocation, whereas fog and edge environments require greater emphasis on heterogeneity, energy efficiency, latency, QoS, and computational overhead. Algorithm variants should therefore be selected according to specific operational requirements and adaptability rather than the underlying metaheuristic family alone. Technique selection should be viewed as a requirement-driven decision process rather than being based solely on algorithmic popularity. The evidence in Table 8 indicates that the appropriate metaheuristic depends jointly on the computing paradigm, scheduling objectives, workload structure, resource constraints, QoS requirements, and computational overhead. In cloud environments, GA, NSGA-III, ACO, and PSO are well suited to large-scale and multi-objective scheduling, whereas fog and edge environments favor adaptive or lightweight variants that can address latency, energy, heterogeneity, and resource limitations. Similarly, cost- and energy-aware algorithms are preferable when economic or sustainability constraints dominate, while deadline- and QoS-sensitive applications require explicit constraint-handling mechanisms. For workflow scheduling, algorithms capable of preserving task dependencies, such as GA and local-search approaches, are particularly appropriate. Overall, the selection process should prioritize alignment between algorithmic capabilities and application requirements**,** rather than assuming that the most frequently used technique is universally superior.

**Table 8: Decision framework for choosing metaheuristic-based scheduling algorithms based on requirements and application contexts**

| Requirement | Recommended Family | Preferred Techniques | Best Environment | Avoid When |
|---|---|---|---|---|
| Minimize makespan | Bio-inspired / Evolutionary | GA, PSO, ACO, WOA, HHO | Cloud, Fog–Cloud | Scheduling decisions must be generated instantaneously |
| Minimize execution time | Fast-converging Metaheuristic | PSO, HHO, SA | Cloud, Fog, Edge | Very large search spaces require extensive exploration |
| Minimize cost | Cost-aware Metaheuristic | GA, ACO, PSO | Cloud, Fog–Cloud | Cost is not a primary scheduling objective |
| Minimize energy consumption | Energy-aware Metaheuristic | GA, PSO, HHO, WOA | Fog, Edge, Fog–Cloud | Computing resources are not energy constrained |
| Optimize makespan and cost jointly | Multi-objective Metaheuristic | GA, PSO, ACO, NSGA-III | Cloud, Fog–Cloud | Strict real-time constraints limit iterative optimization |
| Optimize makespan and energy jointly | Multi-objective Metaheuristic | GA, PSO, HHO, WOA | Fog, Edge, Cloud | Energy has negligible importance |
| Optimize cost, energy, and makespan simultaneously | Multi-objective Evolutionary/Bio-inspired | NSGA-III, GA–based variants, PSO-based variants | Cloud, Fog–Cloud | Scheduling overhead must be extremely low |
| Meet strict deadlines | Priority-/Constraint-aware Metaheuristic | Priority GA, GA variants, SA | Fog, Fog–Cloud, Edge | Tasks have no deadline or QoS requirements |
| Real-time task scheduling | Adaptive / Fast-converging Metaheuristic | PSO, SA, HHO | Fog, Edge | Offline optimization is sufficient |
| Load balancing | Resource-aware Metaheuristic | GA, PSO, ACO | Cloud, Fog–Cloud | Resource-state information is unavailable |
| Maximize resource utilization | Resource-aware Metaheuristic | GA, PSO, CSA | Cloud, Fog–Cloud | Resources are homogeneous and already well balanced |
| Handle heterogeneous resources | Adaptive Bio-inspired / Evolutionary | GA, PSO, HHO, WOA | Fog, Fog–Cloud | Infrastructure is homogeneous |
| Handle dynamic workloads | Adaptive Metaheuristic | HHO, WOA, adaptive PSO, adaptive SA | Fog, Edge, Cloud–Edge | Workloads are static and predictable |
| Scientific workflow scheduling | Evolutionary / Multi-objective | MPGA, NSGA-III, ILS, AILS | Cloud | Tasks are independent and have no workflow dependencies |
| Workflow dependencies | Dependency-aware Metaheuristic | GA, ACO, ILS, VNS | Cloud, Fog–Cloud | Tasks are completely independent |
| Task prioritization | Priority-aware Metaheuristic | Priority-based GA, SA variants | Fog, Edge, Real-time IoT | All tasks have equal priority |
| Data-locality-aware scheduling | Locality-aware Metaheuristic | ACO-based, locality-aware search | Fog–Cloud, Edge–Cloud | Data-transfer overhead is negligible |
| Low-latency scheduling | Fast-converging / Local-search Metaheuristic | SA, PSO, HHO | Edge, Fog | Solution quality is more important than decision latency |
| Avoid premature convergence | Exploration–exploitation balanced | HHO, WOA, PSO, GA | Cloud, Fog | Computational resources are severely constrained |
| Large-scale scheduling | Population-based / Parallel Metaheuristic | GA, PSO, ACO, parallel variants | Cloud, Distributed Cloud | Edge devices cannot support iterative population search |
| Multi-objective QoS optimization | Multi-objective Evolutionary/Bio-inspired | NSGA-III, GA, PSO, ACO | Cloud, Fog–Cloud | Only one QoS criterion needs optimization |
| Reliability-aware scheduling | Multi-objective Metaheuristic | PSO, SA, GA-based approaches | Fog–Cloud, Edge–Cloud | Reliability is not a system requirement |
| Resource-constrained edge scheduling | Lightweight / Adaptive Metaheuristic | SA, PSO, lightweight HHO variants | Edge | Computationally expensive population sizes are required |
| Energy-efficient IoT scheduling | Energy-aware Bio-inspired | GA, PSO, WOA, HHO | Fog, Edge | IoT devices have sufficient energy and processing capacity |
| Static/offline optimization | Evolutionary / Mathematical-inspired Metaheuristic | GA, ACO, ILS, VNS | Cloud, Small-scale Fog–Cloud | Workloads change rapidly during scheduling |
| Dynamic online optimization | Adaptive Metaheuristic | Adaptive PSO, HHO, WOA, SA | Fog, Edge | Scheduling decisions cannot tolerate iterative search |
| Balance exploration and exploitation | Advanced Bio-inspired | HHO, WOA, PSO | Cloud, Fog | Optimization problem is small and simple |

## 6.3 Learning-based techniques

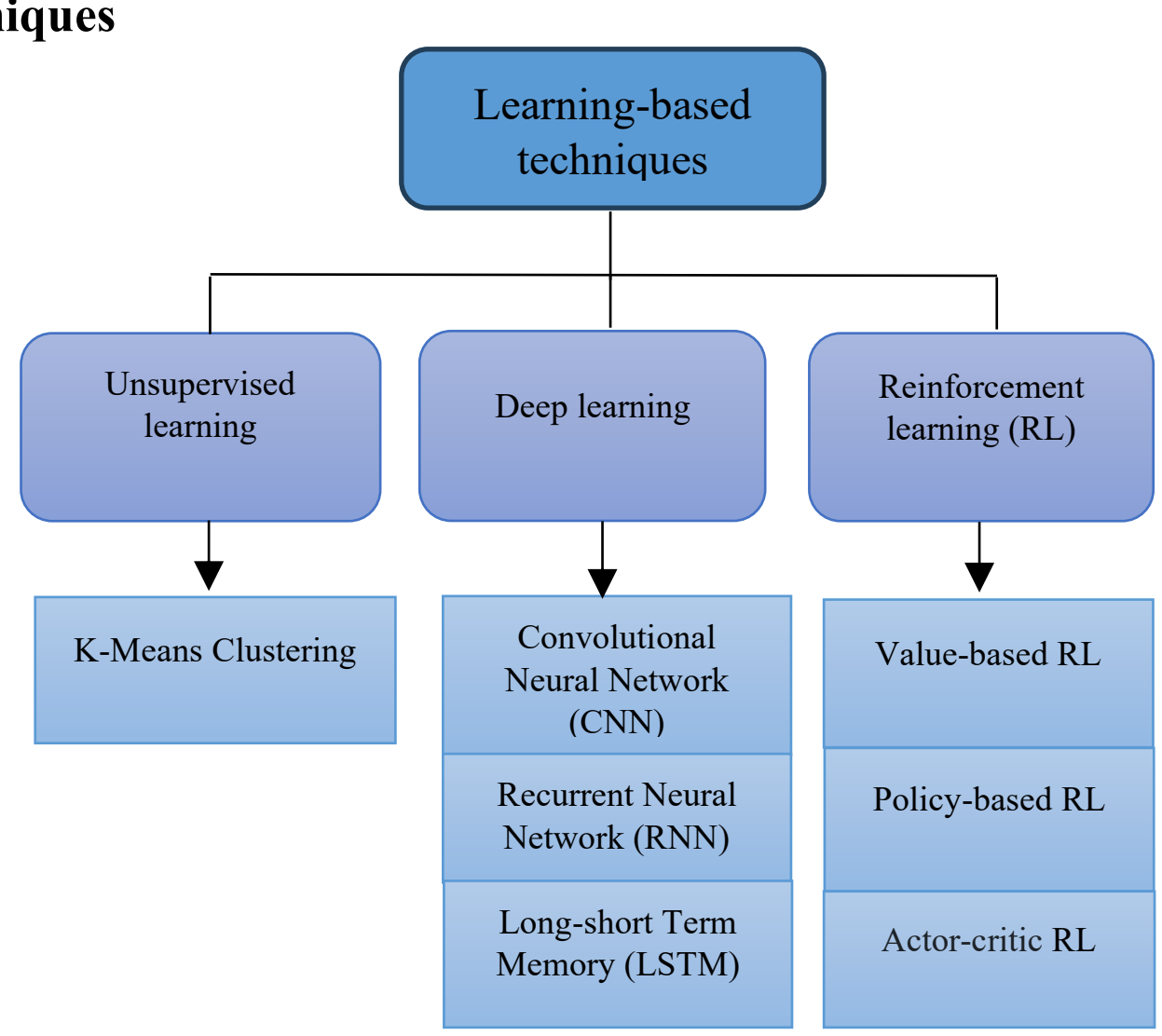

**Figure 7: Classification of learning-based techniques**

Learning-based scheduling techniques use data-driven models to learn task–resource relationships and scheduling policies from observed system states, historical data, or interactions with the computing environment. Unlike rule-based approaches, these methods can adapt scheduling decisions to workload and resource variations, making them attractive for heterogeneous and dynamic cloud–fog–edge environments. However, their applicability depends on factors such as data availability, training cost, state and action dimensionality, model complexity, and the ability to adapt to workload changes.

#### 6.3.1 Classification of learning-based techniques

Based on their learning mechanism, these approaches are classified into unsupervised learning, deep learning, and reinforcement learning (RL), as illustrated in Figure 7. RL is further divided into value-based, policy-based, and actor–critic methods. Unsupervised learning primarily supports task/resource grouping and workload organization; deep learning focuses on feature extraction and prediction; and reinforcement learning learns scheduling policies through interaction with the environment. Within RL, value-based methods estimate action values, policy-based methods directly optimize scheduling policies, and actor–critic methods combine policy and value estimation. Rather than describing every algorithm independently, this section compares representative approaches according to their baseline techniques, extension strategies, objectives, environments, evaluation tools, and application contexts.

##### 6.3.1.1 Unsupervised learning techniques

Unsupervised learning identifies patterns or groups in unlabeled task and resource data. In scheduling, clustering methods such as K-Means can organize tasks according to characteristics such as task length, CPU requirements, memory demand, or resource type before allocation. Their main advantages are relatively low training requirements and simple implementation. However, their effectiveness depends on feature selection, cluster configuration, and the stability of workload patterns.

K-Means-based scheduling has been extended in several directions. *K-means Clustering-based Task Classification and Scheduling (KTCS)* [146] uses task characteristics to classify edge tasks and subsequently distributes them through multiple queues. *Cluster-Based Task Scheduling (CBTS)* [147] clusters tasks and virtual machines according to task length and processing capability to improve workload distribution. *Multi-Objective Optimization Using K-Means Clustering (MOKM)* [148] extends clustering toward multi-objective scheduling by incorporating computation cost, communication cost, earliest finish time, and task length. *DCNS* [149] integrates K-Means with fuzzy logic to provide more flexible task classification in fog environments. These approaches illustrate progression from simple task/resource grouping toward resource-aware, context-aware, and multi-objective clustering.

The principal trade-off is therefore between simplicity and adaptability. Basic K-Means approaches are computationally lightweight but require appropriate feature selection and cluster configuration, whereas fuzzy, adaptive, or multi-objective extensions provide greater flexibility at the cost of additional computational and implementation complexity.

##### 6.3.1.2 Deep Learning Techniques

Deep learning techniques learn nonlinear representations of task and resource characteristics from historical data. CNNs are primarily used for feature extraction from structured task–resource representations, whereas RNN and LSTM architectures capture temporal dependencies in workloads and resource conditions. These methods can reduce dependence on manually designed scheduling rules but generally require larger training datasets and greater computational resources than conventional clustering.

Representative CNN-based approaches include *Dense Neural Network (DenseNet)* [150] and *HunterPlus* [151]. DenseNet extends CNN-based scheduling through dense connectivity and feature reuse, supporting task clustering, resource utilization, and fault-tolerant scheduling. HunterPlus combines CNN processing with bidirectional GRU-based temporal modeling to capture time-dependent task and resource characteristics in cloud–fog environments.

Thus, the principal extension from conventional CNN scheduling is toward stronger feature propagation, temporal awareness, and QoS-oriented decision making.

RNN and LSTM approaches similarly extend predictive scheduling by incorporating historical system states. RNNs can model temporal workload and resource patterns, whereas LSTM uses gated memory to retain longer-term dependencies. Their principal benefit is predictive adaptation to changing workloads; however, training complexity, data requirements, inference overhead, and sensitivity to input quality increase with model complexity.

Overall, deep learning is most appropriate when scheduling decisions depend on complex nonlinear or temporal patterns that can be learned from sufficiently representative historical data. Its principal limitation is that increased predictive capability is accompanied by higher training and deployment requirements.

#### 6.3.1.3 Reinforcement Learning Techniques

Reinforcement learning formulates scheduling as sequential decision making in which an agent observes the system state, selects an action, and receives a reward associated with scheduling performance. States may represent task queues, resource availability, workload characteristics, network conditions, or QoS requirements, while actions correspond to task-to-resource assignments or resource-management decisions. The reward can incorporate makespan, latency, energy, cost, utilization, reliability, or other objectives.

RL is particularly attractive for dynamic environments because scheduling policies can be learned through interaction rather than relying exclusively on labeled historical data. Nevertheless, training time, state–action-space dimensionality, reward design, convergence stability, and computational overhead remain important limitations.

- **Value-based Reinforcement Learning Techniques:** Value-based RL estimates the expected cumulative reward of state–action pairs and selects actions according to their estimated value. Q-Learning is appropriate for relatively small discrete state–action spaces, whereas DQN uses neural networks to approximate the value function and therefore extends value-based learning to larger state spaces.

  *Q-leaning for energy efficiency in cloud computing (QEEC)* [152] extends Q-Learning through hierarchical task ordering and a reward function incorporating response time, CPU utilization, and service requirements. *Q-Learning with Predictive Priority Task Scheduling (QL-PPTS)* [153] combines Q-Learning with predictive priority scheduling and look-ahead execution-time information. These extensions demonstrate that conventional Q-Learning can be improved through more informative reward functions, task prioritization, and predictive scheduling information.

  DQN-based approaches address scalability limitations of tabular Q-Learning. *Extended Prioritized Experience Replay Deep Q-Network (EPERDQN)* [154] improves DQN through prioritized experience replay and richer state representation, whereas *Clipped Double Deep Q-Learning (CDDQL)* [155] focuses on reducing Q-value overestimation and improving training stability. *Clipped Double Deep Q-Learning (CDDQL)* [156] introduces recurrent processing to exploit historical observations under partial observability. *Deep Reinforcement Learning-based Multi-Objective Task Scheduler in Cloud Fog Environment (DRLMOTS)* [158] extends DQN-based scheduling toward multi-objective fog–cloud scheduling through enriched state representation and adaptive reward mechanisms. *Deep Reinforcement learning- enhanced Two-Stage Scheduling (DRL-TSS)* [159] incorporates heuristic task presorting to reduce the search space, while *DQN-Edge* [160] combines attention and dual-value estimation to address workflow dependencies and energy–makespan trade-offs.

  These extensions address different limitations of standard DQN: sample efficiency, value-estimation stability, temporal awareness, search-space reduction, and multi-objective optimization. However, each additional mechanism increases architectural complexity and training requirements.

- **Policy-based Reinforcement Learning Techniques:** Policy-based methods directly optimize a parameterized scheduling policy rather than selecting actions exclusively through an estimated value function. This makes them attractive for complex or continuous decision spaces but can introduce high gradient variance, sample inefficiency, and substantial training requirements.

  REINFORCE-based approaches directly optimize policy parameters using cumulative rewards. The reviewed *DRL-based scheduling* [161] applies policy-based deep RL to task scheduling, while *Cloud manufacturing scheduling problem (CMfg-SP)* [162] extends the approach through transformer-based representations, multi-head attention, graph information, context embeddings, and action masking. The latter illustrates a transition from manually designed representations toward richer task–resource representations and constraint-aware action selection.

Proximal Policy Optimization (PPO)-based approaches improve conventional policy-gradient learning by constraining policy updates through a clipped objective. In scheduling, *IRATS* [163] incorporates mobility, task priority, and communication conditions for vehicular fog environments, whereas *Deep Reinforcement Learning-based IoT application Scheduling algorithm (DRLIS)* [164] uses PPO with multi-objective reward and context-aware state representations for edge–fog scheduling. Thus, PPO-based approaches primarily extend policy learning through training stability, context awareness, mobility adaptation, and multi-objective optimization.

Compared with REINFORCE, PPO generally provides more controlled policy updates but requires repeated optimization and additional computational resources. Therefore, policy-based methods are more suitable when decision flexibility and complex action spaces justify the additional learning cost.

- **Actor-critic (AC) Reinforcement Learning Techniques:** Actor–critic methods combine direct policy optimization with value estimation. The actor selects scheduling actions, while the critic estimates their quality and provides feedback for policy improvement. This structure can reduce policy-gradient variance and support more stable learning in complex environments.

  The reviewed *actor–critic scheduling framework* [165] applies policy optimization to energy- and execution-time-aware scheduling. Hybrid task scheduling technique in fog computing using fuzzy logic and deep reinforcement learning (HTSFFDRL) [166] extends actor–critic scheduling with fuzzy reasoning, asynchronous multi-agent learning, and multi-objective reward design to address uncertainty in fog environments. *RTATSA2C* [167] extends A2C through trust-aware scheduling, task segmentation, dynamic VM scaling, and multi-objective state/reward representations. *MCWS-A3C* [168] applies asynchronous multi-agent A3C to multi-cloud workflow scheduling and introduces workflow-specific value estimation.

  These approaches demonstrate several extension strategies: uncertainty handling, decentralized learning, trust and reliability awareness, workflow-specific value estimation, and multi-objective optimization. The resulting adaptability is accompanied by increased architectural complexity, training cost, and parameter-tuning requirements.

### 6.3.2 Cross-technique comparison of learning-based scheduling families across computing paradigms

The reviewed learning-based techniques exhibit a progressive trade-off between simplicity, predictive capability, adaptability, and computational requirements. Unsupervised learning provides a relatively lightweight mechanism for grouping tasks or resources but offers limited autonomous adaptation after workload patterns change. Deep learning provides stronger nonlinear representation and prediction capabilities but requires representative training data and greater computational resources.

RL addresses dynamic decision making through interaction with the environment. Within RL, value-based methods provide a relatively direct approach for discrete scheduling actions. Q-Learning is appropriate for smaller state–action spaces, whereas DQN extends value estimation to larger state spaces at the cost of neural-network training. DQN variants further address specific limitations through prioritized experience replay, improved value estimation, recurrent processing, heuristic guidance, or multi-objective reward design.

Policy-based RL directly optimizes scheduling policies and is therefore attractive for complex decision spaces. REINFORCE provides a relatively simple policy-gradient formulation but suffers from high gradient variance and sample inefficiency. PPO constrains policy updates and therefore provides a more stable alternative, although it introduces additional optimization overhead.

Actor–critic methods combine the advantages of policy and value learning. They can provide more stable policy optimization for complex scheduling environments but generally require greater architectural and training complexity. Advanced variants such as A2C and A3C further introduce synchronized or asynchronous learning, while recent approaches incorporate fuzzy reasoning, multi-agent learning, trust, reliability, and workflow-specific representations.

Consequently, increasing algorithmic sophistication does not necessarily imply greater practical suitability. The appropriate family depends on workload dynamics, state and action dimensionality, data availability, optimization objectives, computational resources, and deployment constraints.

## Table 10: Summary of the learning-based scheduling algorithm

| Learning-based method | Publish year | Baseline technique | Scheduling issue | Computing environment | Performance parameters | User constraint | Improvement strategy | Simulation tools | Use case |
|---|---|---|---|---|---|---|---|---|---|
| K-means Clustering-based Task Classification and Scheduling (KTCS) [146] | 2020 | K-Means | Task scheduling | Edge computing | Resource utilization and execution time | N/A | Classifies tasks according to resource type and uses multi-queue allocation | CloudSim | Suitable for edge computing with diverse task types. |
| Cluster-Based Task Scheduling (CBTS) [147] | 2021 | K-Means | Task scheduling | Cloud computing | Makespan, execution time, and load balancing | Deadline | Clusters tasks and VMs according to task length and processing capacity | CloudSim | Suitable for load balancing in cloud data centers. |
| Multi-Objective Optimization Using K-Means Clustering (MOKM) [148] | 2024 | K-Means | Task scheduling | Cloud computing | Makespan, resource utilization, and cost. | N/A | Incorporates multiple task and cost-related features for multi-objective clustering | CloudSim | Suitable for computationally intensive scientific tasks by considering multiple conflicting objectives in cloud computing. |
| DCNS with K-means clustering and fuzzy logic [149] | 2023 | K-Means | Task scheduling | Fog computing | Response time and execution time | N/A | Integrates fuzzy logic to provide flexible task clustering | iFogSim | Suitable for task clustering and the subsequent allocation of these optimal clusters to fog devices to identify more suitable virtual machines for allocation. |
| Dense Neural Network (DenseNet) [150] | 2021 | CNN | Task scheduling | Cloud computing | Energy consumption, delay, and cost | Fault tolerance | Uses dense connectivity and feature reuse for task classification and scheduling | CloudSim | Suitable for task clustering and scheduling in a fault-tolerant cloud. |
| HunterPlus [151] | 2023 | CNN | Task scheduling | Fog-cloud computing | Energy consumption, waiting time, and completion time | QOS and SLA violation | Integrates bidirectional GRU for temporal dependency modeling | Python | Suitable for energy-efficient task scheduling for cloud–fog computing environments. |
| Q-leaning for energy efficiency in cloud computing (QEEC) [152] | 2020 | QL | Task scheduling | Cloud computing | Response time, resource utilization, and energy consumption | User constraint in SLA | Adds hierarchical task ordering and multi-objective reward design | CloudSim | Suitable for dynamic task scheduling in a heterogeneous cloud computing environment |
| Q-Learning with Predictive Priority Task Scheduling (QL-PPTS) [153] | 2023 | QL | Task scheduling | Edge computing | Makespan and execution time | N/A | Integrates predictive priority information and look-ahead scheduling | Python | Suitable for static task scheduling problem in edge environments. |
| Extended Prioritized Experience Replay Deep Q-Network (EPERDQN) [154] | 2024 | DQN | Task scheduling | Cloud computing | Response time, waiting time, throughput, and MakeSpan | SLA violations | Uses prioritized experience replay and enriched state representation | CloudSim and Python | suitable for use in environments with complex situations and large scales. |
| Clipped Double Deep Q-Learning (CDDQL) [156] | 2021 | DQN | Task scheduling | Fog computing | Delay, cost, resource utilization, and energy consumption | Deadline | Uses clipped double Q-learning and dual target networks to reduce overestimation | Python | Suitable for a parallel scheduler that is equipped in each fog node. |
| Deep recurrent Q-network (DRQN) [157] | 2022 | DQN | Task scheduling | Fog computing | Resource utilization and energy consumption | N/A | Incorporates recurrent processing to exploit historical states | Python | Suitable for Task Scheduling in Vehicular Fog Computing |
| Deep Reinforcement Learning-based Multi-Objective Task Scheduler in Cloud Fog Environment (DRLMOTS) [158] | 2025 | DRN | Task scheduling | Fog-cloud computing | Energy consumption, makespan, and Fault tolerance | N/A | Enriches state representation and reward design for multi-objective scheduling | Python | Suitable for computationally intensive and time-sensitive tasks in heterogeneous cloud fog environments. |
| Deep Reinforcement learning- enhanced Two-Stage Scheduling (DRL-TSS) [159] | 2023 | DRN | Task scheduling | Edge-cloud computing | Makespan | N/A | Uses heuristic task presorting to reduce the search space | N/A | Suitable for intensive tasks in end–edge–cloud-enabled IoE applications. |
| DQN-Edge [160] | 2026 | DQN | Workflow scheduling | Edge computing | Makespan and energy consumption | N/A | Integrates attention, dual value estimation, and workflow-aware ranking | Python | Suitable for an efficient workflow scheduling method for edge computing environments. |
| DRL-based task scheduling [161] | 2021 | Policy-based DRL | Task scheduling | Edge computing | N/A | N/A | Uses MDP formulation, state representation, reward design, and neural policy learning | Python | Suitable for computationally intensive task scheduling problems in the IoT edge system. |
| Cloud manufacturing scheduling problem (CMfg-SP) [162] | 2022 | Policy-based DRL | Task scheduling | Cloud computing | Makespan, cost, reliability, | QOS | Integrates transformer/MHA, graph representation, | Python | Suitable for task scheduling problems in cloud manufacturing. |
| | | | | | and execution time | | context embedding, and action masking | | |
| IRATS [163] | 2023 | Policy-based DRL (PPO) | Task scheduling | Fog computing | Delay, waiting time, and completion time | Deadline | Incorporates mobility, priority queues, and communication conditions | SUMO, Omnet++ | Suitable for task scheduling in a vehicular fog network. |
| Deep Reinforcement Learning-based IoT application Scheduling algorithm (DRLIS) [164] | 2024 | Policy-based DRL (PPO) | Task scheduling | Edge-fog computing | Load balancing, response time, and cost | N/A | Uses multi-objective reward and context-aware state representation | Fog-Bus2 and Python | Suitable for adaptive scheduling of IoT applications in heterogeneous edge/fog computing environments. |
| Actor–critic based scheduling framework with proximal policy Optimization [165] | 2022 | Actor–critic | Workflow scheduling | Edge-cloud computing | energy consumption and execution time | Deadline | Combines policy optimization with value estimation | CloudSim | Suitable for energy-efficient resource scheduling in edge–cloud environment. |
| Hybrid task scheduling technique in fog computing using fuzzy logic and deep reinforcement learning (HTSFFDRL) [166] | 2024 | Actor–critic | Task scheduling | Fog computing | makespan, energy consumption, and cost | N/A | Adds fuzzy reasoning, asynchronous learning, and multi-objective reward design | SimPy | Suitable for uncertainties and dynamic challenges of task scheduling in fog computing. |
| RTATSA2C [167] | 2024 | Advantage Actor-Critic (A2C) | Task scheduling | Fog-cloud computing | Makespan and reliability | N/A | Adds trust-aware scheduling, task segmentation, and dynamic VM scaling | Simpy | Suitable for a reliability and trust-aware Task Scheduler for cloud-fog computing |
| MCWS-A3C [168] | 2024 | advantage actor–critic algorithm (A3C) | Workflow scheduling | Cloud computing | Makespan, cost, and resource utilization | N/A | Uses distributed agents and workflow-specific critic/value estimation | Python and workflowSim | Suitable for workflow scheduling in a multi-cloud environment |

### 6.3.3 Statistical analysis of learning-based strategies

Figure 8 summarizes the distribution of the reviewed learning-based scheduling studies according to scheduling problem, computing environment, optimization objectives, and evaluation methodology. Task scheduling represents 86% of the reviewed studies, whereas workflow scheduling accounts for 14%. This distribution indicates that learning-based scheduling has been applied predominantly to task-level decision making, while workflow-oriented scheduling remains less represented in the reviewed sample.

Across computing environments, fog computing accounts for 36% of the reviewed studies, while cloud and edge computing each account for 32%. This distribution reflects the suitability of learning-based methods for environments characterized by heterogeneous resources, changing workloads, and latency requirements. However, these percentages describe the reviewed sample and should not be interpreted as measurements of the entire learning-based scheduling literature.

The reviewed studies also demonstrate a progression from conventional machine learning toward deep reinforcement learning. Unsupervised learning is mainly used for task/resource grouping, whereas deep learning is employed for prediction and representation learning. RL and DRL are increasingly used when scheduling requires sequential adaptation to changing system states. This progression reflects the increasing complexity of scheduling environments rather than establishing that one learning paradigm is universally preferable.

The main optimization objectives remain makespan, execution time, energy consumption, and resource utilization. Other requirements, including reliability, fault tolerance, security, scalability, cost, and QoS, appear less frequently. This distribution indicates an opportunity for future studies to consider broader multi-objective formulations rather than focusing predominantly on operational efficiency.

Evaluation methodology also shows substantial dependence on simulation. CloudSim is frequently used for cloud-oriented evaluation, while Python-based implementations are common for RL and deep-learning approaches. Fog- and edge-oriented studies additionally employ platforms such as iFogSim, SimPy, SUMO, OMNeT++, and Fog-Bus2. Although simulation facilitates controlled comparison and reproducibility, the limited use of real-world testbeds and operational workloads restricts conclusions about deployment robustness, workload drift, inference overhead, and practical scalability.

Overall, the statistical analysis suggests three principal characteristics of the reviewed learning-based literature: strong emphasis on task scheduling, increasing adoption of adaptive RL/DRL methods, and substantial reliance on simulation-based evaluation. These observations motivate greater attention to workflow-aware scheduling, real-world validation, adaptation to workload changes, and deployment-oriented performance measures.

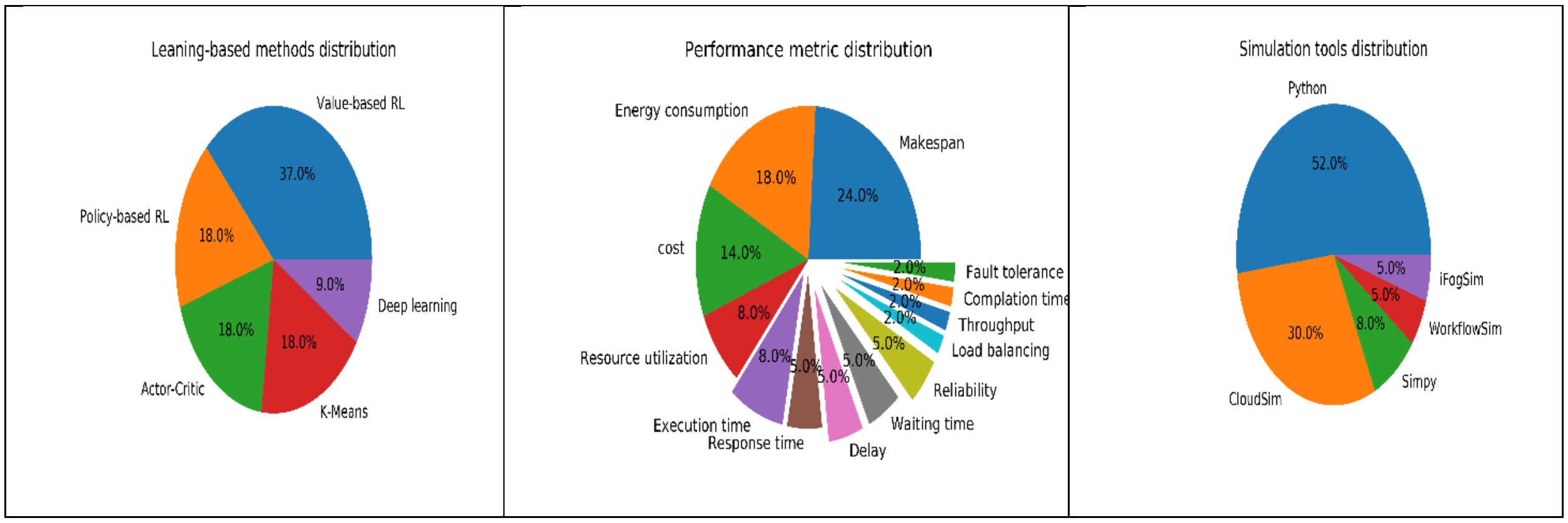

**Figure 8: Statistical information on learning-based techniques**

### 6.3.4 Guideline for selecting learning-based techniques

The selection of learning-based scheduling techniques in cloud–fog–edge computing should be guided by application requirements, workload characteristics, optimization objectives, system dynamics, and the structure of scheduling decisions, rather than by algorithmic popularity alone. Table 11 summarizes the proposed selection framework by relating learning paradigms to representative scheduling requirements and application conditions. Clustering-based approaches are primarily applicable when workload classification, task grouping, or resource organization is required, particularly in relatively stable environments. Deep learning methods can be considered when scheduling relies on prediction or representation learning from historical data and when complex nonlinear relationships must be captured. Value-based reinforcement learning, including Q-Learning and DQN variants, is applicable to discrete scheduling decisions in which policies must adapt to changing system states. Policy-based methods can be considered for problems involving continuous or large action spaces and direct policy optimization, while PPO-based approaches provide a practical option when stable policy updates are required. Actor–critic methods can be considered for complex scheduling problems involving large state spaces or multiple objectives because they combine policy optimization with value estimation.

No single learning-based technique is therefore universally suitable across cloud–fog–edge environments. Technique selection should instead consider the trade-offs among adaptability, computational complexity, state and action-space requirements, training demands, and optimization scope. Combining complementary learning mechanisms may be beneficial for highly dynamic or multi-objective scheduling problems, although such architectures generally introduce additional training and implementation complexity. Accordingly, Table 12 provides condition-based guidance rather than a universal ranking of learning algorithms, supporting the selection of learning-based techniques according to the characteristics and requirements of the scheduling problem.

**Table 11: Decision framework for choosing learning-based scheduling algorithms based on requirements and application contexts**

| Requirement / scheduling characteristic | Suitable family | Representative techniques | Appropriate context | Main consideration |
|---|---|---|---|---|
| Group similar tasks/resources | Unsupervised learning | K-Means, KTCS, CBTS | Task classification and workload organization | Requires meaningful features and cluster configuration |
| Multi-attribute task clustering | Unsupervised learning | MOKM, DCNS | Heterogeneous cloud/fog workloads | Additional clustering/fuzzy complexity |
| Predict complex nonlinear task–resource relationships | Deep learning | CNN, DenseNet | Cloud and cloud-assisted scheduling | Requires representative training data |
| Capture temporal workload patterns | Deep learning | RNN, LSTM, HunterPlus | Dynamic cloud–fog workloads | Higher training and inference cost |
| Adaptive scheduling with discrete actions and small state spaces | Value-based RL | Q-Learning, QEEC, QL-PPTS | Relatively small dynamic scheduling problems | State–action space should remain manageable |
| Large state spaces | Deep value-based RL | DQN, EPERDQN | Large-scale cloud/fog scheduling | Training stability and computational cost |
| Partial observability or temporal dependence | Recurrent value-based RL | DRQN | Dynamic fog/edge environments | Requires historical state information |
| Stable value estimation | Enhanced DQN | CDDQL | Dynamic environments with value-estimation instability | Additional network/training complexity |
| Multi-objective value-based scheduling | DQN variants | DRLMOTS, DQN-Edge | Heterogeneous cloud–fog–edge systems | Reward design becomes more complex |
| Reduce scheduling search space | Hybrid DQN | DRL-TSS | Makespan-oriented scheduling | Depends on embedded heuristic knowledge |
| Direct policy optimization | Policy-based RL | REINFORCE | Complex policy decision spaces | High variance and sample requirements |
| More stable policy learning | PPO | IRATS, DRLIS | Dynamic edge/fog environments | Additional optimization overhead |
| Mobility-aware scheduling | PPO-based RL | IRATS | Vehicular fog and mobile environments | Requires mobility/network information |
| Multi-objective IoT scheduling | PPO-based RL | DRLIS | Edge–fog IoT applications | Reward formulation is critical |
| Complex policy + value learning | Actor–critic | A2C/A3C | Dynamic heterogeneous scheduling | Higher implementation complexity |
| Scheduling under uncertainty | Enhanced actor–critic | HTSFFDRL | Fog environments with uncertain conditions | Fuzzy and asynchronous components increase complexity |
| Trust/reliability-aware scheduling | A2C-based | RTATSA2C | Cloud–fog systems with trust requirements | Additional state/reward dimensions |
| Distributed multi-cloud workflow scheduling | A3C-based | MCWS-A3C | Heterogeneous multi-cloud workflows | Requires distributed training and coordination |
| Resource-constrained edge deployment | Lightweight learning models | Simplified CNN/RL variants | Edge devices with limited resources | Model size and inference latency |
| Rapid adaptation to changing workloads | RL / adaptive DRL | DQN, PPO, actor–critic variants | Highly dynamic cloud–fog–edge environments | Training/adaptation overhead |
| Large-scale multi-objective scheduling | Advanced RL / hybrid learning | DQN, PPO, actor–critic variants | Complex heterogeneous infrastructures | High computational and implementation complexity |

### 6.4 Hybrid techniques

As discussed earlier, heuristics are computationally efficient but highly problem-dependent; metaheuristics provide broader search capabilities but may incur considerable optimization overhead; and learning-based methods offer adaptability but generally require training data and computational resources. To address these limitations, hybrid techniques integrate multiple scheduling paradigms within a unified framework to exploit their complementary strengths. Hybrid scheduling aims to enhance solution quality, computational efficiency, or both, while supporting the simultaneous optimization of multiple performance metrics. This adaptability makes hybrid approaches particularly well-suited to dynamic and heterogeneous environments such as cloud, fog, and edge computing, where scheduling problems are inherently complex and multi-objective.

#### 6.4.1 Classification of hybrid-based techniques

In this context, hybrid scheduling techniques can be classified into six categories based on the combination of underlying methods: Heuristic + Heuristic, Heuristic + Metaheuristic, Heuristic + Learning-based, Metaheuristic + Metaheuristic, Metaheuristic + Learning-based, and Learning-based + Learning-based. In the following sections, representative studies within each category are reviewed, and their respective advantages and limitations are comparatively analyzed.

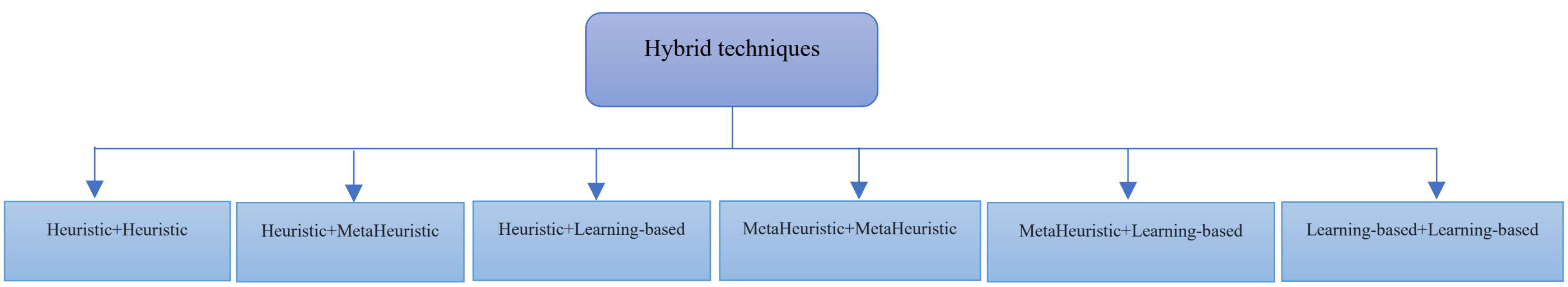

**Figure 9: Classification of hybrid techniques**

##### 6.4.1.1 Heuristic + Heuristic

Heuristic–heuristic methods combine lightweight scheduling rules to exploit complementary decision mechanisms while maintaining relatively low computational overhead. *Hybrid Min-Min and Round Robin (HMMRR)* [169] combines Min-Min and Round Robin to balance execution efficiency and fairness, reducing makespan, response time, and waiting time while mitigating starvation. *Hybrid Algorithm of Min-Min and Max-Min (HAMM)* [170] integrates Min-Min and Max-Min to improve the scheduling of mixed-size tasks by combining efficient short-task allocation with improved handling of long tasks. Thus, HMMRR is more oriented toward fairness and responsiveness, whereas HAMM emphasizes workload balancing and concurrent execution. Overall, heuristic–heuristic hybrids are suitable when fast decisions, simple implementation, and low computational cost are more important than global optimization. Their dependence on predefined rules, however, limits their adaptability to highly dynamic and heterogeneous environments.

#### 6.4.1.2 Heuristic + Metaheuristic

Heuristic–metaheuristic methods combine problem-specific heuristics with the global search capability of metaheuristics. Heuristics are generally used for initialization, task prioritization, solution construction, or local refinement, while metaheuristics improve exploration and help avoid local optima.

Several cloud-oriented studies illustrate different levels of integration. *MaxGen* [171] combines Min-Min, Max-Min, and Genetic Algorithm (GA), using GA to select between heuristic strategies. *Adaptive Decoding Biased Random Key Genetic Algorithm (ADBRKGA)* [172] integrates biased random-key encoding, adaptive decoding, and heuristic-informed initialization to improve workflow optimization. *Hybrid HEFT-PSO-GA algorithm (HEPGA)* [173] combines HEFT, PSO, and GA to exploit high-quality heuristic solutions while maintaining evolutionary exploration. These approaches demonstrate an increasing degree of hybridization, from heuristic selection in MaxGen to tightly integrated multi-metaheuristic search in HEPGA.

Fog-oriented studies provide complementary designs. *PSO-RR* [174] combines PSO with Round Robin to improve fairness, prevent starvation, and satisfy real-time requirements, whereas *Hybrid Genetic, MaxMin, MinMin, and Random Algorithms (GAMMR)* [175] combines GA with Max-Min, Min-Min, and random initialization strategies to improve population diversity and convergence. Collectively, these methods demonstrate that heuristic–metaheuristic hybrids can improve optimization quality while retaining problem-specific scheduling knowledge. They are therefore particularly relevant when solution quality and multi-objective optimization are required but heuristic guidance is available to reduce search effort. Their higher computational cost, however, may limit their application in highly resource-constrained edge environments.

#### 6.4.1.3 Heuristic + Learning- based

Heuristic–learning-based methods combine deterministic scheduling knowledge with data-driven adaptation. Heuristics can provide feasible solutions, task priorities, or system constraints, while learning models adapt scheduling decisions to workload and resource conditions.

*Deep-Q-Learning-based Heterogeneous Earliest Finish time (DQ-HEFT)* [176] integrates Deep Q-Learning with HEFT for cloud workflow scheduling, using learning to guide task prioritization while HEFT provides processor-assignment decisions. *Random Round Robin with clustering* [177] combines K-Means++ clustering with randomized Round Robin and provides a lightweight solution for delay-sensitive fog applications. *Fog-Optimized Deadline-Adaptive Scheduling (FODAS)* [178] integrates Earliest Deadline First (EDF) with deep multi-agent reinforcement learning based on PPO and recurrent neural networks to support deadline-aware and energy-efficient scheduling. *EDF with control* [179] combines SVM-based traffic classification, preemptive EDF, and admission control for mobile edge computing, emphasizing traffic differentiation and system stability.

These studies show a progression from lightweight heuristic combinations toward learning-enhanced and control-aware scheduling. Compared with static heuristics, such methods provide greater adaptability to changing workload and resource conditions, but introduce additional requirements for training data, model execution, memory, and convergence. Therefore, they are most appropriate when dynamic conditions justify learning-based adaptation while domain-specific scheduling rules remain important, particularly in fog and edge environments.

#### 6.4.1.4 Metaheuristic + Metaheuristic

Metaheuristic–metaheuristic methods combine multiple optimization mechanisms to improve exploration, exploitation, convergence, or population diversity. They are primarily intended for complex and multi-objective scheduling problems involving conflicting objectives such as makespan, energy, cost, and resource utilization.

Several cloud studies employ GA-centered hybridization. *hybrid scheduling algorithms Genetic and Thermodynamic Simulated Annealing Algorithm (GATSA)* [180] combines GA with thermodynamic simulated annealing to strengthen global and local search. *Ant Particle Swarm Genetic Algorithm (APSGA)* [181] integrates GA, PSO, and ACO for task prioritization and VM allocation. *GAGELS* [182] combines GA with gravitational-emulation local search to improve makespan and resource utilization. *GAGWO* [183] integrates GA and Grey Wolf Optimization to improve exploration and multi-objective optimization. *HTSA* [184] combines GA with Gravitational Search Algorithm to balance energy, throughput, and resource utilization. *Genetically Modified Multi-Population PSO (GMPSO)* [185] embeds GA operators into a multi-population PSO framework to improve convergence and solution diversity. These studies demonstrate that deeper hybridization can improve search quality, although additional mechanisms increase computational and parameter-tuning requirements.

Other cloud-oriented hybrids include *hybrid scheduling approaches Cuckoo Crow Search Algorithm (CCSA)* [186], which combines Cuckoo Search and Crow Search to improve VM selection; *Fuzzy Hybrid Particle Swarm Parallel Ant Colony Optimization (FHPSPACO)* [187], which integrates heuristic scheduling, PSO, parallel ACO, and fuzzy logic; *hybrid oppositional differential evolution-enabled WOA (h-DEWOA)* [188], which combines WOA with Differential Evolution, opposition-based learning, and chaotic mechanisms; *GWO-MSOA* [189], which combines Grey Wolf Optimization with Modified Sunflower Optimization; and *DECWOA* [190], which incorporates chaotic initialization, adaptive inertia, and differential-variance mechanisms into WOA. These approaches emphasize convergence improvement, search diversity, and multi-objective optimization, although their increasing hybridization also increases algorithmic complexity.

Fog-oriented studies include *Improved Hybrid heuristics Based Scheduling (IHBS)* [191], which combines PSO and Crow Search for low-latency and energy-aware scheduling; *Opposition-based Chaotic Whale Optimization Algorithm (OppoCWOA)* [192], which integrates WOA with opposition-based learning, chaotic mechanisms, and an ILP formulation; *GA-PSO* [193], which combines evolutionary and swarm search for multi-objective fog scheduling; and *HETSVP* [194], which jointly addresses task scheduling and VM placement using improved binary PSO and resource-aware allocation. These approaches illustrate the importance of incorporating system-level constraints into optimization, particularly in fog environments.

For fog–cloud systems, *DE-GWO* [195] combines Differential Evolution and Grey Wolf Optimization for workflow scheduling; *Hybrid monarch butterfly optimization and improved ant colony optimization algorithm (HMA)* [196] combines Monarch Butterfly Optimization and ACO with task-priority awareness for latency- and energy-sensitive industrial applications; *Particle Whale Optimization Algorithm (PWOA)* [197] combines PSO and WOA to improve global search and reduce premature convergence; and *PSO-SA* [198] combines PSO with Simulated Annealing for priority- and deadline-aware scheduling. In edge-oriented environments, *PSO-GA* [199] combines PSO and GA for multi-objective serverless edge workflow scheduling, whereas *Cloud-Edge Smart Manufacturing Architecture (CESMA)* [200] integrates NSGA-II and Improved Monarch Butterfly Optimization within a cloud–edge smart-manufacturing architecture.

Overall, these studies show that metaheuristic–metaheuristic hybrids are most relevant when complex search spaces, multiple conflicting objectives, and high solution quality are the dominant requirements. Their principal limitation is increased computational and implementation complexity, which can restrict their suitability for latency-sensitive or resource-constrained edge environments.

#### 6.4.1.5 Metaheuristic + Learning-based

Metaheuristic–learning-based methods combine global optimization with data-driven prediction or adaptive decision making. Learning components can predict workload and resource states, tune optimization parameters, or guide the search, while metaheuristics provide broader exploration.

Cloud-based studies demonstrate different degrees of integration. *ARS-ATS* [201] uses a genetic algorithm to generate scheduling data for a neural-network-based decision model, emphasizing energy efficiency and makespan reduction. *Particle Swarm Optimization- Bandwidth Aware Divisible Task (PSO-BATS)* [202] combines PSO with bandwidth-aware divisible-task scheduling and regression-based host selection to improve load balancing and throughput. *scheduling with Multi-Layered Regression Host Employment (MLRHE)* [203] employs multilayer regression for host-employment decisions. *FTTHDRL* [204] combines Harris Hawks Optimization with deep reinforcement learning to provide fault-tolerant and SLA-aware scheduling. *GA-DQN* [205] and *SA-DQN* [206] combine GA and simulated annealing, respectively, with DQN to improve workflow scheduling under dynamic conditions.

Fog and edge studies further extend this integration. *Effective Dynamic Load Balancing Technique (EDLB)* [207] combines a convolutional neural network with modified PSO for dynamic load balancing. *Fuzzy-based Reinforcement Learning (FRL)* [208] integrates fuzzy logic and reinforcement learning within an optimization-based scheduling framework for deadline- and energy-aware decisions. *Energy-efficient and deadline-Aware Task scheduling in Fog Computing (ETFC)* [209] combines SVM-based traffic prediction, reinforcement learning, and NSGA-III for traffic-aware fog scheduling. *REACT* [210] integrates reinforcement learning, metaheuristic optimization, and LSTM-based prediction for hierarchical task offloading and resource management in ultra-dense edge–cloud networks.

Collectively, these studies demonstrate that metaheuristic–learning hybrids can provide stronger adaptability than conventional optimization methods, particularly under uncertain and time-varying workloads. However, they also require greater model complexity, training or prediction resources, and computational overhead. They are therefore most suitable when environmental dynamics and adaptation requirements justify the additional learning complexity.

#### 6.4.1.6 Learning-based + Learning-based

Learning–learning hybrids combine multiple learning paradigms to jointly model workload characteristics and optimize scheduling decisions. Typically, deep or supervised learning captures task, resource, or temporal characteristics, while reinforcement or online learning adapts scheduling policies based on system feedback.

*Deep Reinforcement Learning combined with LSTM (DRL-LSTM)* [211] combines deep reinforcement learning with LSTM-based temporal modeling to capture long-term dependencies between workload and resource availability in cloud scheduling. *A3C-R2N2* [212] combines asynchronous advantage actor–critic learning with a residual recurrent neural network for integrated edge–cloud scheduling, enabling temporal modeling and adaptive decision making under stochastic and mobile conditions.

Compared with simpler hybrid categories, learning–learning approaches provide strong potential for self-adaptive scheduling under non-stationary workloads. However, this benefit comes with greater requirements for training data, computational resources, model coordination, convergence stability, and generalization. Consequently, these methods are most appropriate when workload dynamics are substantial and sufficient computational resources are available to support learning and inference.

### 6.4.2 Cross-technique comparison of hybrid-based scheduling families across computing paradigms

This section provides a detailed examination and comparative evaluation of hybrid-based strategies for addressing scheduling challenges. Table 12 summarizes the principal characteristics of hybrid-based algorithms, including the type of scheduling problem (task, job, or workflow level scheduling), their impact on key performance metrics, user-imposed constraints, potential application domains of these techniques and identifies commonly employed simulation tools for performance evaluation across different computing environments.

**Table 12: Summary of the hybrid scheduling algorithm**

| Hybrid method | Publish year | Combine techniques | Scheduling issue | Computing environment | Performance parameters | User constraint | Improvement strategy | Simulation tools | Use Case |
|---|---|---|---|---|---|---|---|---|---|
| Hybrid Min-Min and Round Robin (HMMRR) [169] | 2020 | MinMin + RR | Task scheduling | Cloud computing | Makespan, execution time, resource utilization, and response time | QOS | Combines Min-Min efficiency with Round-Robin fairness to reduce makespan, response/waiting time, and starvation. | C++ | Suitable for satisfying QoS metrics in reducing the makespan of VMs, average response time of processing tasks, as well as average response time, while increasing resource utilization and overall system performance. |
| Hybrid Algorithm of Min-Min and Max-Min (HAMM) [170] | 2021 | MinMin+ MaxMin | Task scheduling | Cloud computing | Makespan, resource utilization, load balancing, and waiting time | N/A | Combines Min-Min and Max-Min to balance short- and long-task execution and improve resource utilization. | CloudSim | Suitable for best concurrent execution between long tasks and short tasks. |
| MaxGen [171] | 2022 | MaxMin+ GA | Task scheduling | Cloud computing | Makespan and resource utilization | N/A | Uses GA to select between Min-Min and Max-Min, improving makespan and load balancing. | MATLAB | Suitable for optimizing MaxMin in cloud computing. |
| Adaptive Decoding Biased Random Key Genetic Algorithm (ADBRKGA) [172] | 2022 | GA+HEFT | Workflow scheduling | Cloud computing | Makespan | N/A | Uses biased random-key encoding, adaptive decoding, and heuristic initialization to improve search quality and convergence for workflows. | C++ | Suitable for static workflow scheduling in cloud computing |
| Hybrid HEFT-PSO-GA algorithm (HEPGA) [173] | 2024 | PSO+GA+ HEFT | Workflow scheduling | Cloud computing | Makespan and Resource Utilization | N/A | Uses HEFT-based initialization with PSO and GA to improve exploration, makespan, and resource utilization. | CloudSim | Suitable for large-scale data-intensive workflows |
| HLWOA [174] | 2025 | HEFT+ WOA | Workflow scheduling | Cloud computing | Makespan and cost | QOS | Uses PSO for global search and Round Robin for fairness and starvation prevention in real-time fog scheduling. | WorkflowSim | Suitable for QOS-aware workflow scheduling in cloud computing. |
| PSO-RR [175] | 2023 | PSO+RR | Task scheduling | Fog computing | Energy consumption and cost | N/A | Uses heuristic-guided GA population initialization and non-dominated selection to improve diversity and convergence. | Python | Suitable for IoT-enabled environments, industrial automation, and autonomous systems, in addition to traditional task scheduling fields. |
| Hybrid Genetic, MaxMin, MinMin, and Random Algorithms (GAMMR) [176] | 2023 | GA+Min-Min+Max-Min+ Random | Task scheduling | Fog-cloud computing | Response time and energy consumption | N/A | Uses Deep Q-Learning to adapt task prioritization while HEFT provides efficient processor assignment. | Python | Suitable for independent task scheduling in fog–cloud computing environment. |
| Deep-Q-Learning-based Heterogeneous Earliest Finish time (DQ-HEFT) [177] | 2020 | HEFT+ DRL | Workflow scheduling | Cloud computing | Makespan and execution time | N/A | Uses K-Means++ clustering and randomized Round Robin to reduce waiting time with low computational overhead. | CloudSim | Suitable for scientific workflow scheduling in heterogeneous cloud computing. |
| Random Round Robin with clustering [178] | 2021 | K-Means +RR | Task scheduling | Fog computing | Delay and waiting time | N/A | Combines EDF with deep multi-agent RL to improve deadline satisfaction, energy efficiency, and adaptability. | CloudSim | Suitable for delay-sensitive IoT applications. |
| Fog-Optimized Deadline-Adaptive Scheduling (FODAS) [179] | 2024 | EDF+DRL | Task scheduling | Fog-cloud computing | Energy consumption, makespan | Priority and deadline | Uses SVM-based traffic classification with EDF and admission control to improve traffic differentiation, throughput, and packet-loss performance. | COSCO simulator | Suitable for aperiodic tasks in heterogeneous fog-cloud environments, |
| EDF scheduling with admission control [180] | 2023 | EDF+SVM | Task scheduling | Edge computing | Delay and throughput | QOS | Combines GA global exploration with simulated-annealing-based local exploitation to improve search efficiency and scheduling performance. | N/A | Suitable for computing-aware scheduling in a mobile edge computing system |
| Genetic And Thermodynamic Simulated Annealing Algorithm (GATSA) [181] | 2021 | GA+ SA | Workflow scheduling | Cloud computing | Execution time, and makespan | Dependency | Combines GA, PSO, and ACO to improve task prioritization, VM allocation, makespan, and CPU utilization. | N/A | Suitable for scheduling of communication-intensive workflows in heterogeneous cloud computing. |
| Ant Particle Swarm Genetic Algorithm (APSGA) [182] | 2022 | GA+PSO+ ACO | Task scheduling | Cloud computing | Makespan, execution time | N/A | Integrates GA with local search to improve makespan and resource utilization. | CloudSim | Suitable for dynamically scheduling tasks on virtual machines |
| GAGELS [183] | 2023 | GA+GSA | Task scheduling | Cloud computing | Execution time and resource utilization | N/A | Combines GA and GWO to mitigate premature convergence and improve multi-objective optimization. | CloudSim | Suitable for effective cloud network resources. |
| GA-GWO [184] | 2023 | GA+GWO | Task scheduling | Cloud computing | Makespan, energy consumption, and cost | N/A | Combines complementary optimization mechanisms to improve energy efficiency, throughput, and resource utilization. | CloudSim | Suitable for heterogeneous cloud computing environments |
| HTSA [185] | 2024 | GA+GSA | Task scheduling | Cloud computing | Energy consumption, makespan, resource utilization and throughput | N/A | Embeds GA operators into multi-population PSO to improve convergence speed and population diversity. | CloudSim | Suitable for a heterogeneous cloud environment. |
| Genetically-Modified Multi-Population PSO (GMPSO) [186] | 2025 | GA+PSO | Workflow scheduling | Cloud computing | Makespan and energy consumption | N/A | Combines Cuckoo Search and Crow Search to improve VM selection, makespan, and cost. | WorkflowSim | Suitable for complex workflow systems that are sensitive to resource utilization, task response time, and energy consumption, such as smart healthcare, scientific computing, and the industrial Internet. |
| Cuckoo Crow Search Algorithm (CCSA) [187] | 2021 | Cuckoo search +CSA | Task scheduling | Cloud computing | Makespan and cost | N/A | Integrates heuristic scheduling, PSO, parallel ACO, and fuzzy control for adaptive parameter selection and improved execution/resource utilization. | CloudSim and JAVA | Suitable for task scheduling issues in cloud computing. |
| Fuzzy Hybrid Particle Swarm Parallel Ant Colony Optimization (FHPSPACO) [188] | 2022 | PSO+ACO | Task scheduling | Cloud computing | Execution time, waiting time, throughput, and resource utilization. | N/A | Combines WOA with differential evolution, opposition learning, chaos, and fitness-aware balancing to improve exploration, convergence, makespan, and energy efficiency. | CloudSim | Suitable for optimizing task scheduling in cloud computing. |
| hybrid oppositional differential evolution-enabled WOA (h-DEWOA) [189] | 2022 | DE+WOA | Job scheduling | Cloud computing | Makespan and energy consumption | N/A | Combines GWO and Modified Sunflower Optimization with elite preservation to reduce premature convergence and improve cost/resource utilization. | CloudSim | Suitable for BoT applications over IaaS clouds. |
| GWO-MSOA [190] | 2025 | GWO+SFO | Task scheduling | Cloud computing | Makespan, cost, and resource utilization | N/A | Enhances WOA using chaotic initialization, adaptive inertia, and differential-variance mechanisms to accelerate | CloudSim | Suitable for the variability of incoming tasks in terms of size, |
| | | | | | | | convergence and reduce execution time/cost. | | processing capacity, and the allocation of these tasks. |
| DECWOA [191] | 2025 | DE+WOA | Task scheduling | Cloud computing | Execution time and cost | N/A | Combines PSO and Crow Search to improve low-latency node selection and energy efficiency in fog environments. | CloudSim | Suitable for task scheduling in cloud data centers. |
| Improved Hybrid heuristics Based Scheduling (IHBS) [192] | 2020 | PSO+CSA | Job scheduling | Fog computing | Delay, completion time, and energy consumption | N/A | Combines WOA, opposition-based learning, and chaotic mechanisms to strengthen global exploration, convergence, and robustness. | mininet simulator | Suitable for utilizing the fog nodes that are limited in terms of computational resources and energy. |
| Opposition-based Chaotic Whale Optimization Algorithm (OppoCWOA) [193] | 2021 | ILP+WOA | Task scheduling | Fog computing | Execution time and energy consumption | N/A | Combines GA exploration with PSO exploitation to improve multi-objective fog scheduling. | Python | Suitable for high IoT task offloading requests and fog resource limitations, |
| GA-PSO [194] | 2024 | GA+PSO | Task scheduling | Fog computing | Execution time, response time, and completion time | N/A | Jointly optimizes task scheduling and VM placement using improved binary PSO and resource-aware allocation. | iFogSim | Suitable for big data applications in fog computing. |
| HETSVP [195] | 2025 | Evolutionary+ PSO | Task scheduling | Fog computing | Makespan, execution time, load balancing, and energy consumption | QOS | Uses Differential Evolution to improve population diversity/convergence while GWO balances exploration and exploitation. | iFogSim | Suitable for effective VM placement. |
| DE-GWO [196] | 2022 | DE+GWO | Workflow scheduling | Fog-cloud computing | Makespan, cost, and energy consumption | N/A | Combines Monarch Butterfly Optimization and ACO with task-priority awareness to improve latency, energy efficiency, and task completion. | FogWorkflowSim | Suitable for multi-objective workflow scheduling for a heterogeneous fog-cloud environment. |
| Hybrid monarch butterfly optimization and improved ant colony optimization algorithm (HMA) [197] | 2022 | MBO+ACO | Task scheduling | Fog-cloud computing | Energy consumption, delay | N/A | Combines PSO and WOA to mitigate premature convergence and improve global search, execution time, and cost. | C++ | Suitable for the task scheduling problem in intelligent production lines. |
| Particle Whale Optimization Algorithm (PWOA) [198] | 2023 | PSO+WOA | Workflow scheduling | Fog-cloud computing | Execution time and cost | N/A | Uses PSO for task prioritization and Simulated Annealing for allocation refinement, improving deadline- and priority-aware scheduling. | WorkflowSim | Suitable for workflow scheduling in a cloud–fog environment. |
| PSO-SA [199] | 2024 | PSO+SA | Workflow scheduling | Fog-cloud computing | Energy consumption and makespan | Deadline | Combines PSO convergence with GA diversity preservation to improve multi-objective edge workflow scheduling. | MATLAB | Suitable for energy-efficient and deadline-aware workflow scheduling algorithms in the fog and cloud environment. |
| PSO-GA [200] | 2022 | PSO+GA | Workflow scheduling | Edge computing | Energy consumption and execution time | N/A | Combines NSGA-II and improved Monarch Butterfly Optimization to strengthen Pareto-based optimization, convergence, reliability, and QoS. | CloudSim | Suitable for workflow scheduling in Serverless Edge Computing for the Internet of Things. |
| Cloud-Edge Smart Manufacturing Architecture (CESMA) [201] | 2025 | NSGA-II + MBO | Task scheduling | Edge-cloud computing | Resource utilization, energy consumption, completion time, and reliability | QOS | Uses GA-generated scheduling knowledge to train a neural model, improving energy efficiency, makespan, and active-rack reduction. | MATLAB | Suitable for optimizing cloud-edge integration in smart manufacturing, addressing latency, resource usage, and energy efficiency. |
| ARS-ATS [202] | 2020 | GA+RNN | Task scheduling | Cloud computing | makespan, energy consumption, execution overhead | N/A | Combines PSO with bandwidth-aware divisible-task scheduling and regression-based host selection to improve load balancing and throughput. | CloudSim | Suitable for energy-efficient task scheduling in the cloud environment. |
| Particle Swarm Optimization-Bandwidth Aware Divisible Task (PSO-BATS) scheduling with Multi-Layered Regression Host Employment | 2023 | PSO+ Regression | Task scheduling | Cloud computing | Makespan, response time, and resource utilization | N/A | Uses multilayer regression to improve host-employment decisions and scheduling efficiency. | CloudSim | Suitable for bandwidth-aware scheduling to eliminate the faults in scheduling the tasks to the Virtual Machines. |
| (MLRHE) [203] | | | | | | | | | |
| FTTHDRL [204] | 2023 | HHO+DRL | Task scheduling | Cloud computing | Makespan and cost | N/A | Combines Harris Hawks Optimization with DRL to improve fault tolerance, SLA satisfaction, and scheduling robustness. | CloudSim | Suitable for Fault tolerant trust aware task scheduling in multi cloud environment. |
| GA-DQN [205] | 2023 | GA+DRL | Workflow scheduling | Cloud computing | Response time and cost | N/A | Uses GA-derived workflow features to guide DQN decision making, improving adaptability and scheduling quality. | Python | Suitable for Cost-aware scheduling systems for real-time workflows in the cloud. |
| SA-DQN [206] | 2024 | SA+DQN | Workflow scheduling | Cloud computing | Response time and cost | N/A | Uses simulated annealing to enhance DQN-based scheduling decisions, improving adaptability and workflow scheduling performance. | Python | Suitable for cost-aware cloud workflow scheduling. |
| Effective Dynamic Load Balancing Technique (EDLB) [207] | 2022 | CNN+MPSO | Task scheduling | Fog computing | Makespan, load balancing, cost, and resource utilization | N/A | Combines CNN-based prediction with modified PSO to improve dynamic load balancing, response time, and resource utilization. | N/A | Suitable for load balancing in fog computing via a dynamic real-time scheduling algorithm. |
| Fuzzy-based Reinforcement Learning (FRL) [208] | 2023 | MINLP + RL | Task scheduling | Fog computing | Execution time, delay, and energy consumption | Deadline | Combines fuzzy logic and RL to provide adaptive and deadline-aware scheduling under energy and resource constraints. | Python | Suitable for a delay and energy-aware task scheduling mechanism for fog-enabled IoT Applications. |
| Energy-efficient and deadline-Aware Task scheduling in Fog Computing (ETFC) [209] | 2024 | GA+SVM | Task scheduling | Fog computing | Energy consumption and cost | Deadline | Integrates traffic prediction, RL, and NSGA-III to adapt scheduling to traffic conditions while jointly improving response time, energy, cost, and deadline satisfaction. | MATLAB | Suitable for energy-efficient and deadline-aware task scheduling in fog computing. |
| REACT [210] | 2025 | GA+RL | Task scheduling | Edge computing | Completion time and resource utilization | N/A | Combines RL, metaheuristic optimization, and LSTM prediction to jointly optimize offloading, resource allocation, and energy efficiency. | CloudSim and Python | Suitable for multi-objective task scheduling in ultra-dense edge networks. |
| Deep Reinforcement Learning combined with LSTM (DRL-LSTM) [211] | 2020 | DRL+LSTM | Task scheduling | Cloud computing | Cost | N/A | Combines DRL with LSTM temporal modeling to capture long-term workload/resource dependencies and improve waiting time and resource cost. | Python | Suitable for automated task scheduling in large-scale cloud computing. |
| A3C-R2N2 [212] | 2022 | RNN+A3C | Task scheduling | Edge-cloud computing | energy consumption, response time, and cost | QOS and SLA | Combines asynchronous actor–critic learning with recurrent modeling to improve temporal adaptation, scalability, responsiveness, energy efficiency, and SLA compliance. | iFogSim and CloudSim | Suitable for dynamic scheduling for stochastic edge-cloud computing environments. |

Heuristic–heuristic hybrids provide the lowest-complexity form of hybridization. However, this simplicity also limits their ability to address complex constraints. Consequently, heuristic–heuristic hybrids are preferable when low scheduling overhead and interpretable resource allocation are more important than extensive optimization across heterogeneous objectives. Their use becomes less appropriate as workload uncertainty, objective conflicts, and resource dynamics increase.

Conventional heuristic and metaheuristic hybrids can operate without extensive training datasets and therefore remain attractive when historical scheduling data are unavailable. Their limitation is that adaptation must occur through algorithmic search rather than learned experience.

Metaheuristic–metaheuristic hybrids provide greater global search capability and are particularly suitable for complex multi-objective optimization problems. The principal benefit of combining evolutionary, swarm, or population-based algorithms is the complementary balance between exploration and exploitation. These hybrids are particularly attractive when the search space is large, the objectives are conflicting, and near-optimal global solutions are more important than instantaneous scheduling decisions. Nevertheless, their iterative population-based search generally requires more computation than lightweight heuristics and can introduce additional parameter sensitivity. They are therefore less suitable for severely resource-constrained edge devices or applications requiring extremely low decision latency.

Metaheuristic–learning hybrids provide a different advantage by combining global search with adaptive decision making. Unlike purely metaheuristic approaches, these methods can exploit historical or real-time information about workload and resource states, making them more appropriate for dynamic and uncertain environments. The reviewed evidence demonstrates a progression from relatively lightweight learning–optimization approaches toward more deeply integrated architectures. The main trade-off of metaheuristic–learning hybrids is therefore adaptability versus computational overhead. Learning components are particularly valuable when resource states and workloads change over time because the scheduler can modify its decisions using accumulated experience rather than repeatedly solving the complete optimization problem from scratch. This advantage is evident in fog and edge environments. This progression suggests that learning becomes increasingly valuable as workload uncertainty and resource dynamics increase; however, the associated training, inference, memory, and coordination costs become progressively more problematic as computation moves toward resource-constrained edge devices.

Learning-based hybrids can exploit historical or streaming information to improve decisions under changing conditions, but their effectiveness depends on the availability, representativeness, and quality of training data. This distinction is particularly important when moving from cloud simulation to real-world edge environments, where workload distributions can differ substantially from training scenarios. Learning–learning hybrids offer the strongest potential for continuous adaptation but also impose the highest methodological requirements. The comparison suggests that the principal advantage of learning–learning hybrids is not simply improved optimization, but adaptation to non-stationary and stochastic environments. However, learning–learning approaches require sufficient training data, computational resources, careful model coordination, and stable learning processes. Their interpretability and generalization to previously unseen workloads also remain concerns. Consequently, they should be considered when workload dynamics justify the learning overhead rather than simply because they employ more advanced learning architectures.

### 6.4.3 Statistical analysis of hybrid-based strategies

Figure 9 provides an integrated statistical view of the hybrid-based scheduling literature across computing paradigms, scheduling problems, performance objectives, simulation environments, and hybridization patterns. The predominance of cloud computing (54%), followed by fog computing (34%) and edge computing (12%), indicates that hybridization has developed most extensively in environments where the computational resources required by complex optimization procedures can be accommodated. Combining two or more algorithms generally introduces additional search, coordination, parameter-tuning, training, or inference overhead; consequently, cloud infrastructures provide a comparatively favorable setting for exploiting sophisticated hybrid mechanisms. The substantial representation of fog computing suggests, however, that the motivation for hybridization is shifting from computationally intensive optimization toward balancing multiple constraints, particularly latency, energy consumption, resource heterogeneity, and workload variability. The comparatively smaller share of edge computing should therefore not be interpreted as lower research relevance. Rather, it reflects a fundamental deployment tension: the environments that potentially benefit most from adaptive and multi-objective scheduling are also those least able to accommodate computationally expensive hybrid algorithms. This observation suggests that the future progression of hybrid scheduling is likely to depend less on increasing algorithmic complexity and more on developing resource-aware and adaptive hybridization, in which the computational cost of the scheduling mechanism itself becomes part of the optimization problem. In addition, task scheduling dominates the reviewed studies (68%), followed by workflow scheduling (27%) and job scheduling (5%), reflecting the broad applicability of task-level resource allocation across cloud, fog, and edge environments. Workflow scheduling is less prevalent because task dependencies, communication overhead, and heterogeneous execution paths introduce greater modeling and optimization complexity. The limited focus on job scheduling (5%) indicates a comparatively underexplored area and suggests opportunities for future research on dynamic job coordination, queueing, and resource sharing.

Based on Figure 6, Metaheuristic + Metaheuristic (48%) is the most popular because one method provides strong global exploration, while the second improves local exploitation or convergence. This is particularly useful for cloud, fog, and edge scheduling, where the search space is highly nonlinear, dynamic, and constrained. Moreover, metaheuristics are relatively easy to integrate because they operate at the optimization level without requiring large training datasets or extensive model-training procedures. Consequently, hybrid metaheuristics can improve solution quality, avoid premature convergence, and maintain reasonable computational cost while preserving the flexibility of the original optimization framework. Metaheuristic + Learning (20%) is the second most common combination because it addresses a different limitation of conventional optimization. The metaheuristic performs the combinatorial search, while the learning component provides prediction, adaptation, or guidance based on workload and resource characteristics. This combination is particularly attractive for dynamic scheduling environments because learning can capture patterns from changing workloads, whereas the metaheuristic can optimize the resulting scheduling decisions. Thus, the hybrid can achieve a better balance between adaptability and optimization capability than either technique alone. In contrast, less frequent combinations such as Learning + Learning or other multi-technique configurations generally require greater model complexity, larger training datasets, higher computational resources, and more careful parameter coordination. Therefore, their lower adoption does not necessarily indicate inferior performance; rather, it reflects a higher implementation and validation burden.

The distribution of performance metrics provides further evidence of how the research community conceptualizes scheduling quality. The three most frequently evaluated metrics are makespan (22%), energy consumption (18%), and cost (15%), indicating that hybrid scheduling research is primarily concerned with balancing temporal efficiency against resource and operational efficiency. The prominence of makespan demonstrates that completing tasks or workflows rapidly remains the fundamental benchmark for determining whether a hybrid strategy improves scheduling performance. However, the substantial attention given to energy consumption and cost indicates that minimizing completion time alone is increasingly regarded as insufficient. A schedule that minimizes makespan may achieve this by allocating additional computational resources, increasing energy consumption or monetary cost; conversely, aggressive energy or cost minimization may increase completion time. The relatively close distribution among these three metrics therefore reflects an underlying multi-objective trade-off rather than three independent evaluation criteria. In this context, hybridization becomes valuable because different algorithmic components can contribute complementary search behaviors or decision mechanisms for navigating conflicting objectives. The statistical pattern consequently suggests a shift from asking whether a scheduler produces a faster solution toward asking whether it produces a balanced solution across performance, energy, and economic efficiency.

The distribution of simulation tools further reveals an important methodological characteristic of the field, with CloudSim accounting for 43% of the reported evaluation environments and Python for 18%. The strong presence of CloudSim can be interpreted as evidence of the community's preference for controlled, extensible, and relatively reproducible experimental environments. CloudSim provides abstractions for modeling computational resources, virtual machines, task allocation, and scheduling policies, making it particularly suitable for comparing alternative hybrid algorithms while maintaining consistent infrastructure assumptions. Its dominance is also consistent with the high proportion of cloud-oriented studies, creating a methodological alignment between the most frequently investigated computing paradigm and the most frequently used evaluation platform. Python's substantial representation reflects a complementary trend toward flexible implementation and experimentation, particularly for learning-based and data-driven components that require machine-learning libraries, custom models, and rapid algorithmic prototyping. Nevertheless, the concentration of evaluations around CloudSim and Python should not be interpreted as evidence that these environments are universally superior. Rather, it reflects their accessibility and flexibility for controlled experimentation.

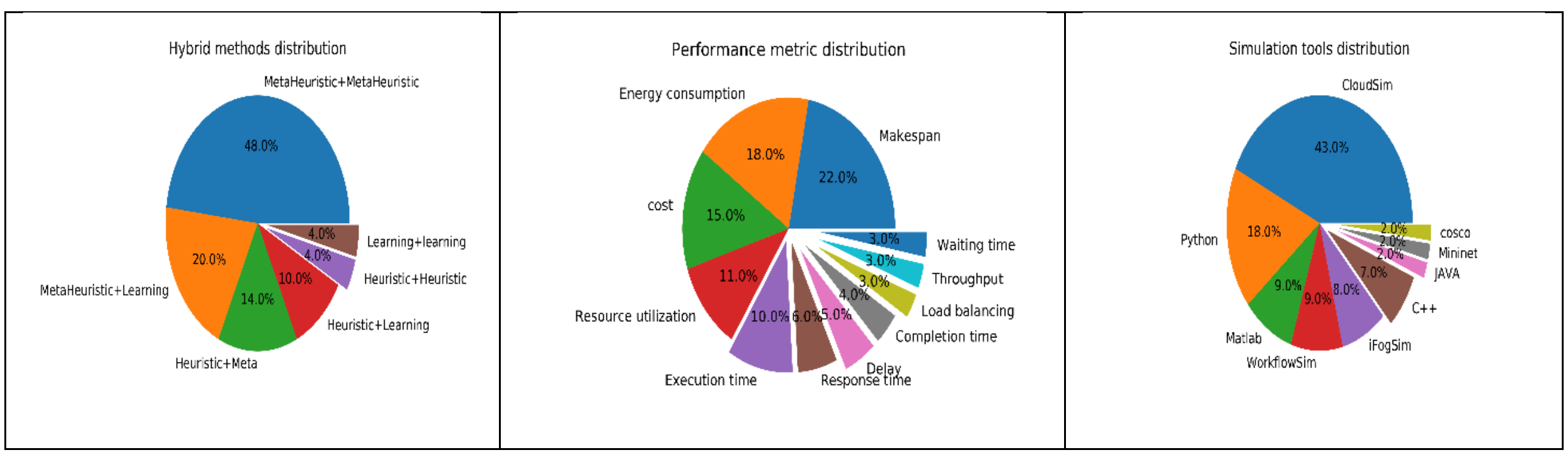

**Figure 9: Statistical information of hybrid-based techniques**

### 6.4.4 Guideline for selecting hybrid-based techniques

The comparative findings presented in Table 12 provide a basis for developing a decision-oriented guideline for selecting hybrid scheduling techniques according to application requirements, scheduling characteristics, and computing environments. Rather than identifying a universally superior hybrid algorithm, the evidence indicates that the suitability of a technique depends on the interaction among optimization objectives, workload dynamics, resource constraints, scheduling granularity, and the computational environment in which the scheduler is deployed. This perspective is particularly important because hybridization introduces a trade-off between solution quality and algorithmic complexity: lightweight heuristic hybrids are generally preferable when rapid and interpretable decisions are required, whereas metaheuristic and learning-based hybrids become more appropriate when the scheduling problem is highly complex, multi-objective, dynamic, or difficult to model analytically. The following guidelines therefore provide a requirement-driven framework for selecting hybrid scheduling approaches across cloud, fog, edge, and integrated edge–cloud environments.

For cloud task scheduling, hybrid heuristic–heuristic approaches (like HMMRR and HAMM) are appropriate when the primary requirements are low computational complexity, improved resource utilization, reduced response or waiting time, and relatively straightforward decision making. When higher global optimization capability is required, particularly for large and multi-objective cloud workloads, metaheuristic–metaheuristic hybrids such as GA–PSO, GA–GWO, APSGA, and HTSA are more appropriate because they combine complementary exploration and exploitation mechanisms. When the workflow environment is highly dynamic or uncertain and adaptive decisions are required, learning-enhanced approaches such as DQ-HEFT, GA-DQN, and SA-DQN provide a stronger alternative because learning can adapt scheduling decisions while heuristic or metaheuristic components guide the search. Lightweight hybrids such as clustering combined with Round Robin are preferable for delay-sensitive IoT workloads and resource-constrained fog nodes because they provide computationally efficient decisions without the training and iterative optimization costs of more sophisticated approaches. Thus, for fog environments, the choice should progressively move from lightweight heuristic hybrids toward learning-enhanced hybrids as workload uncertainty, deadline requirements, and optimization objectives increase; however, the additional computational complexity should be justified by a corresponding improvement in scheduling quality. For edge computing, the selection criterion should shift from maximizing optimization sophistication toward minimizing decision latency, computational overhead, and resource consumption while maintaining sufficient adaptability. Lightweight or moderately complex hybrids are preferable when edge nodes have limited processing and memory capacity. When the environment is highly dynamic, mobile, or stochastic and continuous adaptation is essential, learning-based hybrids become more attractive. REACT, for example, combines reinforcement learning, metaheuristic optimization, and predictive modeling within a hierarchical architecture and is designed for highly dynamic edge–cloud conditions, where task offloading, resource allocation, and energy efficiency must be jointly managed. Nevertheless, its greater architectural and computational complexity means that it is more appropriate for environments capable of supporting such overhead rather than severely resource-constrained edge devices. For integrated edge–cloud environments with highly variable workloads, learning–learning hybrids provide the greatest potential when adaptability, robustness, and continuous self-optimization are primary requirements. These approaches can combine predictive models with reinforcement learning so that workload characteristics and resource states are learned while scheduling policies are dynamically adapted. A3C-R2N2 illustrates this design by combining asynchronous actor–critic learning with recurrent modeling to exploit temporal workload patterns and support rapid adaptation in stochastic and mobile environments, improving energy efficiency, response time, SLA compliance, and operational cost. However, such approaches should not be selected solely because they provide higher adaptability. Their data requirements, training and inference overhead, convergence stability, interpretability, and generalization limitations make them less appropriate when training data are scarce, computational resources are limited, or highly predictable workloads can already be handled effectively by simpler methods.

From a requirement-oriented perspective, the evidence suggests several general selection rules. When low latency and computational simplicity are dominant, heuristic–heuristic or lightweight heuristic–metaheuristic hybrids should be preferred. When global optimization and multi-objective scheduling are required, metaheuristic–metaheuristic hybrids provide a stronger alternative because they can improve exploration, exploitation, convergence, and solution diversity, although at greater computational cost. When workload dynamics, uncertainty, or changing resource states dominate, metaheuristic–learning or learning–learning hybrids become more suitable because learning mechanisms provide adaptation that static optimization cannot readily achieve. The resulting decision principle is therefore requirement, environment, and complexity matching rather than algorithm ranking. In cloud environments, where computational resources are relatively abundant, sophisticated metaheuristic and learning-based hybrids can be justified when optimization objectives are numerous or workloads are highly dynamic. In fog environments, the selection should place greater emphasis on the balance between optimization quality and scheduling overhead, with lightweight hybrids preferred for delay-sensitive workloads and learning-enhanced methods reserved for situations requiring adaptation, deadline awareness, or energy optimization. In edge environments, computational efficiency and responsiveness should generally dominate algorithmic sophistication, while deeper learning-based hybrids should be considered only when their adaptability provides sufficient benefit to offset their training and inference costs. Finally, for heterogeneous fog–cloud and edge–cloud infrastructures, hybrid methods that explicitly model cross-layer resource allocation, offloading, energy, and workload dynamics are more appropriate than techniques designed for a single computational layer.

Overall, the cross-technique evidence indicates that hybrid scheduling families exhibit a clear complexity, adaptability, and optimization trade-off rather than a simple performance hierarchy. Heuristic–heuristic hybrids are preferable when low overhead, simplicity, and rapid scheduling decisions are dominant requirements; metaheuristic–metaheuristic hybrids are more appropriate for large, complex, and multi-objective search spaces where global optimization is prioritized; metaheuristic–learning hybrids become advantageous when global search must be combined with adaptation to dynamic workloads and uncertain resource states; and learning–learning hybrids are most promising for highly stochastic, non-stationary environments requiring continuous adaptation and self-optimization. This progression, however, is accompanied by increasing data dependency, computational overhead, convergence and stability challenges, and reduced interpretability. Therefore, technique selection should be based on requirement–environment matching rather than algorithmic sophistication: the appropriate hybrid is the one that provides sufficient optimization capability and adaptability while remaining computationally feasible under the latency, energy, resource, scalability, and constraint requirements of the target environment. Table 13 summarizes this framework and provides guidance for selecting suitable hybrid-based scheduling algorithms according to requirements, environment characteristics and optimization objectives.

**Table 13: Decision framework for choosing learning-based scheduling algorithms based on requirements and application contexts**

| Requirement | Recommended Family | Preferred Techniques | Best Environment | Avoid When |
|---|---|---|---|---|
| Minimize makespan with low scheduling overhead | Heuristic–Heuristic | HMMRR, HAMM | Cloud, relatively stable fog | Highly dynamic workloads requiring continuous adaptation |
| Maximize resource utilization | Heuristic–Metaheuristic | HEPGA, EDLB, HETSVP | Cloud, Fog | Resource information is incomplete or highly uncertain |
| Balance exploration and exploitation | Metaheuristic–Metaheuristic | GA–PSO, GA–GWO, APSGA, DE–GWO | Cloud, Fog–Cloud | Strict real-time scheduling with limited computational resources |
| Minimize energy consumption | Metaheuristic–Heuristic / Metaheuristic–Metaheuristic | PSO–RR, DE–GWO, GA–GWO, HETSVP | Fog, Fog–Cloud, Edge | Energy is not a primary optimization objective |
| Reduce execution/response time | Heuristic–Heuristic / Heuristic–Learning | HMMRR, Random RR + K-Means, DQ-HEFT | Cloud, Fog, Workflow systems | Long-term workload adaptation is the dominant requirement |
| Meet strict deadlines | Heuristic–Learning / Learning–Optimization | FODAS, FRL, ETFC | Fog, Fog–Cloud, real-time IoT | Workloads are static and deadline constraints are weak |
| Handle dynamic workloads | Learning–Optimization | DQ-HEFT, FODAS, GA-DQN | Cloud, Fog–Cloud | Training data or computational resources are insufficient |
| Adapt to highly dynamic/stochastic environments | Learning–Learning | A3C-R2N2, DRL–LSTM | Edge–Cloud, highly dynamic environments | Resource-constrained edge devices with limited training/inference capacity |
| Improve load balancing | Learning–Metaheuristic | EDLB (CNN–MPSO) | Fog | Global multi-objective optimization is required |
| Optimize cost and makespan jointly | Metaheuristic–Metaheuristic / Metaheuristic–Learning | GA–GWO, HLWOA, FTTHDRL | Cloud | Extremely low-latency scheduling is required |
| Optimize energy and cost jointly | Heuristic–Metaheuristic / Learning–Optimization | PSO–RR, ETFC | Fog, Fog–Cloud | Large-scale exploration is unnecessary |
| Handle workflow dependencies | Heuristic–Metaheuristic | HEFT-based hybrids, HEPGA, HLWOA | Cloud, Workflow systems | Tasks are independent and workflow structure is absent |
| Optimize workflow makespan and execution time | Heuristic–Learning | DQ-HEFT | Cloud, Scientific workflows | Training overhead cannot be tolerated |
| Optimize workflow energy and makespan | Metaheuristic–Metaheuristic | DE–GWO, PSO–SA | Fog–Cloud | Very limited computational resources |
| Handle heterogeneous resources | Heuristic–Learning / Heuristic–Metaheuristic | DQ-HEFT, HEPGA, HETSVP | Cloud, Fog, Fog–Cloud | Resource characteristics are homogeneous and predictable |
| Optimize multiple conflicting objectives | Metaheuristic–Metaheuristic | GA–GWO, HTSA, DE–GWO, APSGA | Cloud, Fog–Cloud | Real-time decisions must be generated with minimal overhead |
| Real-time delay-sensitive scheduling | Lightweight Heuristic Hybrid | Random RR + K-Means, HMMRR | Fog, Edge, IoT | High-quality global optimization is more important than response speed |
| Optimize edge resource utilization and reliability | Multi-objective Metaheuristic Hybrid | CESMA (NSGA-II + MBO) | Edge–Cloud | Edge devices cannot support iterative population-based optimization |
| Task offloading and resource allocation | Deep Hybrid Learning–Optimization | REACT | Edge–Cloud, ultra-dense environments | Training/inference resources are severely constrained |
| Continuous self-optimization and workload prediction | Learning–Learning | A3C-R2N2, DRL–LSTM | Dynamic Fog, Edge–Cloud | Limited historical/streaming data or strict interpretability requirements |
| Minimize algorithmic complexity | Lightweight Heuristic Hybrid | HMMRR, Random RR + K-Means, PSO–RR | Fog, Edge | Complex multi-objective optimization is required |
| Maximize adaptability under workload uncertainty | Learning–Optimization | FODAS, REACT, A3C-R2N2 | Fog–Cloud, Edge–Cloud | Workloads are predictable and static |
| QoS-aware scheduling | Heuristic–Heuristic / Learning–Optimization | HMMRR, HLWOA, FODAS | Cloud, Fog–Cloud | QoS constraints are absent |
| Deadline- and energy-aware scheduling | Learning–Optimization | FRL, ETFC, FODAS | Fog, Fog–Cloud | Computational overhead must be extremely low |
| Large-scale multi-objective scheduling | Metaheuristic–Metaheuristic | GA–GWO, DE–GWO, HTSA | Cloud, Fog–Cloud | Small-scale or highly resource-constrained edge environments |

## 7. Simulation Platforms for Scheduling Evaluation: Comparative Analysis, Research Trends, and Open Challenges

Simulation platforms play a fundamental role in evaluating scheduling strategies across cloud, fog, and edge computing environments. However, simulation tools should not be viewed merely as implementation environments, since their underlying architectural assumptions directly influence the type of scheduling problems that can be modeled and the validity of experimental outcomes. Consequently, simulator selection represents a methodological decision that can significantly affect performance evaluation, reproducibility, and the generalizability of scheduling results.

Existing simulation platforms can be broadly categorized into three groups according to their primary modeling objectives. The first group comprises cloud-centric simulators, including CloudSim, WorkflowSim, DE3C, and COSCO, which primarily focus on virtualized resource management, workflow execution, energy-aware scheduling, and resource provisioning. Owing to their extensibility and mature ecosystems, these platforms remain the dominant evaluation environments for cloud scheduling research. Nevertheless, their centralized architectural assumptions limit their ability to accurately capture network dynamics, device mobility, and latency-sensitive behaviors commonly observed in fog and edge environments.

The second group consists of fog- and edge-oriented simulators, such as iFogSim, FogWorkflowSim, and FogBus2, which extend cloud-based models by incorporating hierarchical architectures composed of cloud servers, fog nodes, and edge devices. These platforms facilitate the evaluation of latency-aware task placement, workflow orchestration, and IoT-driven scheduling scenarios. Although they provide a more realistic representation of distributed infrastructures, many of these simulators inherit architectural constraints from their cloud-centric predecessors and may not adequately capture large-scale heterogeneity, dynamic mobility, or highly volatile network conditions.

The third category includes network- and mobility-oriented simulators such as OMNeT++, Mininet, and SUMO. Unlike cloud and fog simulators, these platforms emphasize communication behavior, mobility patterns, software-defined networking (SDN), and vehicular edge computing scenarios. Consequently, they are particularly suitable for evaluating network-aware scheduling, task offloading strategies, and communication-delay-sensitive applications. However, their limited support for workflow execution, resource allocation policies, and application-level scheduling semantics often necessitates integration with complementary simulation frameworks.

General-purpose modeling frameworks, including SimPy and MATLAB, provide substantial flexibility for designing custom scheduling experiments and optimization studies. While these platforms enable researchers to rapidly prototype novel scheduling mechanisms, the absence of standardized infrastructure models often limits experimental comparability and reproducibility across studies. Furthermore, programming languages such as Java, Python, and C++ should be regarded as implementation environments rather than simulation platforms.

A notable observation emerging from the reviewed literature is that the evolution of simulation platforms closely mirrors the evolution of scheduling research itself. Early studies predominantly relied on CloudSim-based environments because scheduling objectives were largely centered on makespan minimization, resource utilization, and cost optimization within centralized cloud infrastructures. However, the rapid proliferation of IoT applications, cyber-physical systems, and latency-sensitive services has progressively shifted research attention toward fog and edge computing. Consequently, simulation platforms have evolved from cloud-centric resource management tools into multi-layer environments capable of modeling latency, mobility, distributed execution, and heterogeneous resource coordination. More recently, the increasing adoption of artificial intelligence, reinforcement learning, and adaptive scheduling frameworks has created demand for simulation environments capable of supporting dynamic decision-making processes and continuously changing workloads.

The comparative analysis presented in Tables 14 and 15 further reveals that no existing simulator comprehensively supports all scheduling objectives and evaluation requirements. CloudSim-based frameworks remain highly effective for evaluating resource allocation, workflow scheduling, and energy-aware optimization, whereas fog-oriented simulators provide superior support for latency-sensitive and IoT-driven applications. Similarly, network simulators are indispensable for communication-aware scheduling studies, while general-purpose platforms offer flexibility for custom algorithm development. Consequently, the suitability of a simulation platform should be determined by the target computing paradigm, scheduling objectives, and evaluation metrics rather than by tool popularity or familiarity alone.

Another important trend identified in the literature is the growing adoption of hybrid simulation environments. Increasingly complex cloud–fog–edge ecosystems have motivated researchers to combine multiple platforms, such as CloudSim with OMNeT++ or iFogSim with SUMO, to simultaneously model computational, networking, and mobility characteristics. This trend reflects a broader transition from isolated infrastructure evaluation toward holistic modeling of distributed computing ecosystems.

Despite these advances, several significant challenges remain unresolved. First, the lack of standardization across simulation platforms continues to hinder fair comparison among scheduling algorithms. Differences in workload generation, infrastructure configuration, performance metrics, and experimental assumptions often produce inconsistent results across studies.

**Table 14: Comparison of simulation tools**

| Simulator / Library | Programming Language | Computing Environment | User-Friendliness | Complexity | Flexibility | Scalability | Typical Scheduling Use Case |
|---|---|---|---|---|---|---|---|
| **CloudSim** | Java | Cloud | High | Medium | High (extensible) | High | VM allocation, task scheduling, resource provisioning, energy-aware scheduling |
| **WorkflowSim** | Java | Cloud | Medium | Medium | High | Medium-High | DAG/workflow scheduling, fault-tolerant workflow execution, resource provisioning |
| **DE3C** | Java | Cloud | Medium | High | Medium | Medium | Energy-efficient task scheduling, distributed workload balancing |
| **COSCO** | Java | Cloud | Medium | High | Medium | Medium | Coordinated multi-cloud/federated scheduling, workload migration |
| **iFogSim** | Java | All | Medium | Medium | High | Medium | Latency-aware task placement, IoT application scheduling, energy-efficient fog scheduling |
| **FogWorkflowSim** | Java | All | Medium | Medium-High | High | Medium | Workflow scheduling across heterogeneous fog and edge nodes, DAG-based task placement |
| **FogBus2** | Python / Java | All | Medium | Medium | High | Medium | End-to-end IoT–fog–cloud orchestration, task distribution, latency, and resource-aware scheduling |
| **OMNeT++** | C++ | Network simulation for fog and edge | Medium | High | High | Medium-High | Network-aware scheduling, offloading evaluation, and communication-delay modeling |
| **Mininet** | Python | Fog, and edge | High | Medium | Medium | Medium | SDN-based network-aware task scheduling, real-time offloading validation |
| **SUMO** | C++ / Python | Edge (Vehicular) | Medium | Medium | Medium | High | Mobility-aware scheduling, vehicular edge computing task placement, and dynamic workload simulation |
| **SimPy** | Python | All (custom models) | High | Medium | Very High | Medium | Custom scheduling algorithm testing, queuing, and resource contention modeling |
| **MATLAB** | MATLAB | All (abstract models) | Medium | Medium | High | Medium | Analytical, optimization-based scheduling, control-strategy evaluation, and hybrid simulation |

Second, most simulators rely on simplified workload, network, and mobility models that inadequately represent the complexity of real-world deployments, particularly in highly dynamic edge environments. This limitation contributes to a persistent simulation-to-reality gap, whereby algorithms demonstrating promising simulation performance frequently exhibit degraded effectiveness in practical deployments.

A third challenge concerns reproducibility. Many scheduling studies employ customized simulator extensions, proprietary workloads, and undocumented experimental settings, making independent validation difficult. The absence of standardized benchmark datasets and publicly available evaluation frameworks further exacerbates this issue. Consequently, improving reproducibility has emerged as an important research priority for the scheduling community.

**Table 15: Summary of support simulation tools for evaluating performance parameters**

| Simulator / Library | Makespan | Execution Time | Cost | Resource Utilization | Energy Consumption | Response Time | Throughput | Delay | Reliability | Security |
|---|---|---|---|---|---|---|---|---|---|---|
| **CloudSim** | ✓ | ✓ | ✓ | ✓ | ✓ | ✓ | ✓ | ✕ | ✕ | ✕ |
| **WorkflowSim** | ✓ | ✓ | ✓ | ✓ | ✓ | ✓ | ✓ | ✕ | ✓ | ✕ |
| **DE3C** | ✓ | ✓ | ✓ | ✓ | ✓ | ✓ | ✓ | ✕ | ✓ | ✕ |
| **COSCO** | ✓ | ✓ | ✓ | ✓ | ✓ | ✓ | ✓ | ✕ | ✓ | ✕ |
| **iFogSim** | ✓ | ✓ | ✕ | ✓ | ✓ | ✓ | ✓ | ✓ | ✕ | ✕ |
| **FogWorkflowSim** | ✓ | ✓ | ✕ | ✓ | ✓ | ✓ | ✓ | ✓ | ✓ | ✕ |
| **FogBus2** | ✓ | ✓ | ✓ | ✓ | ✓ | ✓ | ✓ | ✓ | ✓ | ✕ |
| **OMNeT++** | ✕ | ✕ | ✕ | ✕ | ✕ | ✕ | ✕ | ✓ | ✕ | ✕ |
| **Mininet** | ✕ | ✕ | ✕ | ✕ | ✕ | ✓ | ✕ | ✓ | ✕ | ✕ |
| **SUMO** | ✕ | ✕ | ✕ | ✕ | ✕ | ✓ | ✕ | ✓ | ✕ | ✕ |
| **SimPy** | ✓ | ✓ | ✓ | ✓ | ✓ | ✓ | ✓ | ✕ | ✕ | ✕ |
| **MATLAB** | ✓ | ✓ | ✓ | ✓ | ✓ | ✓ | ✓ | ✕ | ✕ | ✕ |

Finally, existing simulation platforms provide limited native support for emerging scheduling paradigms, including reinforcement learning, multi-agent optimization, digital-twin-enabled scheduling, and sustainable resource management. As algorithmic innovation continues to accelerate, simulation capabilities are increasingly becoming a bottleneck for experimental validation. Future research should therefore focus on developing integrated and intelligent simulation ecosystems that combine realistic infrastructure modeling, standardized benchmarking, AI-native capabilities, and support for large-scale cloud–fog–edge continuum environments. Such developments will be essential for improving the reliability, reproducibility, and practical relevance of future scheduling research.

## 8. Discussion

The comparative analysis indicates that heuristic, metaheuristic, learning-based, and hybrid scheduling techniques involve distinct trade-offs among computational efficiency, optimization quality, adaptability, and implementation complexity**.** Heuristic methods are suitable for stable workloads and latency-sensitive environments requiring rapid decisions with low overhead, whereas metaheuristics are more appropriate for complex, NP-hard, and multi-objective problems where solution quality is prioritized over computational cost. Learning-based approaches offer greater adaptability to dynamic and uncertain workloads but require adequate training data, computational resources, and mechanisms to address convergence and interpretability challenges. Hybrid techniques provide broader capabilities by combining complementary optimization and learning mechanisms, although this often increases implementation complexity and computational overhead. Overall, technique selection should follow a requirement–environment matching principle rather than an algorithm-ranking principle: heuristics are preferable for simplicity and rapid decisions, metaheuristics for complex global optimization, learning-based methods for dynamic and adaptive scheduling, and hybrid approaches when multiple competing requirements must be addressed simultaneously. In particular, cloud environments can accommodate computationally intensive techniques more readily, while fog and edge environments require greater emphasis on latency, energy efficiency, and computational feasibility; heterogeneous cloud–fog–edge systems therefore benefit from hybrid approaches that jointly consider optimization, adaptation, offloading, and resource constraints. Based on the comparative analysis of heuristic, metaheuristic, learning-based, and hybrid scheduling techniques, Table 16 summarizes a decision framework for selecting an appropriate scheduling technique family according to application requirements. The framework considers key factors, including optimization objectives, workload dynamics, computational constraints, adaptability requirements, and the characteristics of the target computing environment. It therefore provides practical guidance for matching scheduling requirements with the strengths and limitations of each technique family.

**Table 16: Decision framework for selecting scheduling technique families according to application requirements**

| Requirement / Application Context | Recommended Technique | Rationale | Preferred Environment | Main Trade-off |
|---|---|---|---|---|
| Very low scheduling overhead | Heuristic | Fast execution and simple decision rules | Cloud, Fog, Edge | Lower optimization quality |
| Stable and predictable workload | Heuristic | Does not require training and provides predictable decisions | Cloud, Fog | Limited adaptability |
| Deadline-sensitive or real-time scheduling | Heuristic / Lightweight Hybrid | Rapid decisions and explicit priority/deadline handling | Fog, Edge | May sacrifice global optimality |
| Large number of tasks/resources | Heuristic | Low computational complexity and good scalability | Cloud | May produce less optimal allocations |
| Complex NP-hard scheduling problem | Metaheuristic | Broader search capability for large solution spaces | Cloud, Fog | Higher computational cost |
| Multi-objective optimization | Metaheuristic | Effective exploration of competing objectives | Cloud, Fog | Parameter tuning and convergence cost |
| Strong global optimization requirement | Metaheuristic | Better exploration of large search spaces | Cloud | Longer scheduling time |
| Heterogeneous resources | Metaheuristic / Hybrid | Can jointly optimize resource-related objectives | Cloud, Fog, Edge | Increased complexity |
| Historical workload data available | Learning-based | Learns relationships between workload/resource states and scheduling decisions | Cloud, Fog, Edge | Training and generalization requirements |
| Complex nonlinear workload patterns | Deep Learning | Extracts nonlinear patterns from historical data | Cloud, Fog | High training cost |
| Dynamic and uncertain workload | RL / DRL | Learns scheduling policies through interaction with the environment | Fog, Edge, Cloud | Training and convergence overhead |
| Continuous adaptation required | Learning-based / Hybrid | Can adapt decisions to changing system states | Fog, Edge | Computational and implementation complexity |
| Global optimization + workload adaptation | Metaheuristic + Learning | Combines global search with adaptive decision-making | Cloud, Fog, Edge | High complexity |
| Strong exploration–exploitation requirement | Metaheuristic + Metaheuristic | Combines complementary search mechanisms | Cloud, Fog | Higher computational overhead |
| Low overhead + improved optimization | Heuristic + Metaheuristic | Preserves heuristic efficiency while improving search | Cloud, Fog | More complex than standalone heuristic |
| Highly stochastic/non-stationary environment | Learning + Learning | Supports continuous learning and adaptation | Fog, Edge | High data/computational requirements |
| Multiple conflicting objectives and constraints | Hybrid | Integrates complementary optimization capabilities | Cloud–Fog–Edge | Highest design complexity |
| Resource-constrained edge devices | Lightweight Heuristic / Lightweight Hybrid | Minimizes inference and scheduling overhead | Edge | Limited optimization/adaptation |
| Large-scale cloud with abundant resources | Metaheuristic / Learning / Hybrid | Greater computational budget allows sophisticated optimization | Cloud | Higher implementation cost |
| Cross-layer cloud–fog–edge scheduling | Hybrid | Coordinates heterogeneous resources and competing objectives | Cloud–Fog–Edge | High coordination complexity |

### 8.1 Future Research Directions

The comprehensive analysis of scheduling research across cloud, fog, and edge environments indicates a clear transition from static and single-objective approaches toward adaptive, multi-objective, heterogeneous, and intelligent scheduling frameworks. Nevertheless, several challenges remain unresolved, particularly regarding workload dynamics, scalability, conflicting objectives, and the increasing complexity of hybrid approaches. Future research should therefore investigate adaptive and continual scheduling mechanisms capable of responding to workload drift and changing resource conditions without requiring complete model retraining. Reinforcement learning, online learning, and predictive models can support such adaptability, but their evaluation should consider adaptation speed, data requirements, generalization to unseen workloads, and computational overhead. At the same time, scalable scheduling methods should explicitly examine execution time, memory consumption, communication overhead, convergence behavior, and scheduling quality as system size increases, with particular attention to distributed, parallel, and hierarchical implementations of metaheuristic and hybrid algorithms.

Another important direction concerns the development of multi-objective, constraint-aware, and principled hybrid scheduling frameworks. Future schedulers should jointly consider conflicting objectives, including makespan, execution time, energy consumption, cost, resource utilization, latency, QoS, and reliability, while accounting for constraints such as deadlines, budgets, task dependencies, fault tolerance, and SLA requirements. Adaptive objective weighting and Pareto-based optimization could enable scheduling decisions to reflect changing application and system requirements. Moreover, hybrid approaches should move beyond combining techniques solely to improve performance and instead provide explicit justification for the roles of exploration, exploitation, prediction, and adaptation within each component. In particular, metaheuristic–learning combinations should quantify the trade-off between optimization gains and training or inference overhead. For resource-constrained edge environments, lightweight intelligence based on model compression, knowledge distillation, decentralized learning, and lightweight reinforcement learning is also required to balance decision quality, latency, memory, energy consumption, and adaptability.

Finally, future research should strengthen the practical validity, reliability, security, interpretability, and autonomy of cloud–fog–edge scheduling. The strong dependence on simulation highlights the need for real-world testbeds, realistic workloads, and simulation–experimental validation pipelines that incorporate resource contention, network variability, mobility, failures, and communication delays. Standardized benchmarks, workload repositories, infrastructure configurations, baseline algorithms, and reproducible evaluation protocols are also necessary to enable fair comparisons and distinguish genuine algorithmic improvements from favorable experimental settings. In addition, scheduling research should increasingly integrate reliability, fault tolerance, security, privacy, explainability, fairness, robustness, and generalization into optimization objectives and constraints. Decentralized and collaborative approaches, including multi-agent reinforcement learning, federated learning, and decentralized optimization, can further reduce dependence on centralized controllers. Ultimately, these developments point toward autonomous and context-aware scheduling systems capable of monitoring system conditions, characterizing workloads, predicting future states, selecting suitable scheduling strategies, and dynamically adjusting optimization objectives, potentially through meta-scheduling mechanisms that select heuristic, metaheuristic, learning-based, or hybrid techniques according to workload characteristics and system conditions.

## 9. Conclusion

This review presented a comparative analysis of scheduling techniques across the cloud–fog–edge computing continuum, with particular emphasis on understanding the conditions under which different technique families are appropriate. By jointly examining task, job, and workflow scheduling, and analyzing heuristic, metaheuristic, learning-based, and hybrid approaches, the study provides a unified perspective on the relationships among scheduling requirements, computing environments, optimization objectives, constraints, and algorithmic characteristics. The analysis shows that no single technique family is universally superior. Heuristic methods are particularly appropriate when low computational overhead, rapid decision-making, and relatively stable workloads are required, whereas metaheuristic approaches are more suitable for complex and multi-objective optimization problems where solution quality is prioritized. Learning-based methods offer greater adaptability to dynamic and uncertain environments, but their applicability depends on training requirements, computational resources, convergence, and model reliability. Hybrid approaches can combine complementary optimization and adaptation capabilities and are therefore promising for heterogeneous and highly constrained environments, although this benefit is accompanied by increased computational and implementation complexity. These findings support a requirement–environment matching perspective in which scheduling techniques are selected according to the characteristics of the scheduling problem rather than ranked solely by reported performance improvements.

The principal contribution of this review is therefore the development of a requirement-oriented decision framework that translates the comparative evidence into practical guidance for technique selection. The framework considers scheduling granularity, workload dynamics, computing environment, optimization objectives, resource and user constraints, adaptability requirements, and computational budget, thereby helping researchers and practitioners identify suitable technique families for specific scheduling scenarios. The review also highlights several unresolved challenges, including the limited real-world validation of proposed methods, the need for scalable and reliable scheduling under dynamic workloads, the difficulty of balancing latency, energy, cost, QoS, and other conflicting objectives, and the computational and interpretability challenges associated with learning-based and hybrid approaches. Future research should consequently move beyond isolated algorithmic improvements toward standardized benchmarking, realistic experimental validation, adaptive and resource-aware scheduling, and integrated frameworks capable of operating across the cloud–fog–edge continuum. In this context, the proposed decision framework provides a structured basis for guiding future research and supporting more transparent and context-aware selection of scheduling techniques.